\documentclass[preprint,article,accept,moreauthors,pdftex,algorithms]
{Definitions/mdpi}

\setitemize{parsep=6pt,itemsep=0pt,leftmargin=*,labelsep=5.5mm}
\setenumerate{parsep=6pt,itemsep=0pt,leftmargin=*,labelsep=5.5mm}
\setlist[description]{itemsep=0mm}   

\usepackage{nicefrac}

\usepackage{ifthen}
\usepackage{comment}

\usepackage[nameinlink,capitalise]{cleveref}
\usepackage{amsmath,amsfonts,amssymb}
\usepackage{mathtools}
\usepackage{mathrsfs}
\usepackage{wrapfig}
\usepackage{xspace}

\newcommand{\mc}[1]{{\mathcal{#1}}}

\newcommand{\eps}{{\varepsilon}}

\DeclareMathOperator{\cost}{cost}

\DeclareMathOperator{\diam}{diam}

\newcommand{\OPT}{\ensuremath\textsc{Opt}}
\newcommand{\Opt}{\textsc{Opt}}
\newcommand{\cU}{\mathcal{U}}

\newcommand{\hy}{\hbox{-}\nobreak\hskip0pt}

\usepackage{tikz}
\usetikzlibrary{calc, arrows.meta, intersections}

\newcommand{\pname}[1]{\textsc{#1}}

\newcommand{\SAT}{\text{CNF-SAT}\xspace}

\newcommand{\W}{\text{W}}
\newcommand{\FPT}{\text{FPT}}
\newcommand{\Wone}{\mbox{$\W[1]\neq \FPT$}\xspace}

\newcommand{\N}{\mathbb{N}}
\newcommand{\Z}{\mathbb{Z}}

\newcommand{\hpacking}{\textsc{$H$-Packing}\xspace}
\newcommand{\ksp}{\textsc{$k$-Set Packing}\xspace}

\newcommand{\kmeans}{\textsc{$k$-Means}\xspace}
\newcommand{\kmedian}{\textsc{$k$-Median}\xspace}
\newcommand{\kcenter}{\textsc{$k$-Center}\xspace}
\newcommand{\ckmedian}{\textsc{Capacitated $k$-Median}\xspace}

\newcommand{\ekmedian}{\textsc{Euclidean $k$-Median}\xspace}
\newcommand{\ekmeans}{\textsc{Euclidean $k$-Means}\xspace}

\newcommand{\fdeletion}{\textsc{$\mathcal{F}$-Deletion}\xspace}
\newcommand{\wfdeletion}{\textsc{Weighted $\mathcal{F}$-Deletion}\xspace}
\newcommand{\tkd}{\textsc{Treewidth $k$-Deletion}\xspace}
\newcommand{\wtkd}{\textsc{Weighted Treewidth $k$-Deletion}\xspace}
\newcommand{\phd}{\textsc{Planar $\mathcal{H}$-Deletion}\xspace}
\newcommand{\wphd}{\textsc{Weighted Planar $\mathcal{H}$-Deletion}\xspace}
\newcommand{\planarization}{\textsc{Minimum Planarization}\xspace}

\newcommand{\perfectdeletion}{\textsc{Perfect Deletion}\xspace}
\newcommand{\chordaldeletion}{\textsc{Chordal Deletion}\xspace}

\newcommand{\hmd}{\textsc{$\mathcal{H}$-Minor Deletion}\xspace}
\newcommand{\hsd}{\textsc{$H$-Subgraph Deletion}\xspace}

\newcommand{\fvs}{\textsc{Feedback Vertex Set}\xspace}
\newcommand{\wfvs}{\textsc{Weighted Feedback Vertex Set}\xspace}
\newcommand{\dfvs}{\textsc{Directed Feedback Vertex Set}\xspace}

\newcommand{\oct}{\textsc{Odd Cycle Transversal}\xspace}
\newcommand{\doct}{\textsc{Directed Odd Cycle Transversal}\xspace}

\newcommand{\kvs}{\textsc{$k$-Vertex Separator}\xspace}
\newcommand{\wkvs}{\textsc{(Weighted) $k$-Vertex Separator}\xspace}
\newcommand{\balsep}{\textsc{Balanced Seprator}\xspace}
\newcommand{\ksvs}{\textsc{$k$-Subset Vertex Separator}\xspace}

\newcommand{\stcut}{\textsc{$s$-$t$ Cut}\xspace}

\newcommand{\umc}{\textsc{Undirected Multicut}\xspace}
\newcommand{\umwc}{\textsc{Undirected Multiway cut}\xspace}
\newcommand{\dmc}{\textsc{Directed Multicut}\xspace}
\newcommand{\dmwc}{\textsc{Directed Multiway cut}\xspace}
\newcommand{\kcut}{\textsc{$k$-Cut}\xspace}

\newcommand{\meb}{\textsc{Minimum Edge Bisection}\xspace}
\newcommand{\mvb}{\textsc{Minimum Vertex Bisection}\xspace}
\newcommand{\bes}{\textsc{Balanced Edge Separator}\xspace}
\newcommand{\bvs}{\textsc{Balanced Vertex Separator}\xspace}

\newcommand{\is}{\textsc{Independent Set}\xspace}

\newcommand{\vc}{\textsc{Vertex Cover}\xspace}
\newcommand{\wvc}{\textsc{Weighted Vertex Cover}\xspace}
\newcommand{\setcover}{\textsc{Set Cover}\xspace}

\newcommand{\si}{\textsc{Subgraph Isomorphism}\xspace}

\newcommand{\etsp}{\textsc{Euclidean Traveling Salesman}\xspace}

\newcommand{\tw}{\mathbf{tw}}
\newcommand{\genus}{\mathbf{genus}}

\newcommand{\calc}{\mathcal{C}}
\newcommand{\calf}{\mathcal{F}}
\newcommand{\calh}{\mathcal{H}}
\newcommand{\calP}{\mathcal{P}}

\newcommand{\bbn}{\mathbb{N}}
\newcommand{\bbr}{\mathbb{R}}
\newcommand{\ptas}{PTAS\xspace}
\newcommand{\epas}{EPAS\xspace}
\newcommand{\fptas}{EPAS\xspace}
\newcommand{\pas}{PAS\xspace}

\newcommand{\poly}{\text{poly}}
\newcommand{\polylog}{\text{polylog}}
\renewcommand{\tilde}{\widetilde}

\newcommand{\F}{\mathbb{F}}

\newcommand{\polyn}{n^{O(1)}}

\newcommand{\gapeth}{\ensuremath{\operatorname{\mathsf{Gap-ETH}}}\xspace}
\renewcommand{\eth}{\ensuremath{\operatorname{\mathsf{ETH}}}\xspace}

\newcommand{\NPcoNP}{NP$\subseteq$coNP/poly}
\newcommand{\E}{\mathbb{E}}

\usepackage{xparse}
\DeclareDocumentCommand\mypar{om}
 {%
  \IfValueTF{#1}
   {\paragraph[#1]{#2}}
   {\vspace{0.3cm}\noindent\textbf{#2.}\ }%
 }
\firstpage{1} 
\pubvolume{xx}
\issuenum{1}
\articlenumber{5}
\pubyear{2020}
\copyrightyear{2020}
\history{Received: 1 October 2019; Accepted: 2 June 2020} %
\updates{yes} %

\pdfoutput=1

\Title{A Survey on Approximation in Parameterized Complexity: Hardness and Algorithms}
\Author{Andreas Emil Feldmann $^{1,*}$, Karthik C. S. $^{2,*}$, Euiwoong Lee 
$^{3,*}$ and Pasin Manurangsi $^{4,*}$}
\AuthorNames{Andreas Emil Feldmann, Karthik C. S., Euiwoong Lee, and Pasin 
Manurangsi}

\address{%
$^{1}$ \quad Department of Applied Mathematics (KAM), Charles University, 
118~00 Prague, Czechia\\
$^{2}$ \quad Department of Computer Science, Tel Aviv University, 6997801 
Tel Aviv, Israel  \\
$^{3}$ \quad Courant Institute of Mathematical Sciences, New York 
University, 10012 New York, USA \\
$^{4}$ \quad Google Research, 94043 Mountain View, USA}
\corres{Correspondence: feldmann.a.e@gmail.com (A.E.F.); 
karthik0112358@gmail.com (K.C.S.); euiwoong@cims.nyu.edu (E.L.) and 
pasin@google.com (P.M.)}
\abstract{Parameterization and approximation are two popular ways of coping with NP-hard problems. More recently, the two have also been combined to derive many interesting results. We~survey developments in the area both from the algorithmic and hardness perspectives, with~emphasis on new techniques and potential future research directions.}


\begin{document}

\section{Introduction}\label{sec:intro}

In their seminal papers of the mid 1960s, \citet{cobham1964intrinsic} and 
\citet{edmonds1965paths} independently phrased what is now known as the 
Cobham--Edmonds thesis. It states that an optimization problem is {feasibly 
solvable} if it admits an algorithm with the following two properties:

\begin{enumerate}
 \item {Accuracy:} the algorithm should always compute the best possible 
(optimum) solution.
\item {Efficiency:} the runtime of the algorithm should be polynomial in 
the input size $n$.
\end{enumerate}

Shortly after the Cobham--Edmonds thesis was formulated, the development of the 
theory of NP\hy{}hardness and reducibility identified a whole plethora of 
problems that are seemingly intractable, i.e., for which algorithms with the 
above two properties do not seem to exist. Even though the reasons for this 
phenomenon remain elusive up to this day, this has not hindered the development 
of algorithms for such problems. To obtain an algorithm for an NP-hard problem, 
at least one of the two properties demanded by the Cobham--Edmonds thesis needs 
to be relaxed. Ideally, the properties are relaxed as little as possible, in 
order to stay close to the notion of feasible solvability suggested by the~thesis.

A very common approach is to relax the accuracy condition, which means aiming 
for {approximation algorithms}~\cite{Vazirani01book,williamson2011design}. 
The idea here is to use only polynomial time to compute an 
{$\alpha$-approximation}, i.e., a solution that is at most a factor 
$\alpha$ times worse than the optimum solution obtainable for the given input 
instance. Such an algorithm may also be randomized, i.e., there is either a high 
probability that the output is an $\alpha$-approximation, or the runtime is 
polynomial in expectation.

In a different direction, several relaxations of the efficiency condition have  
also been proposed. Popular among these is the notion of {parameterized algorithms}
~\cite{downey2013fpt,pc-book}. Here the input comes together with some parameter 
$k\in\mathbb{N}$, which describes some property of the input and can be expected 
to be small in typical applications. The idea is to isolate the seemingly 
necessary exponential runtime of NP-hard problems to the parameter, while the 
runtime dependence on the input size~$n$ remains polynomial. In particular, the 
algorithm should compute the optimum solution in $f(k)\polyn$ time, for~some 
computable function $f:\mathbb{N}\to\mathbb{N}$ independent of the input size 
$n$. If such an algorithm exists for a problem it is {fixed-parameter 
tractable (FPT)}, and the algorithm is correspondingly referred to as an 
{FPT algorithm}. 

Approximation and FPT algorithms have been studied extensively for the past few 
decades, and~this has lead to a rich literature on algorithmic techniques and 
deep links to other research fields within mathematics. However, in this process 
the limitations of these approaches have also become apparent. Some NP-hard 
problems can fairly be considered to be feasibly solvable in the respective 
regimes, as they admit polynomial-time algorithms with rather small 
approximation factors, or can be shown to be solvable optimally with only a 
fairly small exponential runtime overhead due to the parameter. However, many 
problems can also be shown not to admit any reasonable algorithms in either of 
these regimes, assuming some standard complexity assumptions. Thus considering 
only approximation and FPT algorithms, as has been mostly done in the past, we 
are seemingly stuck in a swamp of problems for which we have substantial evidence
that they cannot be feasibly solved.

To find a way out of this dilemma, an obvious possibility is to lift both the 
accuracy and the efficiency requirements of the Cobham--Edmonds thesis. In this 
way, we obtain a {parameterized $\alpha$-approximation algorithm}, which 
computes an $\alpha$-approximation in $f(k)\polyn$ time for some computable 
function~$f$, given an input of size $n$ with parameter $k$. 
The study of such algorithms had been suggested dating back to the early days of parameterized complexity (cf.~\cite{cai1997fixed, marx2008fpa, downey2013fpt, flum2006parameterized}), and we refer
the readers to an excellent survey of Marx~\cite{marx2008fpa} for discussions
on the earlier results in the area.

Recently this approach has received some increased interest, with many new results 
obtained in the past few years, both in terms of algorithms and hardness of approximation.
The aim of this survey is to give an overview on some of these newer results.
We would like to caution the readers that the goal of this survey is not to
compile all known results in the field but rather to give examples that
demonstrate the flavor of questions studied, techniques used to obtain them, 
and some potential future research directions. Finally, we remark that on the broader theme of approximation in P, there~was an excellent survey recently made available by Rubinstein and Williams~\cite{RW19} focusing on the approximability of popular problems in P which admit simple quadratic/cubic algorithms. 

\subsection*{Organization of the Survey}

The main body of the survey is organized into two sections: one on FPT hardness of approximation (Section~\ref{sec:inapprox})
and the other on FPT approximation algorithms (Section~\ref{sec:algo}). Before these 
two main sections, we list some notations and preliminaries in 
Section~\ref{sec:prelim}. Finally, in Section~\ref{sec:open}, we highlight some open questions 
and future directions (although a large list of open problems have also been detailed throughout Sections~\ref{sec:inapprox}~and~\ref{sec:algo}).

\section{Preliminaries}\label{sec:prelim}

In this section, we review several notions that will appear regularly throughout 
the survey. However, we do not include  definitions of basic concepts such as  
W-hardness, para-NP-hardness, APX-hardness, and so forth;  the interested reader 
may refer to~\cite{downey2013fpt,pc-book,williamson2011design,Vazirani01book} 
for these definitions.  

\mypar{Parameterized approximation algorithms}
We briefly specify the different types of algorithms we will consider. 
As already defined in the introduction, an FPT algorithm computes the optimum 
solution in $f(k)\polyn$ time for some parameter $k$ and computable function 
$f:\mathbb{N}\to\mathbb{N}$ on inputs of size~$n$. The common choices of 
parameters are the standard parameters based on solution size, structural 
parameters, guarantee parameters, and dual parameters.%

An algorithm that computes 
the optimum solution in $f(k)n^{g(k)}$ time for some parameter $k$ and 
computable functions $f,g:\mathbb{N}\to\mathbb{N}$, is called a {slice-wise 
polynomial (XP) algorithm}. If the parameter is the approximation factor, i.e., 
the algorithm computes a $(1+\eps)$-approximation in $f(\eps)n^{g(\eps)}$ time, 
then it is called a {polynomial-time approximation scheme (PTAS)}. The 
latter type of algorithm has been studied {avant la lettre} for quite a 
while. This is also true for the corresponding FPT algorithm, which computes a 
$(1+\eps)$-approximation in $f(\eps)\polyn$ time, and is referred to as an 
{efficient polynomial-time approximation scheme (EPTAS)}. Note that if the standard parameterization of an optimization problem is W[1]-hard, then the
optimization problem does not have an EPTAS (unless FPT = W[1])~\cite{cesati1997efficiency}. 

Some interesting 
links between these algorithms, traditionally studied from the perspective of 
polynomial-time approximation algorithms, and parameterized complexity have been 
uncovered more recently~\cite{cygan2016lower,cai1993fixed,cesati1997efficiency, 
chen2007polynomial,kratsch2012polynomial,marx2008fpa,guo2011safe}.

As also mentioned in the introduction, a {parameterized 
$\alpha$-approximation algorithm} computes an $\alpha$-approximation in 
$f(k)\polyn$ time for some parameter $k$ on inputs of size $n$. If $\alpha$ can 
be set to $1+\eps$ for any~$\eps>0$ and the runtime is $f(k,\eps)n^{g(\eps)}$, 
then we obtain a {parameterized approximation scheme (PAS)} for parameter 
$k$. Note that this runtime is only truly FPT if we assume that $\eps$ is 
constant. If we forbid this and consider $\eps$ as a parameter as well, i.e., 
the runtime should be of the form $f(k,\eps)\polyn$, then we obtain {EPAS}. 

\mypar{Kernelization}
A further topic closely related to the FPT algorithms is kernelization. Here, the 
idea is that an instance is efficiently pre-processed by removing the ``easy 
parts'' so that only the NP-hard core of the instance remains. More concretely, 
a {kernelization algorithm} takes an instance $I$ and a parameter $k$ of 
some problem and computes a new instance~$I'$ with parameter $k'$ of the same 
problem. The runtime of this algorithm is polynomial in the size of the input 
instance $I$ and~$k$, while the size of the output $I'$ and~$k'$ is bounded as 
a function of the input parameter~$k$. For optimization problems, it~should also 
be the case that any optimum solution to~$I'$ can be converted to an optimum 
solution of $I$ in polynomial time. The new instance $I'$ is called the 
{kernel} of $I$ (for parameter~$k$). A fundamental result in 
fixed-parameter tractability is that an (optimization) problem parameterized by 
$k$ is FPT if and only if it admits a kernelization algorithm for the same 
parameter~\cite{CaiCDF97}. However the size of the guaranteed kernel will in 
general be exponential (or worse) in the input parameter. Therefore, an~interesting question is whether an NP-hard problem admits small kernels of 
polynomial size. This~can be interpreted as meaning that the problem has a very 
efficient pre-processing algorithm, which~can be used prior to solving the 
kernel. It also gives an additional dimension to the parameterized complexity 
landscape. 

Kernelization has played a fundamental role in the development of FPT 
algorithms, where  a pre-processing step is often used to simplify the structure 
of the input instance. It is therefore only natural to consider such 
pre-processing algorithms for parameterized approximation algorithms as well. 
\citet{lokshtanov2017lossy} define an {$\alpha$-approximate kernelization 
algorithm}, which computes a kernel~$I'$ such that any $\beta$-approximation in 
$I'$ can be converted into an $\alpha\beta$-approximation to the input 
instance~$I$ in polynomial time. Again, the size of $I'$ and $k'$ need to be 
bounded as a function of the input parameter~$k$, and the algorithm needs to run 
in polynomial time. The instance $I'$ is now called an 
{$\alpha$-approximate kernel}. Analogous to exact kernels, any problem has 
a parameterized $\alpha$-approximation algorithm if and only if it admits an 
$\alpha$-approximate kernel for the same parameter~\cite{lokshtanov2017lossy}, 
which however might be of exponential size in the parameter.

An $\alpha$-approximate kernelization algorithm that computes a polynomial-sized 
kernel, and for which we may set $\alpha$ to $1+\eps$ for any $\eps>0$, is 
called a {polynomial-sized approximate kernelization scheme (PSAKS)}. In 
this case $\eps$ is necessarily considered to be a constant, since any 
kernelization algorithms needs to run in polynomial time.

We remark here that apart from $\alpha$-approximate kernels, there is another 
common workaround for problems with no polynomial kernels, captured using the 
notion of Turing kernels. There is also a lower bound framework for Turing 
kernels~\cite{HKSWW15}, and the question of approximate kernels for problems 
that do not even admit Turing kernels is fairly natural to ask. However we skip 
this discussion for the sake of brevity.

Finally, note that in literature, there is another notion of approximate kernels called  $\alpha$-fidelity kernelization~\cite{fellows2018fidelity} which is different from the one mentioned above. Essentially, an $\alpha$-fidelity kernel is a polynomial time preprocessing procedure such that an optimal solution to the
reduced instance translates to an $\alpha$-approximate solution to the
original. This definition allows a loss of
precision in the preprocessing step, but demands that the reduced
instance has to be solved to optimality. See~\cite{lokshtanov2017lossy} for a detailed discussion on the differences between the two approximate kernel notions.  

\mypar{Complexity-Theoretic Hypotheses}
We assume that the readers have basic knowledge of (classic) parameterized 
complexity theory, including the $\W$-hierarchy, the exponential time hypothesis 
(ETH), and the strong exponential time hypothesis (SETH). The reader may choose 
to recapitulate these definitions by referring to~\cite{pc-book} (Sections 13 and 
14). 

We will additionally discuss two hypotheses that may not be standard to the community.
The~first is the {Gap Exponential Time Hypothesis (Gap-ETH)}, which is a strengthening of ETH.
Roughly speaking, it states that even the approximate version of 3SAT cannot be solved
in subexponential time; a more formal statement of Gap-ETH can be found in 
Hypothesis~\ref{hyp:gap-eth}.
Another hypothesis we will discuss is the {Parameterized Inapproximability Hypothesis (PIH)},
which states that the multicolored version of the \pname{Densest $k$-Subgraph} 
is hard to approximate in FPT time. Once again, we do not define PIH formally 
here; please refer to Hypothesis~\ref{hyp:pih} for a formal statement.

\section{FPT Hardness of Approximation}
\label{sec:inapprox}

In this section, we focus on showing barriers against obtaining good 
parameterized approximation algorithms. The analogous field of study in the 
non-parameterized (NP-hardness) regime is the theory of {hardness of 
approximation}. The celebrated PCP Theorem~\cite{ALMSS98,AS92} and numerous 
subsequent works have developed a rich set of tools that allowed   researchers 
to show tight inapproximability results for many fundamental problems. In the 
context of parameterized approximation, the field is still in the nascent 
stage. Nonetheless, there have been quite a few tools that have already been 
developed, which~are discussed in the subsequent subsections.

We divide this section into two parts. In Section~\ref{sec:noGap}, we discuss the 
results and techniques in the area of hardness of parameterized approximation 
under the standard assumption of $\Wone$.   In~Section~\ref{sec:Gap}, we discuss 
results and techniques in hardness of parameterized approximation under less 
standard assumptions such as the Gap Exponential Time Hypothesis, where the gap 
is inherent in the assumption, and the challenge is to construct gap-preserving 
reductions. 

\subsection{\W[1]-Hardness of Gap Problems}\label{sec:noGap}

 In this subsection, we discuss $\W[1]$-hardness of  approximation of a few fundamental problems. In particular, we discuss the parameterized inapproximability (i.e., $\W[1]$-hard to even approximate) of the \pname{Dominating Set} problem, the \pname{(One-sided) Biclique} problem, the \pname{Even Set} problem, the~\pname{Shortest Vector} problem, and the \pname{Steiner Orientation} problem. 
We emphasize here that the main difficulty that is addressed in this subsection is {gap generation}, i.e., we focus on how to start from a  hard problem (with no gap), say \pname{$k$-Clique} (which is the canonical $\W[1]$-complete problem), and reduce it to one of the aforementioned problems, while generating a non-trivial gap in the process.

\subsubsection{Parameterized Intractability of Biclique and Applications to Parameterized Inapproximability}\label{sec:biclique}

In this subsubsection, we will discuss the parametrized inapproximability   of the one-sided biclique problem, and  show how both that result  and its proof technique lead to more inapproximability results. 

We begin our discussion by formally stating the \pname{$k$-Biclique} problem where we are given as input a  graph $G$ and an integer $k$, and the goal is to determine whether $G$ contains a complete bipartite subgraph with $k$ vertices on each side.
The complexity of \pname{$k$-Biclique} was a long standing open problem and was resolved only recently by Lin~\cite{Lin18} where he showed that it is $\W[1]$-hard. In fact, he~showed a much stronger result and this shall be the focus of attention in this subsubsection.

\begin{Theorem}[\cite{Lin18}]\label{thm:biclique}
Given a bipartite graph $G(L\dot\cup R, E)$ and  $k\in\N$ as input, it is \W[1]-hard to distinguish between the following two cases:
\begin{itemize}
\item {{Completeness:}} There are $k$ vertices in $L$ with at least $n^{\Theta(\frac{1}{k})}$ common neighbors in $R$;
\item {{Soundness:}} Any $k$ vertices in $L$ have at most $(k+6)!$ common neighbors in $R$.
\end{itemize}
\end{Theorem}

We shall refer to the gap problem in the above theorem as the \pname{One-Sided $k$-Biclique} problem. To prove the above result, Lin introduced a technique which we shall refer to as {Gadget Composition}. The gadget composition technique has found more applications since~\cite{Lin18}. We provide below a failed approach (given in~\cite{Lin18}) to prove the above theorem; nonetheless it gives us good insight into how the gadget composition technique works. 

Suppose we can construct a set family $\mathcal{T}=\{S_1,S_2,\ldots,S_{n}\}$ of subsets of $[n]$ for some integers $k$, $n$~and $h>\ell$ (for example, $h={n}^{1/k}$ and $\ell=(k+1)!$) such that:
\begin{description}
\item[]{{Property 1:}} Any $k+1$ distinct subsets in $\mathcal T$ have intersection size at most $\ell$;
\item[]{{Property 2:}} Any $k$ distinct subsets in $\mathcal T$ have intersection size at least $h$.
\end{description}

Then we can combine $\mathcal{T}$ with an instance of \pname{$k$-Clique} to obtain a gap instance of \pname{One-Sided $k$-Biclique} as follows.  Given a graph $G$ and parameter $k$ with $V(G)\subseteq [n]$, we  construct  our instance of \pname{One-Sided $k$-Biclique}, say   $H(L\dot\cup R, E(H))$ by setting $L:=E(G)$ and $R:=[n]$, where for any $(v_i,v_j)\in L$ and $v\in[n]$, we have that $((v_i,v_j),v)\in E(H)$ if and only if $v\in S_i\cap S_j$. Let $s:=k(k-1)/2$. It is easy to check that if $G$ has a $k$-vertex clique, say $\{v_1^*,\ldots ,v_k^*\}$ is a clique in $G$, then Property 2 implies that $|\Delta:=\bigcap_{i\in[k]}S_{v_i^*}|\ge h$. It follows that  the set of $s$ vertices in $L$ given by  $\{(v_i^*,v_j^*) : \text{for all $\{i,j\}\in\binom{[k]}{2}$}\}$ are neighbors of every vertex in $\Delta\subseteq R$. On the other hand, if $G$ contains no $k$-vertex clique, then any $s$ distinct vertices in $L$ (i.e., $s$ edges in $G$) must have at least $k+1$ vertices in $G$ as their end points. Say $V'$ was the set of all vertices contained the $s$ edges. By Property 1, we know that $|\Delta':=\bigcap_{v\in V'}S_{v}|\le \ell$, and thus any $s$ distinct vertices in $L$ have at most $\ell$ common neighbors in $R$.

It is indeed very surprising that this technique can yield non-trivial inapproximability results, as~the gap is essentially produced from the gadget and is oblivious to the input! This also stands in stark contrast to the PCP theorem and hardness of approximation results in NP, where all known results were obtained by global transformations on the input. The key difference between the parameterized and NP worlds is the notion of locality. For example, consider the \pname{$k$-Clique} problem, if a graph does not have a clique of size $k$, then given any $k$ vertices, a random vertex pair in these $k$ vertices does not have an edge with probability at least $1/k^2$. It is philosophically possible to compose the input graph with a simple error correcting code to amplify this probability to a constant, as we are allowed to blowup the input size by any function of $k$.  In contrast, when $k$ is not fixed, like in the NP world, $k$ is of the same magnitude as the input size, and thus we are only allowed to blow up the input size  by $\poly(n)$ factor. Nonetheless, we have to point out that the gadgets typically needed to make the gadget composition technique work must be extremely rich in combinatorial structure (and are typically constructed from random objects or algebraic objects),
and were previously studied extensively in the area of extremal combinatorics. 

Returning to the reduction above from \pname{$k$-Clique} to  \pname{One-Sided $k$-Biclique}, it turns out that we do not know how to construct the set system $\mathcal{T}$, and hence the reduction does not pan out. 
Nonetheless Lin constructed a variant of $\mathcal{T}$, where Property 2 was more refined and the reduction from \pname{$k$-Clique} to  \pname{One-Sided $k$-Biclique}, went through with slightly more effort. 

Before we move on to discussing some applications of Theorem~\ref{thm:biclique} and the gadget composition technique, we remark about known stronger time lower bound for 
\pname{One-Sided $k$-Biclique}  under stronger running time hypotheses.
Lin~\cite{Lin18} showed a lower bound of $n^{\Omega(\sqrt{k})}$ for \pname{One-Sided $k$-Biclique} assuming ETH. We wonder if this can be further improved.

\begin{Open Question}[Lower bound of \pname{One-Sided $k$-Biclique} under ETH and SETH]
Can the running time lower bound on \pname{One-Sided $k$-Biclique}  be improved to $n^{\Omega(k)}$ under ETH? Can it be improved to $n^{k - o(1)}$ under SETH?
\end{Open Question}

We remark that a direction to address the above question was detailed in~\cite{KM19}. While on the topic of the \pname{$k$-Biclique} problem, it is worth noting that the lower bound of $n^{\Omega(\sqrt{k})}$ for \pname{One-Sided $k$-Biclique} assuming ETH yields a running time lower bound of $n^{\Omega(\frac{\log k}{\log\log k})}$ for the \pname{$k$-Biclique} problem (due to the soundness parameters in Theorem~\ref{thm:biclique}). However, assuming {randomized} ETH, the running time lower bound for the \pname{$k$-Biclique} problem can be improved to $n^{\Omega(\sqrt{k})}$~\cite{Lin18}. Can this improved running time lower bound be obtained just under (deterministic) ETH? Finally, we remark that we shall discuss about the hardness of approximation of the \pname{$k$-Biclique} problem in Section~\ref{sec:dense}.

\mypar{Inapproximability of \pname{$k$-Dominating Set} via Gadget Composition} We shall discuss about the inapproximability of \pname{$k$-Dominating Set} in detail in the next subsubsection. We would like to simply highlight here how the above framework was used by Chen and Lin~\cite{chen2016constant} and Lin~\cite{Lin19} to obtain inapproximability results for \pname{$k$-Dominating Set}. 

In~\cite{chen2016constant}, the authors starting from Theorem~\ref{thm:biclique}, obtain the \W[1]-hardness of approximating  \pname{$k$-Dominating Set} to a factor of almost two. Then they amplify the gap to any constant by using a specialized graph product.

We now turn our attention to a recent result of Lin~\cite{Lin19} who provided strong  inapproximability result for \pname{$k$-Dominating Set} (we refer the reader to Section~\ref{sec:domset} to obtain the context for this result). 
Lin's proof of inapproximability of \pname{$k$-Dominating Set} is
a one-step reduction from an instance of \pname{$k$-Set Cover} on a universe of size $O(\log n)$ (where $n$ is the number of subsets given in the collection) to an instance of \pname{$k$-Set Cover} on a universe of size $\poly(n)$ with a gap of $\left(\frac{\log n}{\log\log n}\right)^{1/k}$.
Lin then uses this gap-producing self-reduction to provide running time lower 
bounds (under different time hypotheses) for approximating $k$-set cover to a 
factor of $(1-o(1))\cdot \left(\frac{\log n}{\log\log n}\right)^{1/k}$. Recall 
that \pname{$k$-Dominating Set} is essentially\footnote{Recall that there 
is a pair of polynomial-time $L$-reductions between the minimum dominating set 
problem and the set cover problem~\cite{K92}.} equivalent to \pname{$k$-Set 
Cover}.

Elaborating, Lin designs a gadget by combining the hypercube partition gadget of Feige~\cite{feige1998threshold} with  a derandomizing combinatorial object called {universal set}, to obtain a gap gadget, and then combines the gap gadget with the input \pname{$k$-Set Cover} instance (on small universe but with no gap) to obtain a gap \pname{$k$-Set Cover} instance. This is another success story of the gadget composition technique.

Finally, we remark that Lai~\cite{Lai19} recently extended Lin's inapproximability results for dominating set (using the same proof framework) to rule out constant-depth circuits of size $f(k)n^{o(\sqrt{k})}$ for any
computable function $f$.

\mypar{Even Set}
A recent success story of Theorem~\ref{thm:biclique} is its application to resolve a long standing open problem called \pname{$k$-Minimum Distance Problem} (also referred to as  \pname{$k$-Even Set}), where  we are given as input a generator matrix $\mathbf{A} \in \mathbb{F}_2^{n \times m}$ of a binary linear code and an integer $k$, and the goal is to determine whether the code has distance at most $k$. Recall that the distance of a linear code is $\underset{\vec{0} \ne \mathbf{x} \in \mathbb{F}_2^m}{\min}\ \|\mathbf A\mathbf x\|_0$ where $\| \cdot \|_0$ denote the 0-norm (aka the Hamming norm).

In~\cite{even-set}, the authors showed that \pname{$k$-Even Set} is $\W[1]$-hard under randomized reductions. The~result was obtained by starting from the inapproximability result stated in Theorem~\ref{thm:biclique} followed by a series of intricate reductions. In fact they proved the following stronger inapproximability result.

\begin{Theorem}[\cite{even-set}]\label{thm:even-set}
For any $\gamma \geq 1$, given input $(\mathbf A, k) \in \F^{n \times m} \times \N$, it is \W[1]-hard (under randomized reductions) to distinguish between
\begin{itemize}
\item {{Completeness:}} Distance of the code generated by $\mathbf A$ is at most $k$ , and,
\item {{Soundness:}} Distance of the code generated by $\mathbf A$ is more than $\gamma \cdot k$.
\end{itemize} 
\end{Theorem}

We emphasize that even to obtain the W[1]-hardness of \pname{$k$-Even Set} (with no gap), they needed to start from the gap problem given in Theorem~\ref{thm:biclique}.

The proof of the above theorem proceeds by first showing FPT hardness of approximation of the non-homogeneous variant of \pname{$k$-Minimum Distance Problem} called the \pname{$k$-Nearest Codeword Problem}. In \pname{$k$-Nearest Codeword Problem}, we are given a target vector $\mathbf y$ (in $\F^n$) in addition to $(\mathbf A, k)$, and the goal is to find whether there is any $\mathbf x$ (in $\F^m$) such that the Hamming norm of $\mathbf A\mathbf x - \mathbf y$ is at most $k$.
As an intermediate step of the proof of Theorem~\ref{thm:even-set}, they showed that \pname{$k$-Nearest Codeword Problem} is W[1]-hard to approximate to any constant factor.

An important intermediate problem which was studied by~\cite{even-set} to prove the inapproximability of \pname{$k$-Nearest Codeword Problem}, was the \pname{$k$-Linear Dependent Set} problem where given a set $\mathbf A$ of
$n$ vectors over a finite field $\F_q$ and an integer $k$, the goal  is to decide if there
are $k$ vectors in $\mathbf A$ that are linearly dependent. They ruled out constant factor approximation algorithms for this problem running in FPT time. Summarizing, the high level  proof overview of Theorem~\ref{thm:even-set} follows by reducing  \pname{One-Sided $k$-Biclique} to \pname{$k$-Linear Dependent Set}, which is then reduced to \pname{$k$-Nearest Codeword Problem}, followed by a final randomized reduction to \pname{$k$-Minimum Distance Problem}.

Finally, we note  that there is no reason to define \pname{$k$-Minimum Distance Problem} only for binary code, but can instead be defined over larger fields as well. It turns out that~\cite{even-set}  cannot rule out FPT algorithms for \pname{$k$-Minimum Distance Problem} over $\mathbb{F}_p$ with $p > 2$, when $p$ is fixed and is not part of the input. Thus we have the open problem.

\begin{Open Question}
Is it \W[1]-hard to decide \pname{$k$-Minimum Distance Problem} over $\mathbb{F}_p$ with $p > 2$, when $p$ is fixed and is not part on the input?
\end{Open Question}

\mypar{Shortest Vector Problem} 
Theorem~\ref{thm:biclique} (or more precisely the constant inapproximability of \pname{$k$-Linear Dependent Set} stated above) was also used to resolve the complexity of the parameterized \pname{$k$-Shortest Vector Problem} in lattices, where  the input (in the $\ell_p$ norm) is an integer $k \in \N$ and a matrix $\mathbf A \in \Z^{n \times m}$ representing the basis of a lattice, and we want to determine whether the shortest (non-zero) vector in the lattice has length at most $k$, i.e., whether $\underset{\vec{0} \ne \mathbf x \in \Z^{m}}{\min}\ \|\mathbf A\mathbf x\|_p \leq k$. Again, $k$~is the parameter of the problem. It should also be noted here that (as in ~\cite{DFVW99}), we require the basis of the lattice to be integer valued, which is sometimes not enforced in literature (e.g.,~\cite{VEB,Ajt98}). This is because, if $\mathbf A$ is allowed to be any matrix in $\mathbb{R}^{n \times m}$, then parameterization is meaningless because we can simply scale $\mathbf A$ down by a large multiplicative factor.

In~\cite{even-set}, the authors showed that \pname{$k$-Shortest Vector Problem} is $\W[1]$-hard under randomized reductions.  In fact they proved the following stronger inapproximability result. 

\begin{Theorem}[\cite{even-set}]\label{thm:svp}
For any $p > 1$, there exists a constant $\gamma_p > 1$ such that given input $(\mathbf A, k) \in \Z^{n \times m} \times \N$, it is \W[1]-hard (under randomized reductions) to distinguish between
\begin{itemize}
\item {{Completeness:}} The $\ell_p$ norm of the shortest vector of the lattice generated by $\mathbf A$ is $\leq k$, and,
\item {{Soundness:}} The $\ell_p$ norm of the shortest vector of the lattice generated by $\mathbf A$ is $> \gamma_p \cdot k$.
\end{itemize} 
\end{Theorem}

Notice that Theorem~\ref{thm:even-set} rules out FPT approximation algorithms with {any} constant approximation ratio for \pname{$k$-Even Set}. In contrast, the above result only prove FPT inapproximability with {some} constant ratio for \pname{$k$-Shortest Vector Problem} in $\ell_p$ norm for $p > 1$. As with \pname{$k$-Even Set}, even to prove the W[1]-hardness of \pname{$k$-Shortest Vector Problem} (with no gap), they needed to start from the gap problem given in Theorem~\ref{thm:biclique}.

The proof of the above theorem proceeds by first showing FPT hardness of approximation of the non-homogeneous variant of \pname{$k$-Shortest Vector Problem} called the \pname{$k$-Nearest Vector Problem}. In \pname{$k$-Nearest Vector Problem}, we are given a target vector $\mathbf y$ (in $\Z^n$) in addition to $(\mathbf A, k)$, and~the goal is to find whether there is any $\mathbf x$ (in $\Z^m$) such that the $\ell_p$ norm of $\mathbf A\mathbf x - \mathbf y$ is at most $k$.
As~an intermediate step of the proof of Theorem~\ref{thm:even-set}, they showed that \pname{$k$-Nearest Vector Problem} is W[1]-hard to approximate to any constant factor.
Summarizing, the high level  proof overview of Theorem~\ref{thm:svp} follows by reducing  \pname{One-Sided $k$-Biclique} to \pname{$k$-Linear Dependent Set}, which is then reduced to \pname{$k$-Nearest Vector Problem}, followed by a final randomized reduction to \pname{$k$-Shortest Vector Problem}.

An immediate open question left open from their  work is whether Theorem~\ref{thm:svp} can be extended to \pname{$k$-Shortest Vector Problem} in the $\ell_1$ norm. In other words, 

\begin{Open Question}[Approximation of \pname{$k$-Shortest Vector Problem} in $\ell_1$ norm]Is \pname{$k$-Shortest Vector Problem} in the $\ell_1$ norm in FPT?
\end{Open Question}

\subsubsection{Parameterized Inapproximability of Dominating Set}
\label{sec:domset}

In the \pname{$k$-Dominating Set}  problem we are given an integer $k$ and a graph $G$ on $n$ vertices  as input, and the goal is to determine if there is a dominating set of size at most $k$.  It was a long standing open question to design an algorithm which runs in time $T(k) \cdot \poly(n)$ (i.e., FPT-time), that would find a dominating set of size at most $F(k) \cdot k$ whenever the graph $G$ has a dominating set of size $k$, for~any computable functions $T$ and $F$. 

The first non-trivial progress on this problem was by \citet{chen2016constant} 
who ruled out the existence of such algorithms (under \Wone) for all constant 
functions $F$ (i.e., $F(n)=c$, where $c$ is any universal constant). We discussed their proof technique in the previous subsubsection. A couple of 
years later, \citet{dom-set} completely settled the question, by ruling out the 
existence of such an algorithm (under \Wone) for any computable function $F$. 
Thus, \pname{$k$-Dominating Set} was shown to be {totally inapproximable}. 
We  elaborate on their proof below. 

\begin{Theorem}[\cite{dom-set}]\label{thm:domset}
Let $F:\N\to\N$ be any computable function. Given an instance $(G,k)$ of \pname{$k$-Dominating Set} as input, it is $\W[1]$-hard to distinguish between the following two cases:
\begin{itemize}
\item{{Completeness:}} $G$ has a dominating set of size $k$.
\item{{Soundness:}} Every dominating set of $G$ is of size at least $F(k)\cdot k$.
\end{itemize}
\end{Theorem}

The overall proof follows by reducing  \pname{$k$-Multicolor Clique}  to the gap \pname{$k$-Dominating Set} with parameters as given in the theorem statement. In the \pname{$k$-Multicolor Clique} problem, we are given an integer $k$ and a graph $G$ on vertex set $V:=V_1\dot\cup V_2\dot\cup\cdots \dot\cup V_k$  as input, where each $V_i$ is an independent set of cardinality $n$, and the goal is to determine if there is a clique of size $k$ in $G$. Following a straightforward reduction from the \pname{$k$-Clique} problem, it is fairly easy to see that \pname{$k$-Multicolor Clique} is $\W[1]$-hard. 

The reduction from \pname{$k$-Multicolor Clique} to the gap 
\pname{$k$-Dominating Set} proceeds in two steps. In the first step we reduce 
\pname{$k$-Multicolor Clique} to \pname{$k$-Gap CSP}. This is the step where we 
generate the gap. In the second step, we reduce \pname{$k$-Gap CSP} to  gap 
\pname{$k$-Dominating Set}. This step is fairly standard and mimics ideas from 
Feige's proof of the NP-hardness of approximating the \pname{Max Coverage} 
problem~\cite{feige1998threshold}.

Before we proceed with the details of the above two steps, let us introduce a 
small technical tool from coding theory that we would need. We need codes known 
in literature as {good codes}, these~are binary error correcting codes 
whose rate and relative distances are both constants bounded away from~0  (see 
\cite{G09} (Appendix E.1.2.5) for definitions). The reader may think of them 
as follows: for every $\ell\in\mathbb{N}$, we say that $C_\ell\subseteq 
\{0,1\}^\ell$ is a good code if (i) $|C_\ell|=2^{\rho \ell}$, for some universal 
constant $\rho>0$, (ii) for any distinct $c,c'\in C_\ell$ we have that $c$ and 
$c'$ have different values on at least $\delta\ell $ fraction of coordinates, 
for some universal constant $\delta>0$. An encoding of $C_\ell$ is an 
{injective} function $\mathcal{E}_{C_\ell}:\{0,1\}^{\rho\ell}\to C_\ell$. 
The~encoding is said to be efficient if $\mathcal{E}_{C_\ell}(x)$ can be 
computed in $\poly(\ell)$ time for any $x\in\{0,1\}^{\rho\ell}$.

Let us fix $k\in\N$ and $F:\N\to \N$ as in the theorem statement. 
We further define $$\alpha:=1-\frac{1}{\left(\binom{k}{2}\cdot 
F(\binom{k}{2})\right)^{\binom{k}{2}}}.$$

\mypar{From \pname{$k$-Multicolor Clique} to \pname{$k$-Gap CSP}} Starting from an instance of \pname{$k$-Multicolor Clique}, say $G$ on vertex set $V:=V_1\dot\cup V_2\dot\cup\cdots \dot\cup V_k$, we write down a set of constraints $\mathcal{P}$ on a variable set $X:=\{x_{i,j}\mid i,j\in[k], i\neq j\}$ as follows.
For every $i,j\in[k]$, such that $i\neq j$, define $E_{i,j}$ to be the set of all edges in $G$ whose end points are in $V_i$ and $V_j$. An assignment to 
variable $x_{i,j}$ is an element of $E_{i,j}$, i.e., a pair of vertices, one from $V_i$ and the other from $V_j$.  Suppose that $x_{i,j}$ was assigned the edge $\{v_i,v_j\}$, where $v_i\in V_i$ and $v_j\in V_j$. Then we define the assignment of $x_{i,j}^i$ to be $v_i$ and the  assignment of $x_{i,j}^j$ to be $v_j$. We define $\mathcal{P}:=\{P_1,\ldots ,P_k\}$, where the constraint $P_i$ is defined to be satisfied if the assignment to  all of $x_{1,i}^i,x_{2,i}^i,\ldots ,x_{i-1,i}^i,x_{i+1,i}^i,\ldots ,x_{k,i}^i$ are the same. 
We refer to the problem of determining if there is an assignment to the variables in $X$ such that all the constraints are satisfied  as the \pname{$k$-CSP} problem. Notice that while this is a natural way to write \pname{$k$-Multicolor Clique} as a CSP, where we have tried to check if all variables having a vertex in common, agree on its assignment, there is no gap yet in the \pname{$k$-CSP} problem. In particular, if there was a clique of size $k$ in $G$ then there is an assignment  to the variables of $X$ (by assigning the edges of the clique in $G$ to the corresponding variable in $X$) such that all the constraints in $\mathcal{P}$ are satisfied; however, if every clique in $G$ is of size less than $k$ then there every assignment to the variables of $X$ may violate only one constraint in $\mathcal{P}$ (and not more).

In order to amplify the gap,  we rewrite the set of constraints $\mathcal{P}$  in a different way to obtain the set of constraints $\mathcal{P}'$,   on the same variable set $X$, as follows. Suppose that $x_{i,j}$ was assigned the edge $\{v_i,v_j\}$, where $v_i\in V_i$ and $v_j\in V_j$, then for $\beta\in[\log n]$, we define the assignment of $x_{i,j}^{i,\beta}$ to be the $\beta^{\text{th}}$ coordinate of $v_i$. Recall that $|V_i|=n$ and therefore we can label all vertices in $V_i$ by vectors in $\{0,1\}^{\log n}$.  We define
$\mathcal{P}':=\{P_1',\ldots ,P_{\log n}'\}$, where the constraint $P_\beta'$ is defined to be satisfied if and only if the following holds for all $i\in[k]$: the assignment to  all of $x_{1,i}^{i,\beta},x_{2,i}^{i,\beta},\ldots ,x_{i-1,i}^{i,\beta},x_{i+1,i}^{i,\beta},\ldots ,x_{k,i}^{i,\beta}$ are the same. 
Again notice that there is an assignment to the variables of $X$ such that all the constraints in $\mathcal{P}$ are satisfied if and only if the same assignment also satisfies all the constraints in $\mathcal{P}'$.

However, rewriting $\mathcal{P}$ as $\mathcal{P}'$ allows us to simply apply the error correcting code $C_\ell$ (with parameters $\rho$ and $\delta$, and encoding function $\mathcal{E}_{C_\ell}$) to the constraints in $\mathcal{P}'$, to obtain a gap! In~particular, we choose $\ell$ to be such that $\rho\ell=\log n$. Consider a new set of constraints $\mathcal{P}''$,   on the same variable set $X$, as follows. For any $z\in\{0,1\}^{\log n}$ and $\beta\in[\ell]$, we denote by $\mathcal{E}_{C_\ell}(z)_\beta$, the $\beta^{\text{th}}$ coordinate of $\mathcal{E}_{C_\ell}(z)$. We define
$\mathcal{P}'':=\{P_1'',\ldots ,P_{\ell}''\}$, where the constraint $P_\beta''$ is defined to be satisfied if and only if the following holds for all $i\in[k]$: the assignment to  all of $\mathcal{E}_{C_\ell}(x_{1,i}^{i})_\beta,\mathcal{E}_{C_\ell}(x_{2,i}^{i})_\beta,\ldots ,\mathcal{E}_{C_\ell}(x_{i-1,i}^{i})_\beta,\mathcal{E}_{C_\ell}(x_{i+1,i}^{i})_\beta,\ldots ,\mathcal{E}_{C_\ell}(x_{k,i}^{i})_\beta$ are the same. 

Notice, as before, that there is an assignment to the variables of $X$ such that all the constraints in $\mathcal{P}'$ are satisfied if and only if the same assignment also satisfies all the constraints in $\mathcal{P}''$. However, for~every assignment to $X$ that violates at least one constraint in $\mathcal{P}'$, we have that the same assignment violates at least $\delta$ fraction of the constraints in $\mathcal{P}''$. To see this, consider an assignment that violates the constraint $P_1$ in $\mathcal{P}'$. This implies that there is some $i\in[k]$ such that the assignment to  all of $x_{1,i}^{i,1},x_{2,i}^{i,1},\ldots ,x_{i-1,i}^{i,1},x_{i+1,i}^{i,1},\ldots ,x_{k,i}^{i,1}$ are {not} the same. Let us suppose, without loss of generality, that  the assignment to $x_{1,i}^{i,1}$ and $x_{2,i}^{i,1}$ are different. In other words, we have that $x_{1,i}^{i}\neq x_{2,i}^{i}$, where we think of $x_{1,i}^{i},x_{2,i}^{i}$ as $\log n$ bit vectors. Let $\Delta \subseteq [\ell]$ such that $\beta\in \Delta$ if and only if $\mathcal{E}_{C_\ell}(x_{1,i}^{i})_\beta\neq \mathcal{E}_{C_\ell}(x_{2,i}^{i})_\beta$. By the distance of the code $C_\ell$ we have that $|\Delta|\ge \delta\ell$. Finally, notice that for all $\beta\in\Delta$, we have that the assignment does not satisfy constraint $P_\beta$ in $\mathcal{P}''$. We refer to the problem of distinguishing if there is an assignment to  $X$ such that all the constraints are satisfied or if every assignment to  $X$ does not satisfy a constant fraction of the constraints, as the \pname{$k$-Gap CSP} problem.

In order to rule out $F(k)$ approximation FPT algorithms for 
\pname{$k$-Dominating Set}, we will need that for every assignment to $X$ that 
violates at least one constraint in $\mathcal{P}'$, we have that the same 
assignment violates at least $\alpha$ fraction of the constraints in 
$\mathcal{P}''$ (instead of just $\delta$; note that $\alpha$ is very close to 
1, whereas $\delta$ can be at most half). To boost the gap\footnote{We 
could have skipped this boosting step, had we chosen   a different good code 
with distance $\alpha$ but over a larger alphabet. For example, taking the Reed 
Solomon code over alphabet $\frac{\log n}{1-\alpha}$ would have sufficed. We 
chose not to do so, to keep the proof as elementary as possible.} we apply a 
simple repetition/direct-product trick to our constraint system. Starting from 
$\mathcal{P}''$, we construct a new set of constraints $\mathcal{P}^*$,   on the 
same variable set $X$, as follows. $$\mathcal{P}^*:=\{P_S\mid S\in [\ell]^t\},$$
where $t=\frac{\log (1-\alpha)}{\log (1-\delta)}$.  For every $S\in [\ell]^t$, we define $P_S$ to be satisfied if and only if for all $\beta\in S$, the constraint $P_\beta$ is satisfied.  

It is easy to see that $\mathcal{P}''$ and $\mathcal{P}^*$ have the same set of completely satisfying assignments. However, for every assignment to $X$ that violates $\delta$ fraction of constraints in $\mathcal{P}''$, we have that the same assignment violates at least $\alpha$ fraction of the constraints in $\mathcal{P}^*$. To see this, consider an assignment that violates $\delta$ fraction of constraints in $\mathcal{P}''$, say it violates all constraints $P_\beta\in \mathcal{P}''$, for every $\beta\in\Delta\subseteq[\ell]$ . This implies that the assignment satisfies constraint $P_S$ if and only if  $S\in ([\ell]\setminus \Delta)^t$. This implies that the fraction of constraints in $\mathcal{P}^*$ that the assignment can satisfy is upper bounded by $(1-\delta)^t= 1-\alpha$.

\mypar{From \pname{$k$-Gap CSP} to gap \pname{$k$-Dominating Set}} In the 
second part, starting from the aforementioned instance of \pname{$k$-Gap CSP} 
(after boosting the gap), we construct an instance $H$ of \pname{$k$-Dominating 
Set}. The construction is due to Feige~\cite{feige1998threshold}\footnote{This 
reduction (which employs the hypercube set system) is used 
in~\cite{feige1998threshold} for proving hardness of approximating \pname{Max 
$k$-Coverage}; for \pname{Set Cover}, Feige used a more efficient set system 
which is not needed in our context.} and it proceeds as follows. Let 
$\mathcal{F}$ be the set of all functions from $\{0,1\}^{tk}$ to $\binom{k}{2}$, 
i.e., $\mathcal{F}:=\{f:\{0,1\}^{tk}\to \binom{k}{2}\}$. The graph $H$ is on 
vertex set $U=A\dot\cup B$, where $A=\mathcal{P}^*\times \mathcal{F}$ and 
$B=E(G)$, i.e., $B$ is simply the edge set of $G$. We introduce an edge between 
all pairs of vertices in $B$. We introduce an edge between $a:=(S:=(s_1,\ldots , 
s_t)\in[\ell]^t,f:\{0,1\}^{tk}\to \binom{k}{2})\in A$ and $e:=(v_i,v_j)\in E$ if 
and only if the following~holds. 
\begin{align*}
\exists \tau:=(\tau_1,\ldots,\tau_t)\in\{0,1\}^{kt}, \text{ such that }f(\tau)=\{i,j\} \text{ and }\\
\forall r\in[t], \text{ we have } \mathcal{E}_{C_\ell}(v_i)_{s_r}=(\tau_r)_{i}\text{ and }\mathcal{E}_{C_\ell}(v_j)_{s_r}=(\tau_r)_{j}.
\end{align*} 

Notice that the number of vertices in $H$ is  $|A|+|B|\le (\frac{\log n}{\rho})^{t}\cdot \binom{k}{2}^{2^{tk}}+n^2<\eta(k)\cdot n^{2.01}$, for some computable function $\eta$. It is not hard to check that the following hold:
\begin{itemize}
\item (Completeness) If there is an assignment to $X$ that satisfies all constraints in $\calP^*$, then the corresponding $\binom{k}{2}$ vertices in $B$ dominate all vertices in the graph $H$.
\item (Soundness) If each assignment can only satisfy $(1 - \alpha)$ fraction of constraints in $\calP^*$, then any dominating set of $H$ has size at least $F\left(\binom{k}{2}\right) \cdot \binom{k}{2}$.
\end{itemize}

We skip presenting details of this part of the proof here. The proofs have been derived many times in literature; if needed, the readers may refer to Appendix A of~\cite{dom-set}.
This completes our sketch of the proof of Theorem~\ref{thm:domset}.

A few remarks are in order. First, the \pname{$k$-Gap CSP} problem described in the proof above, is~formalized as the \pname{$k$-Maxcover} problem in~\cite{dom-set} (and was originally introduced in~\cite{param-inapprox}). In particular, the formalism of \pname{$k$-Maxcover} (which may be thought of as the parameterized label cover problem) is generic enough to be used as an intermediate gap problem to reduce to both \pname{$k$-Dominating Set} (as in~\cite{dom-set}) and \pname{$k$-Clique} (as in~\cite{param-inapprox}). Moreover, it was robust enough to capture stronger running time lower bounds (under stronger hypotheses); this will elaborated below.   However, in order to keep the above proof succinct, we skipped introducing the \pname{$k$-Maxcover} problem, and worked with \pname{$k$-Gap CSP}, which was sufficient for the above proof. 

Second, \citet{dom-set} additionally showed that  for every computable 
functions $T,F:\N\to\N$ and every constant $\varepsilon > 0$:

\begin{itemize}
\item Assuming the Exponential Time Hypothesis (ETH), there is no $F(k)$-approximation algorithm for \pname{$k$-Dominating Set} that runs in $T(k) \cdot n^{o(k)}$ time.
\item Assuming the Strong Exponential Time Hypothesis (SETH), for every integer $k \geq 2$, there is no $F(k)$-approximation algorithm for \pname{$k$-Dominating Set} that runs in $T(k) \cdot n^{k - \varepsilon}$ time. 
\end{itemize}

In order to establish Theorem~\ref{thm:domset} and the above two results, 
\citet{dom-set} introduced a framework to prove parameterized hardness of 
approximation results. In this framework, the objective was to start from either 
the $\Wone$ hypothesis, ETH, or SETH, and end up with the gap 
\pname{$k$-Dominating Set}, i.e., they design reductions from instances of 
\pname{$k$-Clique}, 3-$\SAT$, and $\ell$-$\SAT$, to an instance of gap 
\pname{$k$-Dominating Set}. A prototype reduction in this framework has two 
modular parts. In the first part, which is specific to the problem they start  
from, they generate a gap and obtain hardness of gap \pname{$k$-Maxcover}. In 
the second part, they show a gap preserving reduction from gap 
\pname{$k$-Maxcover} to gap \pname{$k$-Dominating Set},  which is essentially 
the same as the reduction from \pname{$k$-Gap CSP} to \pname{$k$-Dominating Set} 
in the proof of Theorem~\ref{thm:domset}. 

The first part of a prototype reduction from the computational problem underlying a hypothesis of interest to gap \pname{$k$-Maxcover} follows by the design of an appropriate communication protocol. In~particular, the computational problem is first reduced to a constraint satisfaction problem (CSP) over $k$ (or some function of $k$) variables  over an alphabet of size $n$.  The predicate of this CSP would depend on the computational problem underlying the hypothesis from which we started. Generalizing ideas from~\cite{ARW17}, they then show how a protocol for computing this predicate in the multiparty (number of players is the number of variables of the CSP) communication model, can be combined with the CSP to obtain an instance of gap \pname{$k$-Maxcover}. For example, for the $\Wone$ hypothesis and ETH, the predicate is a variant of the equality function, and for SETH, the predicate is the well studied disjointness function. The completeness and soundness of the protocols computing these functions translate directly to the completeness and soundness of \pname{$k$-Maxcover}.

Third, we recall that  Lin ~\cite{Lin19} recently provided alternate proofs of 
Theorem~\ref{thm:domset} and the above mentioned stronger running time lower bounds. 
While we  discussed about his proof technique in Section~\ref{sec:biclique}, we would like 
to discuss about his result here. Following the right setting of parameters in 
the proof of Theorem~\ref{thm:domset} (for example set $\alpha=1-\frac{1}{(\log 
n)^{\Omega(1/k)}}$), we can obtain that approximating  \pname{$k$-Dominating 
Set} to a factor of $(\log n)^{1/k^3}$ is $\W[1]$-hard. Lin improved the 
exponent of $1/k^3$ in the approximation factor to $h(k)$ for any computable 
function $h$. 
Can this inapproximability be further improved? On the other hand, can we do better than the simple polynomial time greedy algorithm which provides a (1+$\ln n$) factor approximation? This leads us to the following question:

\begin{Open Question}[Tight inapproximability of \pname{$k$-Dominating Set}]
Is there a $(\log n)^{1-o(1)}$ factor approximation algorithm for  \pname{$k$-Dominating Set} running in time $n^{k-0.1}$?
\end{Open Question}

We conclude the discussion on \pname{$k$-Dominating Set} with an open question 
on $\W[2]$-hardness of approximation. As noted earlier, \pname{$k$-Dominating 
Set} is a $\W[2]$-complete problem, and Theorem~\ref{thm:domset} shows that the problem 
is $\W[1]$-hard to approximate to any $F(k)$ factor. However, is there some 
computable function $F$ for which approximating \pname{$k$-Dominating Set} is in 
$\W[1]$? In other words we have:

\begin{Open Question}[$\W{[2]}$-completeness of approximating \pname{$k$-Dominating Set}]
Can we base total inapproximability of \pname{$k$-Dominating Set} on $\W[2] \ne \FPT$?
\end{Open Question}

\subsubsection{Parameterized Inapproximability of Steiner Orientation by Gap Amplification}
\label{sec:steiner-orientation}

Gap amplification is a widely used technique in the classic literatures on (NP-)hardness of approximation (e.g.~\cite{BermanS92,Raz98,Dinur07}). In fact, the arguably simplest proof of the PCP theorem, due to Dinur~\cite{Dinur07}, is indeed via repeated gap amplification. The overall idea here is simple: we start with a hardness of approximation for a problem with small factor (e.g., $1 + 1/n$). At each step, we perform an operation that transforms an instance of our problem to another instance, in such a way that the gap becomes bigger; usually this new instance will also be bigger than our instance. By repeatedly applying this operation, one can finally arrive at a constant, or even super constant, factor hardness of approximation.

There are two main parameters that determine the success/failure of such an 
approach: how~large the new instance is 
compared to the old instance (i.e., size blow-up) and how large the new gap is 
compared to the old gap, in each operation. To see how these two come into the 
picture, let us first consider a case study where a (straightforward) gap 
amplification procedure {does not} work: \pname{$k$-Clique}. The standard way to 
amplify the gap for \pname{$k$-Clique} is through graph product. Recall that the 
{(tensor) graph product} of a graph $G = (V, E)$ with itself, denoted by 
$G^{\otimes 2}$, is a graph whose vertex set is $V^2$ and there is an edge 
between $(u_1, u_2)$ and $(v_1, v_2)$ if and only if $(u_1, v_1) \in E$ and 
$(u_2, v_2) \in E$. It is not hard to check that, if we can find a clique of 
size $t$ in $G^{\otimes 2}$, then we can find one of size $\sqrt{t}$ in $G$ (and 
vice versa). This implies that, if we have an instance of clique that is hard 
to 
approximate to within a factor of $(1 + \varepsilon)$, then we may take the 
graph product with itself which yields an instance of \pname{Clique} that is hard to 
approximate to within a factor of $(1 + \varepsilon)^2$. 

Now, let us imagine that we start with the hard instance of an exact version of 
\pname{$k$-Clique}. We~may think of this as being hard to approximate to within a factor 
of $(1 - 1/k)$. Hence, we may apply the above gap amplification procedure $\log 
k$ times, resulting in an instance of \pname{Clique} that is hard to approximate to 
within a factor of $\left(1 - 1/k\right)^{2^{\log k}}$, which is a constant 
bounded away from one (i.e., $\approx 1/e$). The bad news here is that the number 
of the vertices of the final graph is $n^{2^{\log k}} = n^k$, where $n$ is the 
number of vertices of the initial graph. This does not give any lower bound, 
because we can solve \pname{$k$-Clique} in the original graph in $n^{O(k)}$ time 
trivially! In the next subsection, we will see a simple way to prove hardness of 
approximating \pname{$k$-Clique}, assuming stronger assumptions. However, it remains an 
interesting and important open question how to prove such hardness from a 
non-gap assumption:

\begin{Open Question}
Is it W[1]-hard or ETH-hard to approximate \pname{$k$-Clique} to within a constant factor in FPT time?
\end{Open Question}

Having seen a failed attempt, we may now move on to a success story. 
Remarkably, Wlodarczyk~\cite{Wlo19} recently managed to use gap amplification to 
prove hardness of approximation for connectivity problems, including the 
\pname{$k$-Steiner Orientation} problem. Here we are given a {mixed graph} $G$, 
whose edges are either directed or undirected, and a set of $k$ terminal pairs 
$\{(s_i, t_i)\}_{i \in [k]}$. The goal is to orient all the undirected edges  in 
such a way that maximizes the number of $t_i$ that can be reached from $s_i$. 
The problem is known to be in XP~\cite{CyganKN13} but is W[1]-hard even when 
all terminal pairs can be connected~\cite{PilipczukW18a}. Starting from this 
W[1]-hardness, Wlodarczyk~\cite{Wlo19} devises a gap amplification step that 
implies a hardness of approximation with factor $(\log k)^{o(1)}$ for the 
problem. Due to the technicality of the gap amplification step, we will not go into the specifics in this survey. However, let us point out the differences between this gap amplification and 
the (failed) one for \pname{Clique} above. The key point here is that the new instance 
of Wlodarczyk's gap amplification has size of the form $f(k) \cdot n$ instead of 
$n^2$ as in the graph product. This means that, even if we are applying 
Wlodarczyk's gap amplification step $\log(k)$ times, or, more generally, $g(k)$ 
times, it only results in an instance of size $\underbrace{f(f(\cdots(f(}_{g(k) 
\text{times}}k))))\cdot n$, which is still FPT! Since the technique is still 
quite new, it is an exciting frontier to examine whether other parameterized 
problems allow such similar gap amplification steps.

\subsection{Hardness from Gap Hypotheses}\label{sec:Gap}

In the previous subsection, we have seen that several hardness of approximation results can be proved based on standard assumptions. However, as alluded to briefly, some basic problems, including \pname{$k$-Clique}, still evades attempts at proving such results. This motivates several researchers in the community to come up with new assumptions that allow more power and flexibility in proving inapproximability results. We will take a look at two of these hypotheses in this subsection; we note that there have also been other assumptions formulated, but we only focus on these two since they arguably have been used most often.

The first assumption, called the Parameterized Inapproximability Hypothesis (PIH) for short, can~be viewed as a gap analogue of the W[1] $\ne$ FPT assumption. There are many (equivalent) ways to state PIH. We choose to state it in terms of an inapproximability of the colored version of \pname{Densest $k$-Subgraph}. In \pname{Multicolored Densest $k$-Subgraph}, we are given a graph $G = (V, E)$ where the vertex set $V$ is partition in to $k$ parts $V_1, \dots, V_k$. The goal is to select $k$ vertices $v_1 \in V_1, v_2 \in V_2, \dots, v_k \in V_k$ such that $\{v_1, \dots, v_k\}$ induces as many edges as possible.

It is easy to see that the exact version of this problem is W[1]-hard, via a straightforward reduction from \pname{$k$-Clique}. PIH postulates that even the approximate version of this problem is hard:

\begin{Hypothesis}[Parameterized Inapproximability Hypothesis 
(PIH)\footnote{We remark that the original conjecture in~\cite{LRSZ20} 
says that the problem is W[1]-hard to approximate. However, we choose to state 
the more relaxed form here.}~\cite{LRSZ20}] \label{hyp:pih}
For some constant $\varepsilon > 0$, there is no $(1 + \varepsilon)$ factor FPT approximation algorithm for \pname{Multicolored Densest $k$-Subgraph}.
\end{Hypothesis}

There are two important remarks about PIH. First, the factor $(1 + \varepsilon)$ 
is not important, and the conjecture remains equivalent even if we state it for 
a factor $C$ for any arbitrarily large constant $C$; this~is due to gap 
amplification via parallel repetition~\cite{Raz98}. Second, PIH implies that 
\pname{$k$-Clique} is hard to approximate to within any constant factor:

\begin{Lemma}
Assuming PIH, there is no constant factor FPT approximation algorithm for 
\pname{$k$-Clique}. 
\end{Lemma}

The above result can be shown via a classic reduction of Feige, Goldwasser, Lov{\'{a}}sz, Safra and Szegedy (henceforth FGLSS)~\cite{FeigeGLSS96}, which was one of the first works connecting proof systems and hardness of approximation. Specifically, the FGLSS reduction transforms $G$ to another graph $G'$ by viewing the edges of $G$ as vertices of $G'$. Then, we connect $\{u_1, v_1\}$ and $\{u_2, v_2\}$ {except} when the union $\{u_1, v_1\} \cup \{u_2, v_2\}$ contains two distinct vertices from the same partition. One can argue that the size of the largest clique in $G'$ is exactly equal to the number of edges in the optimal solution of \pname{Multicolored Densest $k$-Subgraph} on $G$. As a result, PIH implies hardness of approximation of the former. Interestingly, however, it is not known if the inverse is true and this remains an interesting open question:

\begin{Open Question} \label{q:PIH-v-Clique}
Does PIH hold if we assume that \pname{$k$-Clique} is FPT inapproximable to within any 
constant factor?
\end{Open Question}

As demonstrated by the FGLSS reduction, once we have a gap, it is much easier to give a reduction to another hardness of approximation result, because we do not have to create the initial gap ourselves (as in the previous subsection) but only need to preserve or amplify the gap. Indeed, PIH turns out to be a pretty robust hypothesis that gives FPT inapproximability for many problems, including \pname{$k$-Clique}, \pname{Directed Odd Cycle Traversal}~\cite{LRSZ20} and \pname{Strongly Connected Steiner Subgraph}~\cite{chitnis2017parameterized}. We~remark that the current situation here is quite similar to that of the landscape of the classic theory of hardness of approximation {before} the PCP Theorem~\cite{AS92,ALMSS98} was proved. There, Papadimitriou and Yannakakis introduced a complexity class \textsc{MAX-SNP} and show that many optimization problems are hard (or complete) for this class~\cite{PapadimitriouY91}. Later, the PCP Theorem confirms that these problems are NP-hard. In our case of FPT inapproximability, PIH seems to be a good analogy of \textsc{MAX-SNP} for problems in W[1] and, as mentioned before, PIH has been used as a starting point of many hardness of approximation results. However, there has not yet been many reverse reductions to PIH, and this is one of the motivation behind Question~\ref{q:PIH-v-Clique} above.

Despite the aforementioned applications of PIH, there are still quite a few 
questions that seem out of reach of PIH, such as whether there is an $o(k)$ 
factor FPT approximation for \pname{$k$-Clique} or questions related to running time 
lower bounds of approximation algorithms. On this front, another stronger 
conjecture called the Gap Exponential Time Hypothesis (Gap-ETH) is often used 
instead:

\begin{Hypothesis}[Gap Exponential Time Hypothesis (Gap-ETH)~\cite{Dinur16,ManurangsiR17}] \label{hyp:gap-eth}
For some constants $\varepsilon, \delta > 0$, there~is no $O(2^{\delta n})$-time algorithm that can, given a 3CNF formula, distinguish between the following two cases:
\begin{itemize}
\item (Completeness) the formula is satisfiable.
\item (Soundness) any assignment violates more than $\varepsilon$ fraction of the clauses.
\end{itemize}

Here $n$ denotes the number of clauses.\footnote{The version where $n$ 
denotes the number of variables is equivalent to the current formulation, 
because we can always assume without loss of generality that $m = O(n)$ 
(see~\cite{Dinur16,ManurangsiR17}).}
\end{Hypothesis}
 
Clearly, Gap-ETH is a strengthening of ETH, which can be thought of in the above form but with $\varepsilon = 1/n$. Another interesting fact is that Gap-ETH is stronger than PIH. This can be shown via the standard reduction from \pname{3SAT} to \pname{$k$-Clique} that establishes $N^{\Omega(k)}$ lower bound for the latter. The~reduction, due to Chen et al.~\cite{ChenHKX04,ChenHKX06}, proceed as follows. First, we partition the set of clauses $C$ into $C_1, \dots, C_k$ each of size $n/k$. For each $C_i$, we create a partition $V_i$ in the new graph where each vertex corresponds to all partial assignments (to variables that appear in at least one clause of $C_k$) that satisfy all the clauses in $C_k$. Two vertices are connected if the corresponding partial assignments are consistent, i.e., they do not assign a variable to different values. 

If there is an assignment that satisfies all the clauses, then clearly the restrictions of this assignment to each clause corresponds to $k$ vertices from different partitions that form a clique. On the other hand, it is also not hard to argue that, in the soundness case, the number of edges induced by any $k$ vertices from different partitions is at most $1 - \Theta(\varepsilon)$. Thus, Gap-ETH implies PIH as claimed.

Now that we have demonstrated that Gap-ETH is at least as strong as PIH, we may go further and ask how much more can we achieve from Gap-ETH, compared to PIH. The obvious consequences of Gap-ETH is that it can give explicit running time lower bounds for FPT hardness of approximation results. Perhaps more surprising, however, is that it can be used to improve the {inapproximability ratio} as well. The rest of this subsection is devoted to present some of these examples, together with brief overviews of how the proofs of these results work.

\subsubsection{Strong Inapproximability of $k$-Clique}
\label{sec:clique}

Our first example is the \pname{$k$-Clique} problem. Obviously, we can approximate \pname{$k$-Clique} to within a factor of $k$, by just outputting any single vertex. It had long been asked whether an $o(k)$-approximation is achievable in FPT time. As we saw above, PIH implies that a constant factor FPT approximation does not exist, but does not yet resolve this question. Nonetheless, assuming Gap-ETH, this question can be resolved in the negative:

\begin{Theorem}[\cite{param-inapprox}] \label{thm:clique-total-inapprox}
Assuming Gap-ETH, there is no $o(k)$-FPT-approximation for \pname{$k$-Clique}.
\end{Theorem}

The reduction used in~\cite{param-inapprox} to prove the above inapproximability 
is just a simple modification of the above reduction~\cite{ChenHKX04,ChenHKX06} 
that we saw for \pname{$k$-Clique}. Suppose that we would like to rule out a 
$\frac{k}{g}$-approximation, where $g = g(k)$ is a function such that $\lim_{k 
\to \infty} g(k) = \infty$. The only change in the reduction is that, instead 
of letting $C_1, \dots, C_k$ be the partition of the set of clauses $C$, we let 
each $C_i$ be a set of $\frac{D n}{g}$ clauses for some sufficiently large 
constant $D > 0$. The rest of the reduction works similar to before: for each 
$C_i$, we create a vertex corresponding to each partial assignment that 
satisfies all the clauses in $C_i$. Two vertices are joined by an edge if and 
only if they are consistent. This completes the description of the reduction.

To see that the reduction yields Theorem~\ref{thm:clique-total-inapprox}, first note 
that, if there is an assignment that satisfies the CNF formula, then we can 
again pick the restrictions on this formula onto $C_1, \dots, C_k$; these gives 
$k$ vertices that induces a clique in the graph. 

On the other hand, suppose that every assignment violates more than an 
$\varepsilon$ fraction of clauses. We~will argue that there is no clique of size 
$g$ in the constructed graph. The only property we need from the subsets $C_1, 
\dots, C_k$ is that the union of any $g$ such subsets contain at least $(1 - 
\varepsilon)$ fraction of the clauses. It is not hard to show that this is true 
with high probability, when we choose $D$ to be sufficiently large. Now, suppose 
for the sake of contradiction that there exists a clique of size $g$ in the 
graph. Since the vertices corresponding to the same subset $C_i$ form an 
independent set, it must be that these $g$ vertices are from different subsets. 
Let us call these subsets $C_{i_1}, \cdots, C_{i_g}$. Because these vertices 
induce a clique, we can find a global assignment that is consistent with each 
vertex. This~global assignment satisfies all the clauses in $C_{i_1} \cup \cdots 
\cup C_{i_g}$. However, $C_{i_1} \cup \cdots \cup C_{i_g}$ contains at least $1 
- \varepsilon$ fraction of all clauses, which contradicts to our assumption that 
every assignment violates more than $\varepsilon$ fraction of the clauses. 

Now, if we can $o(k)$-approximate \pname{$k$-Clique} in $T(k) \cdot N^{O(1)}$ time. 
Then, we may run this algorithm to distinguish the two cases in Gap-ETH in
$f(k) \cdot (2^{D n/g})^{O(1)} = 2^{o(n)}$ time, which violates Gap-ETH. This 
concludes our proof sketch. We end by remarking that the reduction may also be 
viewed as an instantiation of the randomized graph 
product~\cite{BermanS92,BellareGS98,Zuckerman96,Zuckerman96}, and it can also be 
derandomized. We~omit the details of the latter here. Interested readers may 
refer to~\cite{param-inapprox} for more detail.

\subsubsection{Strong Inapproximability of Multicolored Densest $k$-Subgraph and Label Cover}

For our second example, we go back to the \pname{Multicolored Densest $k$-Subgraph} once again. Recall that PIH asserts that this problem is hard to approximate to some constant factor, and we have seen above that Gap-ETH also implies this. On the approximation front, however, only the trivial $k$-approximation algorithm is known: just pick a vertex that has edges to as many partitions as possible. Then, output that vertex and one of its neighbors from each partition. It is hence a natural question to ask whether it is possible to beat this approximation ratio. This question has been, up to lower order terms, answered in the negative, assuming Gap-ETH:

\begin{Theorem}[\cite{DM18}] \label{thm:2csp-strong-inapprox}
Assuming Gap-ETH, there is no $k^{1 - o(1)}$-approximation for \pname{Multicolored Densest $k$-Subgraph}.
\end{Theorem}

An interesting aspect of the above result is that, even in the NP-hardness 
regime, no NP-hardness of factor $k^{\gamma}$ for some constant $\gamma > 0$ is 
known. In fact, the problem is closely related to (and is a special case of) a 
well-known conjecture in the hardness of approximation community called the 
Sliding Scale Conjecture (SSC)~\cite{BGLR93}.\footnote{See also the related 
Projection Game Conjecture (PGC)~\cite{Moshkovitz15}.} (See~\cite{DM18} for 
more discussion on the relation between the two.) Thus, this is yet another 
instance where taking a parameterized complexity perspective helps us advance 
knowledge even in the classical settings.

To prove Theorem~\ref{thm:2csp-strong-inapprox}, arguably the most natural 
reduction here is the above reduction for Clique! Note that we now view the 
vertices corresponding to each subset $C_i$ as forming a partition $V_i$. The~argument in the YES case is exactly the same as before: if the formula is 
satisfiable, then there is a (multicolored) $k$-clique. However, as the readers 
might have noticed, the argument in the NO case does not go through anymore. In 
particular, even when the graph is quite dense (e.g., having half of the edges 
present), it may not contain any large clique at all and hence it is unclear how 
to recover back an assignment that satisfies a large fraction of constraints.

This obstacle was overcomed in~\cite{DM18} by proving an {agreement testing 
theorem} (i.e., {direct product theorem}), which is of the following form. 
Given $k$ local functions $f_1, \dots, f_k$, where $f_i: S_i \to \{0, 1\}$ is a 
boolean function whose domain $S_i$ is a subset of a universe $\cU$. If some 
(small) $\zeta$ fraction of the pairs agree\footnote{Naturally, we say that 
two functions $f_i$ and $f_j$ {agree} iff $f_i(x) = f_j(x)$ for all $x \in S_i 
\cap S_j$.} with each other, then we can find (i.e., ``decode'') a global 
function $h: [n] \to \{0, 1\}$ that ``approximately agrees'' with roughly 
$\zeta$ fraction of the local functions. The theorem in~\cite{DM18} works when 
$S_1, \dots, S_k$ are sets of size $\Omega(n)$. 

Due to the technical nature of the definitions, we will not fully formalize 
the notions in the previous paragraph. Nonetheless, let us sketch how to apply 
the agreement testing theorem to prove the NO case for our reduction. Suppose 
for the sake of contradiction that the formula is not $(1 - \delta)$-satisfiable 
and that there exists a $k$-subgraph with density $\zeta \geq \frac{1}{k^{1 - 
o(1)}}$. Recall that each selected vertex is simply a partial assignment onto 
the subset of clauses $C_i$ for some $i$; we may view this as a function $f_i: 
S_i \to \{0, 1\}$ where $S_i$ denote the set of variables that appear in $C_i$. 
Here the universe $\cU$ is the set of all variables. With this perspective, we 
can apply the agreement testing theorem to recover a global function $h: \cU \to 
\{0, 1\}$ that ``approximately agrees'' with roughly $\zeta k$ of the local 
functions. Notice that, in this context, $h$ is simply a global assignment for 
the CNF formula. Previously in the proof for inapproximability of Clique, we had 
a global assignment that (perfectly) agrees with $g$ local functions, from which 
we can conclude that this assignment satisfies all but $\delta$ fraction of the 
clauses. It~turns out that relaxing ``perfect agreement'' to ``approximate 
agreement'' does not affect the proof too much, and the latter still implies 
that $h$ satisfies all but $\delta$ fraction of clauses as desired.

As for the proof of the agreement testing theorem itself, we will not delve too much into detail here. However, we note that the proof is based on looking at different ``agreement levels'' and the graph associated with them. It turns out that such a graph has a certain transitivity property, which~allows one to ``decode'' back the global function $h$. This general approach of looking at different agreement levels and their transitivity properties is standard in the direct product/agreement testing literature~\cite{RazS97,ImpagliazzoKW12,DinurN17}. The main challenge in~\cite{DM18} is to make the proof works for $\zeta$ as small as $1/k$, which requires a new notion of transitivity.

To end this subsection, we remark that the \pname{Multicolored Densest $k$-Subgraph} is 
known as the 2-ary Constraint Satisfaction Problem (\pname{2-CSP}) in the classical 
hardness of approximation community. The problem, and in particular its special 
case called {Label Cover}, serves as the starting point of almost all known 
NP-hardness of approximation (see e.g.,~\cite{AroraBSS93,Hastad01,Chan16}). The 
technique in~\cite{DM18} can also be used to show inapproximability for Label 
Cover with strong running time lower bound of the form $f(k) \cdot 
N^{\Omega(k)}$~\cite{M20}. Due to known reductions, this has numerous 
consequences. For example, it implies, assuming Gap-ETH, that approximating 
\pname{$k$-Even Set} to within any factor less than two cannot be done in $f(k) \cdot 
N^{o(k)}$ time, considerably improving the lower bound mentioned in the previous 
subsection.

\subsubsection{Inapproximability of $k$-Biclique and Densest $k$-Subgraph}\label{sec:dense}

While PIH (or equivalently the \pname{Multicolored Densest $k$-Subgraph} problem) can serve as a starting point for hardness of approximation of many problems, there are some problems for which not even a constant factor hardness is known under PIH, but strong inapproximability results can be obtained via Gap-ETH. We will see two examples of this here.

First is the \pname{$k$-Biclique} problem. Recall that in this problem, we are 
given a bipartite graph and we would like to determine whether there is a 
complete bipartite subgraph of size $k$. As stated earlier in the previous 
subsection, despite its close relationship to \pname{$k$-Clique}, 
\pname{$k$-Biclique} turned out to be a much more challenging problem to prove 
intractibility and even its W[1]-hardness was only shown recently~\cite{Lin18}. 
This difficulty is corroborated by its approximability status in the classical 
(non-parameterized) regime; while \pname{Clique} is long known to be NP-hard to 
approximate to within $N^{1 - o(1)}$ factor~\cite{Hastad96}, \pname{Biclique} is 
not even known to be NP-hard to approximate to within say 1.01 
factor.\footnote{We note, however, that strong inapproximability of 
\pname{Biclique} is known under stronger 
assumptions~\cite{Khot06,BhangaleGHKK17,manurangsi2018inapproximability}.} With 
this in mind, it is perhaps not a surprise that \pname{$k$-Biclique} is not 
known to be hard to approximate under PIH. Nonetheless, when we assume Gap-ETH, 
we can in fact prove a very strong hardness of approximation for the problem:

\begin{Theorem}[\cite{param-inapprox}] \label{thm:biclique-total-inapprox}
Assuming Gap-ETH, there is no $o(k)$-FPT-approximation for \pname{$k$-Biclique}.
\end{Theorem}

Note that, similar to \pname{$k$-Clique}, a $k$-approximation for \pname{Biclique} can be easily achieved by outputting a single edge. Hence, in terms of the inapproximability ratio, the above result is tight. 

Due to its technicality, we only sketch an outline of the proof of 
Theorem~\ref{thm:biclique-total-inapprox} here. Firstly, the~reduction starts by 
constructing a graph that is similar (but not the same) to that of $k$-Clique 
that we describe above. The main properties of this graph is that (i) in the YES 
case where the formula is satisfiable, the graph contains many copies of 
\pname{$k$-Biclique}, and (ii) in the NO case where the formula is not even $(1 - 
\delta)$-satisfiable, the graph contains few copies of \pname{$g$-Biclique}. The 
construction and these properties were in fact shown in~\cite{Manurangsi17}. 
In~\cite{param-inapprox}, it was observed that, if we subsample the graph by 
keeping each vertex independently with probability $p$ for an appropriate value 
of $p$, then (i) ensures that at least one of the \pname{$k$-Biclique} survives the 
subsampling , whereas (ii) ensures that no \pname{$g$-Biclique} survives. This indeed 
gives the claimed result in the above theorem.

We remark that, while Theorem~\ref{thm:biclique-total-inapprox} seems to resolve the 
approximability of \pname{$k$-Biclique}, there is still one aspect that is not yet 
completely understood: the running time lower bound. To demonstrate this, recall 
that, for \pname{$k$-Clique}, the reduction that gives hardness of \pname{$k$-vs-$g$ Clique} has 
size $2^{O(n/g)}$; this~means that we have a running time lower bound of $f(k) 
\cdot N^{\Omega(g)}$ on the problem. This is of course tight, because we can 
determine whether a graph has a $g$-clique in $N^{O(g)}$ time.  However, for~\pname{$k$-Biclique}, the known reduction that gives hardness for \pname{$k$-vs-$g$ Biclique} 
has size $2^{O(n/\sqrt{g})}$. This~results in a running time lower bound of only 
$f(k) \cdot N^{\Omega(\sqrt{g})}$. Specifically, for the most basic setting of 
constant factor approximation, Theorem~\ref{thm:biclique-total-inapprox} only 
rules out algorithms with running time $f(k) \cdot N^{o(\sqrt{k})}$. Hence, an 
immediate question here is:

\begin{Open Question} \label{q:biclique-time}
Is there an $f(k) \cdot N^{o(k)}$-time algorithm that approximates \pname{$k$-Biclique} to within a constant factor?
\end{Open Question}

To put things into perspective, we note that, even for {exact} algorithms for \pname{$k$-Biclique}, the best running time lower bound is still $f(k) \cdot N^{\Omega(\sqrt{k})}$~\cite{Lin18} (under any reasonable complexity assumption). This means that, to answer Question~\ref{q:biclique-time}, one has to first settle the best known running time lower bound for exact algorithms, which would already be a valuable contribution to the understanding of the problem.

Let us now point out an interesting consequence of 
Theorem~\ref{thm:biclique-total-inapprox} for the \pname{Denest $k$-Subgraph} 
problem. This is the ``uncolored'' version of the \pname{Multicolored Densest 
$k$-Subgraph} problem as defined above, where there are no partitions $V_1, 
\dots, V_k$ and we can pick any $k$ vertices in the input graph $G$ with the 
objective of maximizing the number of induced edges. The approximability status 
of \pname{Denest $k$-Subgraph} very much mirrors that of \pname{$k$-Biclique}. 
Namely, in the parameterized setting, PIH is not known to imply hardness of 
approximation for \pname{Densest $k$-Subgraph}. Furthermore, in the classic 
(non-parameterized) setting, \pname{Densest $k$-Subgraph} is not 
known\footnote{Again, similar to \pname{Biclique}, \pname{Densest 
$k$-Subgraph} is known to be hard to approximate under stronger 
assumptions~\mbox{\cite{Khot06,raghavendra2010graph,AAMMW11,Manurangsi17}}.} to 
be NP-hard to approximate even to within a factor of say 1.01. Despite these, 
Gap-ETH does give a strong inapproximability for \pname{Densest $k$-Subgraph}, 
as stated below:

\begin{Theorem}[\cite{param-inapprox}] \label{thm:dks-inapprox}
Assuming Gap-ETH, there is no $k^{o(1)}$-FPT-approximation for \pname{Densest $k$-Subgraph}.
\end{Theorem}

In fact, the above result is a simple consequence of 
Theorem~\ref{thm:biclique-total-inapprox}. To see this, recall the following classic 
result in extremal graph theory commonly referred to as the 
K\H{o}v\'{a}ri-S\'{o}s-Tur\'{a}n (KST) Theorem~\cite{KST54}: any $k$-vertex 
graph that does not contain a $g$-biclique as a subgraph has density at most 
$O(k^{-1/g})$. Now, the hardness for \pname{$k$-Biclique} from 
Theorem~\ref{thm:biclique-total-inapprox} tells us that there is no FPT time 
algorithm that can distinguish between the graph containing $k$-biclique from 
one that does not contain $g$-biclique for any $g = \omega(1)$. When the graph 
contains a $k$-biclique, we have a $k$-vertex subgraph with density (at least) 
1/2. On the other hand, when the graph does not even contain a $g$-biclique, 
the KST Theorem ensures us that any $k$-vertex subgraph has density at most 
$O(k^{1/g})$. This indeed gives a gap of $O(k^{1/g})$ in terms of approximation 
Densest $k$-Subgraph and finishes the proof sketch for 
Theorem~\ref{thm:dks-inapprox}.

Unfortunately, Theorem~\ref{thm:dks-inapprox} does not yet resolve the FPT 
approximability of \pname{Densest $k$-Subgraph}. In particular, while the hardness is 
only of the form $k^{o(1)}$, the best known algorithm (which is the same as that 
of the multicolored version discussed above) only gives an approximation ratio 
of $k$. Hence, we~may ask whether this can be improved:

\begin{Open Question} \label{q:dks-time}
Is there an $o(k)$-FPT-approximation algorithm for \pname{Densest $k$-Subgraph}?
\end{Open Question}

This should be contrasted with Theorem~\ref{thm:2csp-strong-inapprox}, for which the 
FPT approximability of \pname{Multicolored Densest $k$-Subgraph} is essentially resolved 
(up to lower order terms).

\section{Algorithms}\label{sec:algo}

In this section we survey some of the developments on the algorithmic side in 
recent years. The~organization of this section is according to problem 
types. We begin with 
basic packing and covering problems in 
Sections~\ref{sec:packing} and \ref{sec:covering}. 
We then move on to 
clustering in Section~\ref{sec:clustering}, 
network design in Section~\ref{sec:network-design},
and cut problems in Section~\ref{sec:cuts}.
In Section~\ref{sec:width-reduction} we present width 
reduction~problems. 

The algorithms in the above mentioned subsections compute 
approximate solutions to problems that are W[1]-hard. Therefore it is necessary 
to approximate, even when using parameterization. However, one may also aim to 
obtain faster parameterized runtimes than the known FPT algorithms, by 
sacrificing in the solution quality. We present some results of this type in 
Section~\ref{sec:faster-smaller}.

\subsection{Packing Problems}
\label{sec:packing}

For a packing problem the task is to select as many combinatorial objects of 
some mathematical structure (such as a graph or a set system) as possible under some constraint, 
which restricts some objects to be picked if others are. A basic example is the 
\pname{Independent Set} problem, for which a maximum sized set of vertices of a 
graph needs to be found, such that none of them are adjacent to each other. 

\subsubsection{Independent Set}

The \pname{Independent Set} problem is notoriously hard in general. Not only is 
there no polynomial time $n^{1-\eps}$\hy{}approximation 
algorithm~\cite{zuckerman2006linear} for any constant $\eps>0$, unless P=NP, but 
also, under \gapeth, no $g(k)$\hy{}approximation can be computed in $f(k)\polyn$ 
time~\cite{param-inapprox} for any computable functions $f$ and~$g$, where $k$ 
is the solution size. On the other hand, for planar graphs a PTAS 
exists~\cite{baker1994approximation}. Hence a natural question is how the 
problem behaves for graphs that are ``close'' to being planar. 

\iffalse
This can be formalized in several ways. For instance it is well-known that planar graphs 
are exactly those excluding $K_5$ and $K_{3,3}$ as minors. Both of these minors 
are {apex graphs}, which are graphs that contain a vertex (the 
{apex}) whose removal leaves a planar graph. Thus a generalization of 
planar graphs is the class of {apex-minor-free} graphs, which exclude some 
fixed apex graph $H$ as a minor (for example bounded {genus} graphs are 
apex-minor-free~\cite{eppstein2002subgraph}.) 
\fi

One way to generalize planar graphs is to consider minor-free graphs, 
because planar graphs are exactly those excluding $K_5$ and $K_{3,3}$ as minors.
When parameterizing by the size of 
an excluded minor, the \pname{Independent Set} problem is paraNP-hard, 
since the problem is NP-hard on planar graphs~\cite{johnson1979computers}. 
Nevertheless a PAS can be obtained for this 
parameter~\cite{demaine2004equivalence}.

\begin{Theorem}[\cite{demaine2004equivalence, grohe2013simple}]\label{thm:apex-minor}
Let $H$ be a fixed graph. For $H$-minor-free graphs, 
\pname{Independent Set} admits an $(1+\eps)$-approximation 
algorithm that runs in $f(H, \eps) \polyn$ time for some function $f$. 
\end{Theorem}

This result is part of the large framework of ``bidimensionality theory'' where any graph in an appropriate minor-closed class has treewidth bounded above in terms of the
problem's solution value, typically by the square root
of that value. These properties lead to efficient, often
subexponential, fixed-parameter algorithms, as well as
polynomial-time approximation schemes, for bidimensional problems in many minor-closed graph classes.
The bidimensionality theory  is based on
algorithmic and combinatorial extensions to parts of the
Robertson--Seymour Graph Minor Theory, in particular
initiating a parallel theory of graph contractions. The
foundation of this work is the topological theory of
drawings of graphs on surfaces.
We refer the reader to the survey of~\cite{demaine2008bidimensionality}
and more recent papers~\cite{fomin2011bidimensionality, demaine2011contraction, grohe2013simple}.

A different way to generalize planar graphs is to consider a {planar 
deletion set}, i.e., a set of vertices in the input graph whose removal leaves a 
planar graph. Taking the size of such a set as a parameter, \pname{Independent 
Set} is again paraNP-hard~\cite{johnson1979computers}. However, by first finding 
a minimum sized planar deletion set, then guessing the intersection of this set 
with the optimum solution to \pname{Independent Set}, and finally using the PTAS 
for planar graphs~\cite{baker1994approximation}, a PAS can be obtained 
parameterized by the size of a planar deletion set~\cite{marx2008fpa}.

\begin{Theorem}[\cite{marx2008fpa}]
For the \pname{Independent Set} problem a $(1+\eps)$-approximation can be 
computed in $2^{k}n^{O(1/\eps)}$ time for any $\eps>0$, where $k$ is the size 
of a minimum planar deletion set.
\label{thm:noisyplanar}
\end{Theorem}

Ideas using linear programming allow us to generalize and handle larger noise at the expense of worse dependence on $\eps$. 
\citet{bansal2017lp} showed that given a graph obtained by adding $\delta n$ 
edges to some planar graph, one can compute a $(1 + O(\eps + 
\delta))$-approximate independent set in time $n^{O(1/\eps^4)}$, 
which is faster than the $2^{k} n^{O(1/\eps)}$ running time of Theorem~\ref{thm:noisyplanar} for large $k = \delta n$.~\citet{magen2009robust}~showed that for every graph $H$ and $\eps > 0$, given a 
graph $G = (V, E)$ that can be made $H$-minor-free after at most $\delta n$ 
deletions and additions of vertices or edges, the size of the maximum 
independent set can be approximately computed within a factor $(1 + \eps + 
O(\delta |H| \sqrt {\log |H|}))$ in time $n^{f(\eps, H)}$. Note~that this
algorithm does not find an independent set. 
Recently, Demaine et al.~\cite{demaine2019structural} presented a general framework to obtain better approximation algorithms for various problems including \pname{Independent Set} and \pname{Chromatic Number}, when the input graph is close to well-structured graphs (e.g., bounded degeneracy, degree, or treewidth). 

It is also worth noting here that \pname{Independent Set} problem can be generalized to the \pname{$d$-Scattered Set} problem where we are given an (edge-weighted) graph and are asked to select at least $k$ vertices, so that the distance between any pair is at least $d$~\cite{KLP18}. Recently in~\cite{KLP19} some lower and upper bounds on the approximation of the  \pname{$d$-Scattered Set} problem have been provided.

A special case of \pname{Independent Set} is the \pname{Independent Set of 
Rectangles} problem, where a set of axis-parallel rectangles is given in the 
two-dimensional plane, and the task is to find a maximum sized subset of 
non-intersecting rectangles. This is a special case, since pairwise 
intersections of rectangles can be encoded by edges in a graph for which the 
vertices are the rectangles. Parameterized by the solution size, the problem is 
W[1]-hard~\cite{marx2005efficient}, and while a QPTAS is 
known~\cite{adamaszek2013approximation}, it is a challenging open question 
whether a PTAS exists. It was shown~\cite{grandoni2019parameterized} however 
that both a PAS and a PSAKS exist for \pname{Independent Set of Rectangles} 
parameterized by the solution size, even for the weighted version.

The runtime of this PAS is $f(k,\eps)n^{g(\eps)}$ for some functions $f$ and 
$g$, where $k$ is the solution size. Note that the dependence on $\eps$ in the 
degree of the polynomial factor of this algorithm cannot be removed, unless 
FPT=W[1], since any {efficient} PAS with runtime $f(k,\eps)\polyn$ could be 
used to compute the optimum solution in FPT time by setting $\eps$ to 
$\frac{1}{k+1}$ in the W[1]-hard unweighted version of the 
problem~\cite{marx2005efficient}. However, in the so-called {shrinking 
model} an efficient PAS can be obtained~\cite{pilipczuk2017approximation} for 
\pname{Independent Set of Rectangles}. The parameter in this case is a factor 
$0<\delta<1$ by which every rectangle is shrunk before computing an approximate 
solution, which is compared to the optimum solution without shrinking. 

\begin{Theorem}[\cite{grandoni2019parameterized,pilipczuk2017approximation}]
For the \pname{Independent Set of Rectangles} problem a $(1+\eps)$-approximation 
can be computed in $k^{O(k/\eps^8)}n^{O(1/\eps^8)}$ time for any $\eps>0$, where 
$k$ is the size of the optimum solution, or in $f(\delta,\eps)\polyn$ time for 
some computable function $f$ and any $\eps>0$ and $0<\delta<1$, where $\delta$ 
is the shrinking factor. Moreover, a $(1+\eps)$\hy{}approximate kernel with 
$k^{O(1/\eps^8)}$ rectangles can be computed in polynomial time.
\end{Theorem}

Another special case of \pname{Independent Set} is the \pname{Independent Set on Unit Disk graph} problem, where  given set of $n$ unit disks in the Euclidean plane, the task is to determine if  there exists a set of $k$ non-intersecting
disks. The problem is NP-hard~\cite{ClarkCJ90} but admits a PTAS~\cite{HuntMRRRS98}. Marx~\cite{marx2005efficient} showed that, when parameterized by the solution size, the problem is 
W[1]-hard; this also rules out EPTAS (and even {efficient} PAS) for the problem, assuming $\FPT \ne \W[1]$. On the other hand, in~\cite{AF04} the authors give an FPT algorithm for a special case of \pname{Independent Set on Unit Disk graph} when there is a lower bound on the distance between any pair of centers.

\iffalse
One more problem loosely related to scheduling is the \pname{2D Geometric 
Knapsack} problem, where a set of axis-parallel rectangles is given together 
with an axis-parallel square. The goal is to find the largest number of 
rectangles that can be packed into the square, i.e., their coordinates can be 
translated so that they are fully contained in the square without overlapping. 
A variant of \pname{2D Geometric Knapsack} also allows us to rotate the rectangles 
by 90 degrees to pack them into the square. While it is known that both variants 
of the problem are NP-hard~\cite{leung1990packing}, it is not known whether a 
PTAS exists. When parameterizing by the solution size, both variants are 
W[1]-hard~\cite{grandoni2019parameterized}, and for the variant that allows 
rotations a PAS and a PSAKS exist~\cite{grandoni2019parameterized}.

\begin{Theorem}[\cite{grandoni2019parameterized}]
For the \pname{2D Geometric Knapsack} problem a $(1+\eps)$-approximation can be 
computed in $k^{O(k/\eps)}n^{O(1/\eps^3)}$ time for any $\eps>0$, where $k$ is 
the size of the optimum solution. Moreover, a $(1+\eps)$\hy{}approximate kernel 
with $k^{O(1/\eps)}$ rectangles can be computed in polynomial time.
\end{Theorem}
\fi

\subsubsection{Vertex Coloring}

A problem related to \pname{Independent Set} is the \pname{Vertex Coloring} 
problem, for which the vertices need to be colored with integer values, such 
that no two adjacent vertices have the same color (which means that each color 
class forms an independent set in the graph). The task is to minimize the number 
of used colors. For planar graphs the problem has a polynomial time 
$4/3$-approximation algorithm~\cite{marx2008fpa} via the celebrated Four Color 
Theorem, and a better approximation is not possible in polynomial 
time~\cite{stockmeyer1973planar}. Using this algorithm, a $7/3$-approximation 
can be computed in FPT time when parameterizing by the size of a planar deletion 
set~\cite{marx2008fpa}. When generalizing planar graphs by excluding any fixed 
minor, and taking its size as the parameter, a $2$-approximation can be computed 
in FPT time~\cite{demaine2005algorithmic}. Due to the NP-hardness for planar 
graphs~\cite{stockmeyer1973planar}, neither of these two parameterizations 
admits a PAS, unless P=NP.

\begin{Theorem}[\cite{marx2008fpa,demaine2005algorithmic}]
For the \pname{Vertex Coloring} problem 
\begin{itemize}
\item a $7/3$-approximation can be computed in $k^{k}\polyn$ time, where $k$ is 
the size of a minimum planar deletion set, and
\item a $2$-approximation can be computed in $f(k)\polyn$ time for some 
function $f$, where $k$ is the size of an excluded minor of the input graph.
\end{itemize}
\end{Theorem}

One way to generalize \pname{Vertex Coloring} is to see each color class as an 
induced graph of degree~$0$. The \pname{Defective Coloring} 
problem\footnote{sometimes called \pname{Improper Coloring}} correspondingly 
asks for a coloring of the vertices, such that each color class induces a graph 
of maximum degree $\Delta$, for some given $\Delta$. The aim again is to 
minimize the number of used colors. In contrast to \pname{Vertex Coloring}, the 
\pname{Defective Coloring} problem is W[1]-hard~\cite{belmonte2018parameterized} 
parameterized by the {treewidth}. This parameter measures how ``tree-like'' a 
graph is, and is defined as follows. 

\begin{Definition}\label{def:treewidth}
A {tree decomposition} of a graph $G=(V,E)$ is a tree $T$ for which every 
node is associated with a {bag} $X\subseteq V$, such that the following 
properties hold:
\begin{enumerate}
\item the union of all bags is the vertex set $V$ of $G$,
\item for every edge $(u,v)$ of $G$, there is a node of $T$ for which the 
associated bag contains $u$ and~$v$, and
\item for every vertex $u$ of $G$, all nodes of $T$ for which the associated 
bags contain $u$, induce a connected subtree of~$T$.
\end{enumerate}

The {width} of a tree decomposition is the size of the largest bag minus 
$1$ (which implies that a tree has a decomposition of width $1$ where each bag 
contains the endpoints of one edge). The treewidth of a graph is the 
smallest width of any of its tree decompositions. 
\end{Definition}

Treewidth is  fundamental parameter of a graph and will be discussed more 
elaborately  in Section~\ref{sec:treewidth}. However, it is worth mentioning 
here that \pname{Vertex Coloring} is in FPT  when parameterized by treewidth. 
 
The strong polynomial-time approximation lower bound of $n^{1-\eps}$ for 
\pname{Vertex Coloring}~\cite{zuckerman2006linear} naturally carries over to the 
more general \pname{Defective Coloring} problem. A much improved approximation 
factor of~$2$ is possible though in FPT time if the parameter is the 
treewidth~\cite{belmonte2018parameterized}. It can be shown however, that a PAS 
is not possible in this case, as there is no $(3/2-\eps)$-approximation 
algorithm for any $\eps>0$ parameterized by the 
treewidth~\cite{belmonte2018parameterized}, unless FPT=W[1].
A natural question is whether the bound $\Delta$ of \pname{Defective Coloring} 
can be approximated instead of the number of colors. For this setting, a 
bicriteria PAS parameterized by the treewidth 
exists~\cite{belmonte2018parameterized}, which computes a solution with the 
optimum number of colors where each color class induces a graph of maximum 
degree at most $(1+\eps)\Delta$.

\begin{Theorem}[\cite{belmonte2018parameterized}]
For the \pname{Defective Coloring} problem, given a tree decomposition of width 
$k$ of the input graph,
\begin{itemize}
\item a solution with the optimum number of colors where each color class 
induces a graph of maximum degree $(1+\eps)\Delta$ can be computed in 
$(k/\eps)^{O(k)}\polyn$ time for any $\eps>0$,
\item a $2$-approximation (of the optimum number of colors) can be computed in 
$k^{O(k)}\polyn$ time, but
\item no $(3/2-\eps)$-approximation (of the optimum number of colors) can be 
computed in $f(k)\polyn$ time for any $\eps>0$ and computable function $f$, 
unless FPT=W[1].
\end{itemize}
\end{Theorem}

The algorithms of the previous theorem build on the techniques 
of~\cite{lampis2014fpas} using approximate addition trees in combination with 
dynamic programs that yield XP algorithms for these problems. This technique can 
be applied to various problems (cf.~Section~\ref{sec:covering}), including a 
different generalization of \pname{Vertex Coloring} called \pname{Equitable 
Coloring}. Here the aim is to color the vertices of a graph with as few colors 
as possible, such that every two adjacent vertices receive different colors, and~all color classes contain the same number of vertices. It is a generalization of 
\pname{Vertex Coloring}, since~one may add a sufficiently large independent set 
(i.e., a set of isolated vertices) to a graph such that the number of colors 
needed for an optimum \pname{Vertex Coloring} solution is the same as for an 
optimum \pname{Equitable Coloring} solution. 

The \pname{Equitable Coloring} problem is W[1]-hard even when combining the 
number of colors needed and the treewidth of the graph as 
parameters~\cite{fellows2011complexity}. On the other hand, a PAS 
exists~\cite{lampis2014fpas} if the parameter is the {cliquewidth} of the 
input graph. This is a weaker parameter than treewidth, as the cliquewidth of a 
graph is bounded as a function of its treewidth. However, while bounded treewidth 
graphs are sparse, cliquewidth also allows for dense graphs (such as complete 
graphs). Formally, a~graph of cliquewidth $\ell$ can be constructed using the 
following recursive operations using $\ell$ labels on the vertices:
\begin{enumerate}
\item Introduce($x$): create a graph containing a singleton vertex labelled 
$x\in\{1,\ldots,\ell\}$.
\item Union($G_1,G_2$): return the disjoint union of two vertex-labelled graphs 
$G_1$ and $G_2$.
\item Join($G,x,y$): add all edges connecting a vertex of label $x$ with a 
vertex of label $y$ to the vertex-labelled graph $G$.
\item Rename($G,x,y$): change the label of every vertex of $G$ with label 
$x$ to $y\in\{1,\ldots,\ell\}$.
\end{enumerate}

A {cliquewidth expression} with $\ell$ labels is a recursion tree 
describing how to construct a graph using the above four operations using only 
labels from the set $\{1,\ldots,\ell\}$.  Notice that the cliquewidth of a complete graph is two and therefore   we have graphs of bounded clique-width but unbounded treewidth.  As stated earlier the cliquewidth of a graph is bounded above exponentially in its treewidth and this dependence is tight for some graph families~\cite{CR05}.  

The PAS for \pname{Equitable Coloring} 
will compute a coloring using at most $k$ colors such that the ratio between the 
sizes of any two color classes is at most $1+\eps$. In this sense it is a 
{bicriteria} approximation algorithm.

\begin{Theorem}[\cite{lampis2014fpas}]
For the \pname{Equitable Coloring} problem, given a cliquewidth expression with 
$\ell$ labels for the input graph, a solution with optimum number of colors 
where the ratio between the sizes of any two color classes is at most $1+\eps$, 
can be computed in $(k/\eps)^{O(k\ell)}\polyn$ 
time\footnote{In~\cite{lampis2014fpas} the runtime of these 
algorithms is stated as $(\log n/\eps)^{O(k)}2^{k\ell}\polyn$, which can be 
shown to be upper bounded by~$(k/\eps)^{O(k\ell)}\polyn$ (see 
e.g.,~\cite[Lemma~1]{katsikarelis2019structural}).} for any~$\eps>0$, where 
$k$ is the optimum number of colors.
\end{Theorem}

A variant of \pname{Vertex Coloring} is the \pname{Min Sum Coloring} problem, 
where, instead of minimizing the number of colors, the aim is to minimize the sum 
of (integer) colors, where the sum is taken over all vertices. This problem is 
FPT parameterized by the treewidth~\cite{salavatipour2003sum}, but the related 
\pname{Min Sum Edge Coloring} problem is NP-hard~\cite{marx2009complexity} on 
graphs of treewidth~$2$ (while being polynomial time solvable on 
trees~\cite{giaro2000edge}). For this problem the edges need to be colored with 
integer values, so that no two edges sharing a vertex have the same color, and 
the aim again is to minimize the total sum of colors. Despite being 
APX-hard~\cite{marx2009complexity} and also paraNP-hard for parameter treewidth, 
\pname{Min Sum Edge Coloring} admits a PAS for this 
parameter~\cite{marx2004minimum}.

\begin{Theorem}[\cite{marx2004minimum}]
For the \pname{Min Sum Edge Coloring} problem a $(1+\eps)$-approximation can be 
computed in $f(k,\eps)n$ time for any $\eps>0$, where $k$ is the treewidth of 
the input graph.
\end{Theorem}

\subsubsection{Subgraph Packing}
\label{sec:subgraphpacking}
A special family of packing problems can be obtained by {subgraph packing}. Let $H$ be a fixed ``pattern'' graph. 
The \hpacking problem, given the ``host'' graph $G$, asks to find the maximum number of vertex-disjoint copies of $H$. One can also let $H$ be a family of graphs and ask the analogous problem. There is another choice whether each copy of $H$ is required to be an induced subgraph or a regular subgraph. We focus on the regular subgraph case here. 

When $H$ is a single graph with $k$ vertices, a simple greedy algorithm that 
finds an arbitrary copy of $H$ and adds it to the packing, guarantees a 
$k$-approximation in time $f(H, n) \cdot n$. Here $f(H, n)$ denotes the time to 
find a copy of $H$ in an $n$-vertex graph. Following a general result for \ksp, 
a~$(k + 1 + \eps)/3$-approximation algorithm that runs in polynomial time for 
fixed $k, \eps$ exists~\cite{cygan2013improved}. When~$H$ is 
$2$-vertex-connected or a star graph, even for fixed $k$, it is NP-hard to 
approximate the problem better than a factor $\Omega(k/ 
\polylog(k))$~\cite{guruswami2015inapproximability}. There is no known connected 
$H$ that admits an FPT (or even~XP) algorithm achieving a 
$k^{1-\delta}$-approximation for some $\delta > 0$; in particular, the 
parameterized approximability of \pname{$k$-Path Packing} is wide open. It is 
conceivable that \pname{$k$-Path Packing} admits a parameterized 
$o(k)$-approximation algorithm, given an $O(\log k)$-approximation algorithm 
for \pname{$k$-Path Deletion}~\cite{lee2017partitioning} and an improved kernel 
for \pname{Induced $P_3$ Packing}~\cite{fomin2019subquadratic}.

When $H$ is the family of all cycles, the problem becomes the \pname{Vertex 
Cycle Packing} problem, for~which the largest number of vertex-disjoint cycles 
of a graph needs to be found. No polynomial time $O(\log^{1/2-\eps} 
n)$-approximation is possible for this 
problem~\cite{friggstad2007approximability} for any $\eps>0$, unless every 
problem in NP can be solved in randomized quasi-polynomial time. Furthermore, 
despite being FPT~\cite{lokshtanov2019packing} parameterized by the solution 
size, \pname{Vertex Cycle Packing} does not admit any polynomial-sized exact 
kernel for this parameter~\cite{bodlaender2011kernel}, unless \NPcoNP. 
Nevertheless, a PSAKS can be found~\cite{lokshtanov2017lossy}.

\begin{Theorem}[\cite{lokshtanov2017lossy}]
For the \pname{Vertex Cycle Packing} problem, a $(1+\eps)$\hy{}approximate kernel 
of size $k^{O(1/(\eps\log\eps))}$ can be computed in polynomial time, where $k$ 
is the solution size.
\end{Theorem}

\subsubsection{Scheduling}

Yet another packing problem on graphs, which, however, has applications in 
scheduling and bandwidth allocation, is the \pname{Unsplittable Flow on a Path} 
problem. Here a path with edge capacities is given together with a set of tasks, 
each of which specifies a start and an end vertex on the path and a demand 
value. The goal is to find the largest number of tasks such that for each edge 
on the path the total demand of selected tasks for which the edge lies between 
its start and end vertex, does~not exceed the capacity of the edge. This problem 
admits a QPTAS~\cite{batra2015new}, but it remains a challenging open question 
whether a PTAS exists. When parameterizing by the solution size, 
\pname{Unsplittable Flow on a Path} is W[1]-hard~\cite{wiese2017unsplit}. 
However a PAS exists~\cite{wiese2017unsplit} for this parameter.

\begin{Theorem}[\cite{wiese2017unsplit}]
For the \pname{Unsplittable Flow on a Path} problem a $(1+\eps)$-approximation 
can be computed in $2^{O(k\log k)}n^{g(\eps)}$ time for some computable function 
$g$ and any $\eps>0$, where $k$ is the solution size.
\end{Theorem}

Another scheduling problem is \pname{Flow Time Scheduling}, for which a set of 
jobs is given, each of which is specified by a processing time, a release date, 
and a weight. The jobs need to be scheduled on a given number of machines, such 
that no job is processed before its release date and a job only runs on one 
machine at a time. Given a schedule, the {flow time} of a job is the 
weighted difference between its completion time and release date, and the task 
for the \pname{Flow Time Scheduling} problem is to minimize the sum of all flow 
times. Two types of schedules are distinguished: in a {preemtive} schedule 
a job may be interrupted on one machine and then resumed on another, while in a 
{non-preemtive} schedule every job runs on one machine until its completion 
once it was started. If pre-emptive schedules are allowed, \pname{Flow Time 
Scheduling} has no polynomial time $O(\log^{1-\eps}p)$-approximation 
algorithm~\cite{garg2008minimizing}, unless P~=~NP, where $p$ is the maximum 
processing time. For the more restrictive non-preemtive setting, no 
$O(n^{1/2-\eps})$\hy{}approximation can be computed in polynomial 
time~\cite{kellerer1999approximability}, unless~P~=~NP, where $n$ is the number 
of jobs. The latter lower bound is in fact even valid for only one machine, and 
thus parameterizing \pname{Flow Time Scheduling} by the number of machines will 
not yield any better approximation ratio in this setting. A natural parameter 
for \pname{Flow Time Scheduling} is the maximum over all processing times and 
weights of the given jobs. It is not known whether the problem is FPT or 
W[1]-hard for this parameter. However, when combining this parameter with the 
number of machines, a PAS can be obtained~\cite{wiese2018fixed} despite the 
strong polynomial time approximation lower bounds.

\begin{Theorem}[\cite{wiese2018fixed}]
For the \pname{Flow Time Scheduling} problem a $(1+\eps)$-approximation can be 
computed in $(mk)^{O(mk^3/\eps)}\polyn$ time in the preemtive setting, and in 
$(mk/\eps)^{O(mk^5)}\polyn$ time in the non-preemtive setting, for any $\eps>0$, 
where $m$ is the number of machines and $k$ is an upper bound on every 
processing time and weight.
\end{Theorem}

\subsection{Covering Problems}
\label{sec:covering}

For a covering problem the task is to select a set of $k$ combinatorial objects in a mathematical structure, such as a graph or set system (i.e., hypergraph), under some constraints that demands certain other objects to be intersected/covered. A basic example is the \pname{Set Cover} problem where we are given a set system, which is simply a collection of subsets of a universe. The goal is to determine whether there are $k$ subsets whose union cover the whole universe.

There are two ways define optimization based on covering problems. First, we may view the covering demands as strict constraints and aim to find a solution that minimize the constraint/cost while covering all objects (i.e., relaxing the size-$k$ constraint); this results in a minimization problem. Second, we may view the size constraint as a strict constraint and aim to find a solution that covers as many objects as possible; this results in a maximization problem. We divide our discussion mainly into two parts, based on these two types of optimization problems. In Section~\ref{subsec:covering-other}, we discuss problems related to covering that fall into neither category. 

\subsubsection{Minimization Variants}

We start out discussion with the minimization variants. For brevity, we overload the problem name and use the same name for the minimization variant (e.g., we use \pname{Set Cover} instead of the more cumbersome \pname{Min Set Cover}). Later on, we will use different names for the maximization versions; hence, there will be no confusion.

\mypar{Set Cover, Dominating Set and Vertex Cover} As discussed in detail in Section~\ref{sec:domset}, \pname{Set Cover} and equivalently \pname{Dominating Set} are very hard to approximate in the general case. Hence, special cases where some constraints are placed on the set system are often considered. Arguably the most well-studied special case of \pname{Set Cover} is the \pname{Vertex Cover} problem, in which the set system is a graph. That is, we would like to find the smallest set of vertices such that every edge has at least one endpoint in the selected set (i.e., the edge is ``covered''). \pname{Vertex Cover} is well known to be FPT~\cite{BussG93} and admit a linear-size kernel~\cite{NemhauserT75}. A generalization of \pname{Vertex Cover} on $d$-uniform hypergraph, where the input is now a hypergraph and the goal is to find the smallest set of vertices such that every hyperedge contain at least one vertex from the set, is also often referred to as \pname{$d$-Hitting Set} in the parameterized complexity community. However, we will mostly use the nomenclature {\pname{Vertex Cover} on $d$-uniform hypergraph} because many algorithms generalizes well from \pname{Vertex Cover} in graphs to hypergraphs. Indeed, branching algorithms for \pname{Vertex Cover} on graphs can be easily generalized to hypergraphs, and hence the latter is also FPT. Polynomial-size kernels are also known for \pname{Vertex Cover} on $d$-uniform hypergraphs~\cite{Abu-Khzam10}.

While \pname{Vertex Cover} both on graphs and $d$-uniform hypergraphs are already tractable, approximation can still help make algorithms even faster and kernels even smaller. We defer this discussion to Section~\ref{sec:faster-smaller}. 

\mypar{Connected Vertex Cover}
A popular variant of \pname{Vertex Cover} that is the 
\pname{Connected Vertex Cover} problem, for which the computed solution is 
required to induce a connected subgraph of the input. Just as \pname{Vertex 
Cover}, the problem is FPT~\cite{cygan2012deterministic}. However, unlike \pname{Vertex Cover}, \pname{Connected Vertex Cover} does not 
admit a polynomial-time kernel~\cite{dom}, unless \NPcoNP. In spite of this, a PSAKS for \pname{Connected Vertex Cover} exists:

\begin{Theorem}[\cite{lokshtanov2017lossy}]
For any $\eps > 0$, an $(1 + \eps)$-approximate kernel with $k^{O(1/\eps)}$ vertices can be computed in polynomial time.
\end{Theorem}

The ideas behind~\cite{lokshtanov2017lossy} is quite neat and we sketch it here. There are two reduction rules: (i) if there exists a vertex with degree more than $\Delta := 1/\varepsilon$ just ``select'' the vertex and (ii) if we see a vertex with more than $k$ {false twins}, i.e., vertices with the same set of neighbors, then we simply remove it from the graph. An important observation for (i) is that, since we have to either pick the vertex or all $\geq \Delta$ neighbors anyway, we might as well just select it even in the second case because it affects the size of the solution by a factor of at most $\frac{1 + \Delta}{\Delta} = 1 + \varepsilon$. For (ii), it is not hard to see that we either select one of the false twins or all of them; hence, if a vertex has more than $k$ false twins, then it surely cannot be in the optimal solution. Roughly speaking, these two observations show that this is an $(1 + \varepsilon)$-kernel. Of~course, in the actual proof, ``selecting'' a vertex needs to be defined more carefully, but we will not do it here. Nonetheless, imagine the end step when we cannot apply these two reduction rules anymore. Essentially speaking, we end up with a graph where some (less than $(1 + \varepsilon)k$) vertices are marked as ``selected'' and the remaining vertices have degree at most $\Delta$. Now, every vertex is either inside the solution, or all of its neighbors must lie in the solution. There are only (at most) $k$ vertices in the first case. For the second case, note that these vertices have degree at most $\Delta$ and they have at most $k$ false twins, meaning that there are at most $k^{1 + \Delta} = k^{1 + 1/\varepsilon}$ such vertices. In other words, the kernel is of size $k^{O(1/\eps)}$ as desired. This constitutes the main ideas in the proof; let us stress again that the actual proof is of course more complicated than this since we did not define rule (i) formally.

Recently, Krithika et al.~\cite{krithika2018revisiting} considered the following 
structural parameters beyond the solution size: split deletion set, clique cover 
and cluster deletion set. In each case, the authors provide a PSAKS for the 
problem. We will not fully define these parameters here, but we note that the 
first parameter (split deletion set) is always no larger than the size of the 
minimum vertex cover of the graph. In~another very recent work, Majumdar et 
al.~\cite{majumdar2020compress} give a PSAKS for each of the following 
parameters, each of which is always no larger than the solution size: the 
deletion distance of the input graph to the class of cographs, the class of 
bounded treewidth graphs, and the class of all chordal graphs. Hence, these 
results may be viewed as a generalization of the aforementioned PSAKS 
from~\cite{lokshtanov2017lossy}.

\mypar{Connected Dominating Set}
Similar to \pname{Connected Vertex Cover}, the \pname{Connected Dominating Set} 
problem is the variant of \pname{Dominating Set} for which the solution 
additionally needs to induce a connected subgraph of the input graph. When 
placing no restriction on the input graph, the problem is as hard to approximate 
as \pname{Dominating Set}. However, for some special classes of graphs, PSAKS or 
bi-PSAKS\footnote{Recall that a bi-kernel is similar to a kernel except 
that its the output need not be an instance of the original problem. Bi-PSAKS 
can be defined analogously to PSAKS, but with bi-kernel instead of kernel. In 
the case of \pname{Connected Dominating Set}, the bi-kernel outputs an  instance 
of an annotated variant of \pname{Connected Dominating Set}, where some 
vertices are marked and do not need to be covered by the solution.} are known; 
these include graphs with bounded expansion, nowhere dense graphs, and 
$d$-biclique-free graphs~\cite{eiben2017lossy}.

\mypar{Covering Problems parameterized by Graph Width Parameters} Several works in literature also study the approximability of variants of \pname{Vertex Cover} and \pname{Dominating Set} parameterized by graph widths~\cite{angel2016parameterized,lampis2014fpas}. These variants include:
\begin{itemize}
\item \pname{Power Vertex Cover (PVC)}. Here, along with the input graph, each edge has an integer demand and 
we have to assign (power) values to vertices, such that each edge has at least one endpoint with a value at least its demand. The goal is to minimize the total assigned power. Note that this is generalizes of \pname{Vertex Cover}, where edges have unit demands.
\item \pname{Capacitated Vertex Cover (CVC)}. The problem is similar to \pname{Vertex Cover}, except that each vertex has a capacity which limits the number of edges that it can cover. Once again, \pname{Vertex Cover} is a special case of \pname{CVC} where each vertex's capacity is $\infty$.
\item \pname{Capacitated Dominating Set (CDS)}. Analogous to \pname{CVC}, this 
is a generalization of \pname{Dominating Set} where each vertex has a capacity 
and it can only cover/dominate at most that many other vertices.
\end{itemize}
All problems above are FPT under standard parameter (i.e., the 
optimum)~\cite{angel2016parameterized,dom2008capacitated}. However, when 
parameterizing by the treewidth\footnote{see Definition~\ref{def:treewidth} for 
the definition of the treewidth}, all three problems become 
W[1]-hard~\cite{angel2016parameterized}. (This is in contrast to \pname{Vertex 
Cover} and \pname{Dominating Set}, both of which admit straightforward dynamic 
programming FPT algorithms parameterized by treewidth.) Despite this, good FPT 
approximation algorithms are known for the problem. In particular, a PAS is 
known for \pname{PVC}~\cite{angel2016parameterized}. For \pname{CVC} and 
\pname{CDS}, a bicriteria PAS exists for the problem~\cite{lampis2014fpas}, 
which in this case computes a solution of size at most the optimum, so that no 
vertex capacity is violated by more than a factor of $1+\eps$.

The approximation algorithms for \pname{CVC} and \pname{CDS} are results of a 
more general approach of Lampis~\cite{lampis2014fpas}. The idea is to execute an 
``approximate'' version of dynamic programming in tree decomposition instead of 
the exact version; this helps reduce the running time from $n^{O(w)}$ to $(\log 
n/\eps)^{O(w)}$, which is FPT. The approach is quite flexible: several 
approximation for graphs problems including covering problems can be achieved 
via this method and it also applies to clique-width. Please refer 
to~\cite{lampis2014fpas} for more details.

\mypar{Packing-Covering Duality and Erd\H{o}s-P\'{o}sa Property}
Given a set system $(V, \calc)$ where $V$ is the universe and $\calc = \{ C_1, \dots, C_m \}$ is a collection of subsets of $V$, \pname{Hitting Set} is the problem of computing the smallest $S \subseteq V$ that intersects every $C_i$, and \pname{Set Packing} is the problem of computing the largest subcollection $\calc' \subseteq \calc$ such that no two sets in $\calc'$ intersect. It can also be observed that the optimal value for \pname{Hitting Set} is at least the optimal value for \pname{Set Packing}, 
while the standard LP relaxations for them ({covering LP} and { packing LP}) have the same optimal value by strong duality. 
Studying the other direction of the inequality (often called the { packing-covering duality}) for natural families of set systems has been a central theme in combinatorial optimization. The gap between the covering optimum and packing optimum is large in general (e.g., \pname{Dominating Set}/\pname{Independent Set}), but~can be small for some families of set systems (e.g., \pname{$s$-$t$ Cut}/\pname{$s$-$t$ Disjoint Paths} and \pname{Vertex Cover}/\pname{Matching} especially in bipartite graphs). 

One notion that has been important for both parameterized and approximation algorithms is the Erd\H{o}s-P\'{o}sa property~\cite{erdos1965}. A family of set systems is said to have the Erd\H{o}s-P\'{o}sa property when there is a function $f : \N \to \N$ such that for any set system in the family, if the packing optimum is $k$, the covering optimum is at most $f(k)$. This immediately implies that the multiplicative gap between these two optima is at most $f(k)/k$, and constructive proofs for the property for various set systems have led to $(f(k)/k)$-approximation algorithms. 
Furthermore, for some problems including \pname{Cycle Packing}, the Erd\H{o}s-P\'{o}sa property gives an immediate parameterized algorithm. We refer the reader to a recent survey~\cite{raymond2017recent} and papers~\cite{lokshtanov2019packing, kim2018erdHos, van2019tight}. 

The original paper of Erd\H{o}s and P\'{o}sa~\cite{erdos1965} proved the property for set systems $(V, \calc)$ when there is an underlying graph $G = (V, E)$ and $\calc$ is the set of cycles, which corresponds to the pair \pname{Cycle Packing}/\pname{Feedback Vertex Set}; every graph either has at least $k$ vertex-disjoint cycles or there is a feedback vertex set of size at most $O(k \log k)$. Many subsequent papers also studied natural set systems arising from graphs where $V$ is the set of vertices or edges and $\calc$ denotes a collection of subgraphs of interest. 
For those set systems, 
Erd\H{o}s-P\'{o}sa Properties are closely related to Set Packing introduced in Section~\ref{sec:subgraphpacking} and \fdeletion problems introduced in Section~\ref{sec:width-reduction}.

\subsubsection{Maximization Variants}
\label{sec:covering-max}

We now move on to the maximization variants of covering problems. To our knowledge, these~covering problems are much less studied in the context of parameterized approximability compared to their minimization counterparts. In particular, we are only aware of works on the maximization variants of \pname{Set Cover} and \pname{Vertex Cover}, which are typically called \pname{Max $k$-Coverage} and \pname{Max $k$-Vertex Cover} respectively.

\mypar{Max $k$-Coverage} Recall that here we are given a set system and the goal is to select $k$ subsets whose union is maximized. It is well known that the simple greedy algorithm yields an $\left(\frac{e}{e - 1}\right)$-approximation~\cite{cornuejols1980worst}. Furthermore, Fiege shows, in his seminal work~\cite{feige1998threshold}, show that this is tight: $\left(\frac{e}{e - 1} - \varepsilon\right)$-approximation is NP-hard for any constant $\varepsilon > 0$. In fact, recently it has been shown that this inapproximability applies also to the parameterized setting. Specifically, under Gap-ETH, $\left(\frac{e}{e - 1} - \varepsilon\right)$-approximation cannot be achieved in FPT time~\cite{cohenaddad2019tight} or even $f(k) \cdot n^{o(k)}$ time~\cite{M20}. In other words, the trivial algorithm is tight in terms of running time, the greedy algorithm is tight in terms of approximation ratio, and there is essentially no trade-off possible between these two extremes. We remark here that this hardness of approximation is also the basis of hardness for \pname{$k$-Median} and \pname{$k$-Means}~\cite{cohenaddad2019tight} (see Section~\ref{sec:clustering}).

Due to the strong inapproximability result for the general case of \pname{Max $k$-Coverage}, different parameters have to be considered in order to obtain a PAS for \pname{Max $k$-Coverage}.  An interesting positive result here is when the parameters are $k$ and the VC dimension of the set system, for which a PAS exists while the exact version of the problem is W[1]-hard~\cite{badanidiyuru2012approximating}. %

\mypar{Max $k$-Vertex Cover} Another special case of \pname{Max $k$-Coverage} is the restriction when each element belongs to at most $d$ subsets in the system. This corresponds exactly to the maximization variant of the \pname{Vertex Cover} problem on $d$-uniform hypergraph, which will refer to as \pname{Max $k$-Vertex Cover}. Note here that, for such set systems, their VC-dimensions are also bounded by $\log d + 1$ and hence the aforementioned PAS of~\cite{badanidiyuru2012approximating} applies here as well. Nonetheless, \pname{Max $k$-Vertex Cover} admits a much simpler PAS (and even PSAKS) compared to \pname{Max $k$-Coverage} parameterized by $k$ and VC-dimension, as we will discuss more below.

\pname{Max $k$-Vertex Cover} was first studied in the context of parameterized 
complexity by Guo et al.~\cite{guo2007parameterized} who showed that the problem 
is W[1]-hard. Marx, in his survey on parameterized approximation 
algorithms~\cite{marx2008fpa}, gave a PAS for the problem with running time 
$2^{\tilde{O}(k^3/\varepsilon)}$. Later, Lokshtanov et 
al.~\cite{lokshtanov2017lossy} shows that Marx's approach can be used to give a 
PSAKS of size $O(k^5/\varepsilon^2)$. Both~of these results mainly focus on 
graphs. Later, Skowron and 
Faliszewski~\cite{skowron2017chamberlin}\footnote{The argument 
of~\cite{skowron2017chamberlin} was later independently rediscovered 
in~\cite{manurangsi2018note} as well.} gave a more general argument that both 
works generally for any $d$-uniform hypergraph and improves the running time and 
kernel size:

\begin{Theorem}[\cite{skowron2017chamberlin}]\label{thm:max-k-vertex-cover}
For the \pname{Max $k$-Vertex Cover} problem in $d$-uniform hypergraphs, a 
\mbox{$(1+\eps)$}-approximation can be 
computed in $O^*\left((d/\eps)^{k}\right)$ time for any $\eps>0$. Moreover, an $(1+\eps)$\hy{}approximate kernel 
with $O(dk/\eps)$ vertices can be computed in polynomial time.
\end{Theorem} 

The main idea of the above proof is simple and elegant, and hence we will include it here. For~convenience, we will only discuss the graph case, i.e., $d = 2$. It suffices to just give the $O(k/\eps)$-vertex kernel; the PAS immediately follows by running the brute force algorithm on the output instance from the kernel. The kernel is as simple as it gets: just keep $2k/\eps$ vertices with highest degrees and throw the remaining vertices away! Note that there is a subtle point here, which is that we do not want to throw away the edges linking from the kept vertices to the remaining vertices. If self-loops are allowed in a graph, this is not an issue since we may just add a self-loop to each vertex for each edge adjacent to it with the other endpoint being discarded. When self-loops are not allows, it is still possible to overcome this issue but with slightly larger kernel; we refer the readers to Section 3.2 of~\cite{manurangsi2018note} for more detail.

Having defined the kernel, let us briefly discuss the intuition as to why it works. Let $V_{2k/\eps}$ denote the set of $2k/\eps$ highest-degree vertices. The main argument of the proof is that, if there is an optimal solution $S$, then we may modify it to be entirely contained in $V_{2k/\eps}$  while preserving the number of covered edges to within $(1 + \eps)$ factor. The modification is simple: for every vertex that is outside of $V_{2k/\eps}$, we replace it with a random vertex from $V_{2k/\eps}$. Notice here that we always replace a vertex with a higher-degree vertex. Naturally, this should be good in terms of covering more edges, but there is a subtle point here: it is possible that the high degree vertices are ``double counted'' if a particular edge is covered by both endpoints. The size $2k/\eps$ is selected exactly to combat this issue; since the set is large enough, ``double counting'' is rare for random vertices. This finishes our outline for the intuition.

We end by remarking that \pname{Max $k$-Vertex Cover} on graphs is already APX-hard~\cite{petrank1994hardness}, and~hence the PASes mentioned above once again demonstrate additional power of FPT approximation algorithms over polynomial-time approximation algorithms.

\subsubsection{Other Related Problems}
\label{subsec:covering-other}

There are several other covering-related problems that do not fall into the two categories we discussed so far. We discuss a couple such problems below.

\mypar{Min $k$-Uncovered} The first is the \pname{Min $k$-Uncovered} problem, where the input is a set system and we would like to select $k$ sets as to minimize the number of uncovered elements. When we are concerned with exact solutions, this is of course the \pname{Set Cover}. However, the optimization version becomes quite different from \pname{Max $k$-Coverage}. In particular, since it is hard to determine whether we can find $k$ subsets that cover the whole universe, the problem is not approximable at all in the general case. However, if restrict ourselves to graphs and hypergraphs (for which we refer to the problem as \pname{Min $k$-Vertex Cover}), it is possible to get a (randomized) PAS for the problem~\cite{skowron2017chamberlin}:

\begin{Theorem}[\cite{skowron2017chamberlin}]
For the \pname{Min $k$-Vertex Uncovered} problem in $d$-uniform hypergraphs, a 
\mbox{$(1+\eps)$}-approximation can be 
computed in $O^*\left((d/\eps)^{k}\right)$ time for any $\eps>0$.
\end{Theorem}

The algorithm is based on the following simple randomized branching: pick a random uncovered element and branch on all possibilities of selecting a subset that contains it. Notice that since an element belongs to only $d$ subsets, the branching factor is at most $d$. The key intuition in the approximation proof is that, when the number of elements we have covered so far is still much less than that in the optimal solution, there is a relatively large probability (i.e., $\varepsilon$) that the random element is covered in the optimal solution. If we always pick such a ``good'' element in most branching steps, then we would end up with a solution close to the optimum. Skowron and Faliszewski~\cite{skowron2017chamberlin} formalizes this intuition by showing that the algorithm outputs an $(1 + \varepsilon)$-approximate solution with probability roughly $\varepsilon^k$. Hence, by repeating the algorithm $(1/\varepsilon)^k$ time, one arrives at the claimed PAS. To the best of our knowledge, it is unknown whether a PSAKS exists for the problem.

\mypar{Min $k$-Coverage} Another variant of the \pname{Set Cover} problem 
studied is \pname{Min $k$-Coverage},\footnote{The problem has also been 
referred to as \pname{Min $k$-Union} and \pname{Small Set Bipartite Vertex 
Expansion} in the literature~\mbox{\cite{ChlamtacDKKR18,ChlamtacDM17}}.} where 
we would like to select $k$ subsets that {minimizes} the number of covered 
elements. We stress here that this problem is {not} a relaxation of \pname{Set 
Cover} but rather is much more closely related to graph expansion problems 
(see~\cite{ChlamtacDM17}).

It is known that, when there is no restriction on the input set system, the problem is (up to a polynomial factor) as hard to approximate as the \pname{Denest $k$-Subgraph} problem~\cite{ChlamtacDKKR18}. Hence, by the inapproximability of the latter discussed earlier in the survey (Theorem~\ref{thm:dks-inapprox}), we also have that there is no $k^{o(1)}$-approximation algorithm for the problem that runs in FPT time.

Once again, the special case that has been studied in literature is when the 
input set system is a graph, in which case we refer to the problem as \pname{Min 
$k$-Vertex Cover}. Gupta, Lee, and Li~\cite{gupta2018fpt,gupta2018faster} used 
the technique of Marx~\cite{marx2008fpa} to give a PAS for the problem with 
running time $O^*((k/\varepsilon)^{O(k)})$. The~running time was later improved 
in~\cite{manurangsi2018note} to $O^*((1/\varepsilon)^{O(k)})$. The algorithm 
there is again based on branching, but the rules are more delicate and we will 
not discuss them here. An interesting aspect to note here is that, while both 
\pname{Max $k$-Vertex Cover} and \pname{Min $k$-Vertex Cover} have PSAKS of the 
(asymptotically) same running time, the former admits a PSAKS whereas the latter 
does not (assuming a variant of the Small Set Expansion 
Conjecture)~\cite{manurangsi2018note}.

To the best of our knowledge, \pname{Min $k$-Vertex Cover} has not been explicitly studied on $d$-uniform hypergraphs before, but we suspect that the above results should carry over from graphs to hypergraphs as well.

\subsection{Clustering}
\label{sec:clustering}

Clustering is a representative task in unsupervised machine learning that has been studied in many fields. In combinatorial optimization communities, it is often formulated as the following: 
Given a set $P$ of points and a set $F$ of candidate centers (also known as facilities), 
and a metric on $X \supseteq P \cup F$ given by the distance $\rho : X \times X \to \bbr^+ \cup \{ 0 \}$, 
choose $k$ centers $C \subseteq F$ to 
minimize some objective function $\cost := \cost(P, C)$. 
To fully specify the problem, the choices to make are the following. Let~$\rho(C, p) := \min_{c \in C} \rho(c, p)$. 
\begin{itemize}
\item Objective function: Three well-studied objective functions are 
\begin{itemize}
\item \kmedian ($\cost(P, C) := \sum_{p \in P} \rho(C, p)$).
\item \kmeans ($\cost(P, C) := \sum_{p \in P} \rho(C, p)^2$).
\item \kcenter ($\cost(P, C) := \max_{p \in P} \rho(C, p)$). 
\end{itemize}

\item Metric space: The ambient metric space $X$ can be
\begin{itemize}
\item A general metric space explicitly given by the distance $\rho : X \times X \to \bbr^+ \cup \{ 0 \}$. 
\item The Euclidean space $\bbr^{d}$ equipped with the $\ell_2$ distance. 
\item Other structured metric spaces including metrics with { bounded doubling dimension} or { bounded highway dimension}.
\end{itemize}
\end{itemize}

While many previous results on clustering focused on non-parameterized polynomial time, there~are at least three natural parameters one can parameterize: The number of clusters $k$, the~dimension $d$ (if defined), and the approximation accuracy parameter $\eps$.
In general metric spaces, parameterized approximation algorithms (mainly with parameter $k$) were considered very recently, 
but in Euclidean spaces, many previous results already give parameterized approximation algorithms with parameters $k, d$, and $\eps$. 

\subsubsection{General Metric Space}
We can assume $X = P \cup F$ without loss of generality. Let $n := |X|$ and note 
that the distance $\rho : X \times X \to \bbr^+ \cup \{ 0 \}$ is explicitly 
specified by $\Theta(n^2)$ numbers. A simple exact algorithm running in time 
$O(n^{k + 1})$ can be achieved by enumerating all $k$ centers $c_1, \dots, c_k 
\in F$ and assign each point $p$ to the closest center. In this setting, the 
best approximation ratios achieved by polynomial time algorithms are $2.611 + 
\eps$ for \kmedian~\cite{byrka2014improved}, $9 + \eps$ for 
\kmeans~\cite{kanungo2004local}, and $3$ for 
\kcenter~\cite{gonzalez1985clustering}.\footnote{A special case that has 
received significant attention assumes $P = F$. In this case, the best 
approximation ratio for \kcenter becomes~$2$.}
From the hardness side, it is NP-hard to approximate 
\kmedian within a factor $1 + 2/e - \eps \approx 1.73 - \eps$, 
\kmedian within a factor $1 + 8/e - \eps \approx 3.94 - \eps$, 
\kcenter within a factor $3 - \eps$~\cite{guha1999greedy}.

While there are some gaps between the best algorithms and the best hardness results for \kmedian and \kmeans, it is an interesting question to ask how parameterization by $k$ changes the approximation ratios for both problems. Cohen-Addad et al.~\cite{cohenaddad2019tight} studied this question and gave exact answers.
\begin{Theorem} [\cite{cohenaddad2019tight}]
For any $\eps > 0$, there is an $(1 + 2/e + \eps)$-approximation algorithm for \kmedian, and~an $(1 + 8/e + \eps)$-approximation algorithm for \kmeans, both running in time $(O(k \log k/\eps^2))^k n^{O(1)}$. 

There exists a function $g : \bbr^+ \to \bbr^+$ such that assuming the \gapeth, for any $\eps > 0$, 
any \mbox{$(1 + 2/e - \eps)$}-approximation algorithm for \kmedian, and 
any $(1 + 8/e - \eps)$-approximation algorithm for \kmeans, must run in time at least $n^{k^{g(\eps)}}$. 
\label{thm:kmedian}
\end{Theorem}

These results show that if we parameterize by $k$, $1+2/e$ (for \kmedian) and $1+8/e$ (for \kmeans) are the exact limits of approximation for parameterized approximation algorithms. Similar reductions also show that no parameterized approximation algorithm can achieve $(3 - \eps)$-approximation for \kcenter for any $\eps > 0$ (only assuming $\W[2] \neq \FPT$), so the power of parameterized approximation is exactly revealed for all three objective functions. 

\mypar{Algorithm for \kmedian}
We briefly describe ideas for the algorithm for \kmedian in Theorem~\ref{thm:kmedian}. 
The main technical tool that the algorithm uses is a {coreset}, which will 
be also frequently used for Euclidean subspaces in the next subsection. 

When $S$ is a set of points with weight functions $w : S \to \bbr^{+}$, let us extend the definition of the objective function $\cost(S, C)$ such that 
\[
\cost(S, C) := \sum_{p \in S} w(p) \cdot \rho(C, s). 
\]

Given a clustering instance $(P, F, \rho, k)$ and $\eps > 0$, a subset $S \subseteq P$ with weight functions $w : S \to \bbr^{+}$ is called a (strong) corset if for any $k$ centers $C = \{ c_1, \dots, c_k \} \subseteq F$, 
\[
|\cost(S, C) - \cost(P, C) | \leq \eps \cdot \cost(P, C).
\]

For a general metric space, Chen~\cite{chen2006k} gave a coreset of cardinality $\tilde{O}(k^2 \log^2 n/\eps^2)$. (In~this subsection, $\tilde{O}(\cdot)$ hides $\poly(\log \log n, \log k, \log (1/\eps)$.) 
This was improved by Feldman and Langeberg~\cite{feldman2011unified} to $O(k \log n/\eps^2)$. We introduce high-level ideas of~\cite{chen2006k} later.

Given the coreset, it remains to give a good parameterized approximation algorithm for the problem for a much smaller (albeit weighted) point set $|P| = O(\poly(\log n, k))$. Note that $|F|$ can be still as large as $n$, so naively choosing $k$ centers from $F$ will take $n^k$ time and exhaustively partitioning $P$ into $k$ sets will take $k^{|P|} = n^{\poly(k)}$ time. (Indeed, exactly solving this small case will give an \epas, which will contradict the \gapeth.) 

Fix an optimal solution, and let $C^* = \{ c^*_1, \dots, c^*_k \}$ are the optimal centers and $P^*_i \subseteq P$ is the cluster assigned to $c^*_i$. 
One information we {can} guess is, for each $i \in [k]$, the point $p_i \in P^*_i$ closest to $c^*_i$ and (approximate) $\rho(c^*_i, p_i)$. Since $|P| = \poly(k \log n)$, guessing them only takes time $(k \log n)^{O(k)}$, which~can be made FPT by separately considering the case $(\log n)^k \leq n$ and the case $(\log n)^k \geq n$, in~which $k = \Omega(\log n/\log \log n)$ and $(\log n)^k = (k \log k)^{O(k)}$. 

Let $F_i \subseteq F$ be the set of candidate centers that are at distance approximately $r_i$ from $p_i$, so that $c^*_i \in F_i$ for each $i$. 
The algorithm chooses $k$ centers $C \subseteq F$ such that $|C \cap F_i| \geq 1$ for each $i \in [k]$. Let~$c_i \in C \cap F_i$. For any point $p \in P$ (say $p \in P^*_j$, though the algorithm doesn't need to know $j$), then 
\[
\rho(C, p) \leq \rho(c_j, p) \leq \rho(c_j, p_j) + \rho(p_j, c^*_j) + \rho(c^*_j, p) \leq 3\rho(c^*_j, p),
\]
where $\rho(c_j, p_j) \approx \rho(p_j, c^*_j) \leq \rho(c^*_j, p)$ by the choice of $p_j$. 

This immediately gives a $3$-approximation algorithm in FPT time, which is worse than the best polynomial time approximation algorithm. To get the optimal $(1 + 2/e)$-approximation algorithm, we further reduce the job of finding $c_i \in F_i$ to { maximizing a monotone submodular function with a partition matroid constraint}, which is known to admit an optimal $(1 - 1/e)$-approximation algorithm~\cite{calinescu2011maximizing}. 
Then we can ensure that for $(1 - 1/e)$ fraction of points, the distance to the chosen centers is shorter than in the optimal solution, 
and for the remaining $1/e$ fraction, the distance is at most three times the distance in the optimal solution. 
We refer the reader to~\cite{cohenaddad2019tight} for further details. 

\mypar{Constructing a coreset} 
As discussed above, a coreset is a fundamental building block for optimal parameterized approximation algorithms for \kmedian and \kmeans for general metrics.
We briefly describe the construction of Chen~\cite{chen2006k}
that gives a coreset of cardinality $\tilde{O}(k^2 \log^2 n/\eps^2)$ for \kmedian. 
Similar ideas can be also used to obtain an \epas for Euclidean spaces parameterized by $k$, though better specific constructions are known in Euclidean spaces. 

We first partition $P$ into $P_1, \dots P_{\ell}$ such that $\ell = O(k \log n)$ and 
\[
\sum_{i = 1}^{\ell} |P_i| \diam(P_i) = O(\OPT).
\]

Such a partition can be obtained by using a known (bicriteria) constant factor approximation algorithm for \kmedian. 
Next, let $t = \tilde{O}(k \log n)$, and for each $i = 1, \dots, \ell$, we let $S_i = \{ s_1, \dots, s_t \}$ be a random subset of $t$ points of $P_i$ 
where each $s_j$ is an independent and uniform sample from $P_i$ and is given weight $|P_i|/t$. 
(If $|P_i| \leq t$, we simply let $S_i = P_i$ with weights $1$.) The final coreset $S$ is the union of all $S_i$'s. 

To prove that it works, we simply need to show that for any set of $k$ centers $C \subseteq F$ with $|C| = k$, 
\[
\Pr[|\cost(S, C) - \cost(P, C)| > \eps \cdot \cost(P, C)] \leq o(1/n^k),
\]
so that the union bound over $\binom{n}{k}$ choices of $C$ works. Indeed, we show that for each $i = 1, \dots, \ell$, 
\begin{equation}
\Pr[|\cost(S_i, C) - \cost(P_i, C)| > \eps \cdot |P_i| \diam(P_i)] \leq o(1/(\ell \cdot n^k)),
\label{eq:coreset}
\end{equation}
so that we can also union bound and sum over $i \in [\ell]$, using the fact that $\sum_{i} |P_i| \diam(P_i) = O(\OPT) \leq O(\cost(P, C))$. 

It is left to prove~\eqref{eq:coreset}. Fix $C$ and $i$ (let $P_i = \{ p_1, \dots, p_{|P_i|} \}$), and recall that 
\[
\cost(P_i, C) = \sum_{j=1}^{|P_i|} \rho(C, p_j). 
\]

When $|P_i| \leq t$, $S_i = P_i$, so \eqref{eq:coreset} holds. Otherwise, recall that $S_i = \{ s_1, \dots, s_t \}$  where each $s_j$ is an independent and uniform sample from $P_i$ with weight $w := |P_i|/t$. For $j = 1, \dots, t$, let $X_j := w \cdot \rho(C, s_j)$.
Note that $\cost(S_i, C) = \sum_j X_j$ and $\cost(P_i, C) = t \cdot \E[X_j] = \E[\cost(S_i, C)]$. 
A crucial observation is that that $|\rho(C, p_j) - \rho(C, p_{j'})| \leq \diam(P_i)$ for any $j, j' \in [|P_i|]$, so that
$|X_{j} - X_{j'}| \leq w \cdot \diam(P_i)$ for any $j, j' \in [t]$. 
If we let $X_{\min} := \min_{j \in [|P_i|]} ( w \rho(C, p_j) )$ and $Y_j := (X_j - X_{\min})/(w \cdot \diam(P_i))$, 
$Y_j$'s are $t$ i.i.d. random variables that are supported in $[0, 1]$. 
The standard Chernoff--Hoeffding inequality gives
\[
\Pr\bigg[| \sum_{j} X_j - t \E[X_j] | > \eps |P_i| \diam(P_i) \bigg] = 
\Pr\bigg[| \sum_{j} Y_j - t \E[Y_j] | > \eps t \bigg] \leq \exp(O(\eps^2 t)) \leq o(1/(\ell \cdot n^k)), 
\]
proving~\eqref{eq:coreset} for $t = \tilde{O}(k \log n)$ and finishing the proof. 
A precise version of this argument was stated in 
Haussler~\cite{haussler1992decision}.

\subsubsection{Euclidean Space}\label{sec:cluster-Euclidean}

For Euclidean spaces, we assume that $X = F = \bbr^d$ for some $d \in \bbn$, 
endowed with the standard $\ell_2$ metric. Let $n = |P|$ in this subsection. 
Now we have $k$ and $d$ as natural structural parameters of clustering tasks. 
Many previous approximation algorithms in \ekmedian and \ekmeans in Euclidean spaces, without explicit mention to parameterized complexity, are parameterized approximation algorithms parameterized by $k$ or $d$ (or both).
The highlight of this subsection is that for both \ekmedian and \ekmeans, an
\epas exists with only one of $k$ and $d$ as a parameter. 
Without any parameterization, both \ekmedian and \kmeans are known to be APX-hard~\cite{lee2017improved-k-means, cohen2019inapproximability}.
We introduce these results in the chronological order, highlighting important ideas. 

\mypar{\ekmedian with parameter $d$} 
The first \ptas for \ekmedian in Euclidean spaces with fixed $d$ appears in Arora et al.~\cite{arora1998approximation}. 
The techniques extend Arora's previous \ptas for the \etsp problem in Euclidean 
spaces~\cite{arora1998polynomial}, 
first proving that there exists a near-optimal solution that interacts with a 
{ quadtree} (a geometric division of $\bbr^d$ into a hierarchy of square 
regions) in a restricted sense,
and finally finding such a tour using dynamic programming. 
The running time is $n^{O(1/\eps)}$ for $d = 2$ and $n^{(\log n/\eps)^{d-2}}$ 
for $d > 2$. 
Kolliopoulous and Rao~\cite{kolliopoulos1999nearly} improved the running time to 
$2^{O((\log (1/\eps)/\eps)^{d - 1})} n \log^{d + 6} n$, which is an \epas 
with parameter $d$. 

\iffalse %
Recall that a $\log^k n$ factor in the running time can be 
converted into an FPT running time, because it becomes $n^{o(1)}$ when $k = 
o(\log n/\log \log n)$ and otherwise $n$ becomes a function of $k$. 
\fi %

\mypar{\ekmedian and \ekmeans with parameter $k$} 

An \epas for \ekmeans even with parameters both $k$ and $d$ took longer to be 
discovered, and first appeared when 
\citet{matouvsek2000approximate} gave an approximation scheme that runs in time 
$O(n \eps^{2k^2 d} \log^k n)$. 
After this, several improvements on the running time followed~\citet{badoiu2002approximate, de2003approximation, har2004coresets}.

\iffalse %
\citet{badoiu2002approximate} gave an approximation scheme running in 
$O(2^{(k/\eps)^{O(1)}} d^{O(1)} n \log^{O(k)} n)$ time for \kmedian.
\citet{de2003approximation} gave an approximation scheme 
running in $O(g(k, \eps) d^{O(1)} \log^{O(k)} n)$ time for \kmeans.
\citet{har2004coresets} gave an approximation scheme running in $O(n + k^{k+2} 
\eps^{-(2d+1)} \log^{k + 1}n)$ time for \kmeans.
\fi %

Kumar et al.~\cite{kumar2004simple, kumar2005linear} gave approximation schemes 
for both \kmedian and \kmeans, running~in time $2^{(k/\eps)^{O(1)}} dn$. 
This shows that an \epas can be obtained by using only $k$ as a parameter.
Using~this result and and improved coresets, more improvements followed
~\cite{chen2006k, feldman2007ptas, feldman2011unified}.
The current best runtime to get $(1 + \eps)$-approximation is
$O(ndk + d \cdot \poly(k/\eps) + 2^{\tilde{O}(k/\eps)})$ for \kmeans~\cite{feldman2007ptas}, and~$O(ndk + 2^{\poly(1/\eps, k)})$ time for \kmedian~\cite{feldman2011unified}.

A crucial property of the Euclidean space that allows an \epas with parameter $k$ 
(which is ruled out for general metrics by Theorem~\ref{thm:kmedian})
is the { sampling property}, which says that for any set $Q \subseteq \bbr^d$ as one cluster, 
there is an algorithm that is given only $g(1/\eps)$ samples from $Q$ and 
outputs $h(1/\eps)$ candidate centers such that one of them is $\eps$-close to 
the optimal center for the entire cluster $Q$ for some functions $g, h$. (For example, for 
\kmeans, the mean of $O(1/\eps)$ random samples $\eps$-approximates the actual 
mean with constant probability.)
This idea leads to an $(1 + \eps$)-approximation algorithm running in time $|P|^{f(\eps, k)}$.
Together even with a general coreset construction
of size $\poly(k, \log n, 1/\eps)$, one already gets an \epas with parameter $k$. 
Better coresets construction are also given in Euclidean spaces. 
Recent developments~\cite{sohler2018strong, becchetti2019oblivious, 
huang2020coresets} construct core-sets of size $\poly(k, 1/\eps)$ (no dependence 
on $n$ or~$d$), which is further extended to the shortest-path metric of an 
excluded-minor graph~\cite{braverman2020coresets}.

\iffalse %

For both \kmedian and \kmeans,~\cite{chen2006k, feldman2007ptas, feldman2011unified} 
constructed (strong) coresets introduced above for general metrics. While the size of such a 
coreset was $\poly(k, \log n, 1/\eps)$ for general metrics, for $\bbr^d$, a 
coreset of size $\poly(k, d, 1/\eps)$ can be also achieved by using $\eps$-nets 
of approximate sizes.

\cite{feldman2007ptas, feldman2011unified} also 
constructed {weak coresets}, which does not preserve the cost of every 
possible center set $C \subseteq \bbr^d$ but still allows a
$(1+\eps)$-approximation for the coreset to imply a $(1+\eps)$-approximation 
for the original instance, and whose size $\poly(k, 1/\eps)$ does not depend on 
$d$ or $\log n$. These coresets give a faster \epas with parameter $k$.
\fi

\mypar{\ekmeans with parameter $d$} 

\citet{cohen2019local} and \citet{friggstad2019local} recently gave 
approximation schemes running in time $n^{f(d, \eps)}$ using local search 
techniques. These results were improved to an \epas in~\cite{cohen2018fast}, and 
also extended to doubling metrics~\cite{cohen2018near}. 

\mypar{Other metrics and \kcenter} 
For the \pname{$k$-Center} problem an \epas exists when parametrizing by both 
$k$ and the doubling dimension~\cite{DBLP:conf/swat/FeldmannM18}, and also for 
planar graphs there is an \epas for parameter $k$, which is implied by the 
EPTAS of \citet{planar-EPTAS} (cf.~\cite{DBLP:conf/swat/FeldmannM18}).

There are also parameterized approximation schemes for metric spaces with {
bounded highway 
dimension}~\cite{becker2018polynomial,DBLP:conf/icalp/Feldmann15, 
DBLP:conf/swat/FeldmannM18} and various graph width 
parameters~\cite{katsikarelis2019structural}. 

\mypar{Capacitated clustering and other variants} 
Another example where the parameterization by $k$ helps is \ckmedian, where each possible center $c \in F$ has a capacity $u_c \in \bbn$ and can be assigned at most $u_c$ points. It is not known whether there exists a constant-factor approximation algorithm, and~known constant factor approximation algorithms either open $(1+\eps)k$ centers~\cite{li2016approximating} or violate capacity constraints by an $(1 + \eps)$ factor~\cite{demirci2016constant}. 
\citet{adamczyk2018constant} gave a $(7 + \eps)$-approximation algorithm in $f(k, \eps) n^{O(1)}$ time, showing that a constant factor parameterized approximation algorithm is possible. 
The approximation ratio was soon improved to $(3 + \eps)$~\cite{cohen2019fixed}. 
For \pname{Capacitated Euclidean $k$-Means}, 
\cite{xu2019constant} also gave a $(69 + \eps)$-approximation algorithm for  in $f(k, \eps) n^{O(1)}$ time. 

While the capacitated versions of clustering look much harder than their uncapacitated counterparts, 
there is no known theoretical separation between the capacitated version and the uncapacitated version in any clustering task. 
Since the power of parameterized algorithms for uncapacitated clustering is well understood,
it is a natural question to understand the ``capacitated VS uncapacitated question'' in the FPT setting. 

\begin{Open Question}
Does \ckmedian admit an $(1 + 2/e)$-approximation algorithm in FPT time with parameter $k$? 
Do \pname{Capacitated Euclidean $k$-Means/$k$-Median} admit an \epas with parameter $k$ or $d$? 
\end{Open Question}

Since clustering is a universal task, like capacitated versions, many variants of clustering tasks have been studied including 
\pname{$k$-Median/$k$-Means with Outliers}~\cite{krishnaswamy2018constant} and 
\pname{Matroid/Knapsack Median}~\cite{swamy2016improved}.
While no variant is proved to harder than the basic versions, it would be interesting to see whether 
they all have the same parameterized approximability with the basic versions.

\subsection{Network design}
\label{sec:network-design}

In network design, the task is to connect some set of vertices in a metric, which 
is often given by the shortest-path metric of an edge-weighted graph. Two very 
prominent problems of this type are the \pname{Travelling Salesperson (TSP)} and 
\pname{Steiner Tree} problems. For \pname{TSP} all vertices need to be connected 
in a closed walk (called a {route}), and the length of the route needs to 
be minimized.\footnote{Sometimes also the non-metric version of
\pname{TSP} is 
considered, which however is much harder than the metric one. We only consider 
the metric version here.} For \pname{Steiner Tree} a subset of the vertices 
(called {terminals}) is given as part of the input, and the objective is to 
connect all terminals by a tree of minimum weight in the metric (or graph). Both 
of these are fundamental problems that have been widely studied in the past, 
both on undirected and directed input graphs.

\mypar{Undirected graphs}
A well-studied parameter for \pname{Steiner Tree} is the number of terminals, 
for~which the problem has been known to be FPT since the early 1970s due to the 
work of \citet{DBLP:journals/networks/DreyfusW71}. Their algorithm is based on 
dynamic programming and runs in $3^k\polyn$ time if $k$ is the number of 
terminals. Faster algorithms based on the same ideas with runtime 
$(2+\delta)^k\polyn$ for any constant $\delta>0$ exist~\cite{fuchs2007dynamic} 
(here the degree of the polynomial depends on~$\delta$). The unweighted 
\pname{Steiner Tree} problem also admits a $2^k\polyn$ time 
algorithm~\cite{DBLP:conf/icalp/Nederlof09} using a different technique based on 
subset convolution. Given any of these exact algorithms as a subroutine, a 
faster PAS can also be found~\cite{fellows2018fidelity} 
(cf.~Section~\ref{sec:faster-smaller}). On the other hand, no exact polynomial-sized 
kernel exists~\cite{dom} for the \pname{Steiner Tree} problem, unless \NPcoNP. 
Interestingly though, a PSAKS can be obtained~\cite{lokshtanov2017lossy}. 

This kernel is based on a well-known fact proved by \citet{borchers-du}, which 
is very useful to obtain approximation algorithms for the \pname{Steiner Tree} 
problem. It states that any Steiner tree can be covered by smaller trees 
containing few terminals, such that these trees do not overlap much. More~formally, a {full-component} is a subtree of a Steiner tree, for which the 
leaves coincide with its terminals. For the optimum Steiner tree $T$ and any 
$\eps>0$, there exist full-components $C_1,\ldots,C_\ell$ of $T$ such that 
\begin{enumerate}
 \item each full-component $C_i$ contains at most $2^{\lceil 1/\eps\rceil}$ 
terminals (leaves),
 \item the sum of the weights of the full-components is at most $1+\eps$ times 
the cost of $T$, and
\item taking any collection of Steiner trees $T_1,\ldots, T_\ell$, such that 
each tree $T_i$ connects the subset of terminals that forms the leaves of 
full-component $C_i$, the union $\bigcup_{i=1}^\ell T_i$ is a feasible solution 
to the input instance.
\end{enumerate}
Not knowing the optimum Steiner tree, it is not possible to know the subsets of 
terminals of the full-components corresponding to the optimum. However, it is 
possible to compute the optimum Steiner tree for every subset of terminals of 
size at most $2^{\lceil 1/\eps\rceil}$ using an FPT algorithm for \pname{Steiner 
Tree}. The time to compute all these solutions is $k^{O(2^{1/\eps})}\polyn$, 
using for instance the \citet{DBLP:journals/networks/DreyfusW71} algorithm. Now 
the above three properties guarantee that the graph given by the union of all 
the computed Steiner trees, contains a $(1+\eps)$-approximation for the input 
instance. In fact, the best polynomial time approximation algorithm known to 
date~\cite{DBLP:journals/jacm/ByrkaGRS13} uses an iterative rounding procedure 
to find a $\ln(4)$-approximation of the optimum solution in the union of these 
Steiner trees. To obtain a kernel, the union needs to be sparsified, since it 
may contain many Steiner vertices and also the edge weights might be very large. 
However, \citet{lokshtanov2017lossy} show that the number of Steiner vertices 
can be reduced using standard techniques, while the edge weights can be encoded 
so that their space requirement is bounded in the parameter and the cost of any 
solution is distorted by at most a $1+\eps$ factor.

\begin{Theorem}[\cite{lokshtanov2017lossy,fellows2018fidelity}]
\label{thm:SteinerTree}
For the \pname{Steiner Tree} problem a $(1+\eps)$-approximation can be computed 
in \mbox{$(2+\delta)^{(1-\eps/2)}\polyn$} time for any constant $\delta>0$ (and 
in $2^{(1-\eps/2)}\polyn$ time in the unweighted case) for any $\eps>0$, where 
$k$ is the number of terminals. Moreover, a $(1+\eps)$-approximate kernel of 
size~$(k/\eps)^{O(2^{1/\eps})}$ can be computed in polynomial time.
\end{Theorem}

A natural alternative to the number of terminals is to consider the vertices 
remaining in the optimum tree after removing the terminals: a folklore result 
states that \pname{Steiner Tree} is W[2]-hard parameterized by the number of 
non-terminals (called {Steiner vertices}) in the optimum solution. At~the 
same time, unless P=NP there is no PTAS for the problem, as it is 
APX-hard~\cite{chlebik2008steiner}. However an approximation scheme is 
obtainable when parametrizing by the number of Steiner vertices $k$ in the 
optimum, and also a PSAKS is obtainable under this parameterization.

To obtain both of these results, \citet{st-pas} devise a reduction rule that is 
based on the following observation: if the optimum tree contains few Steiner 
vertices but many terminals, then the tree must contain (1) a large component 
containing only terminals, or (2) a Steiner vertex that has many terminal 
neighbours. Intuitively, in case (2) we would like to identify a large star with 
terminal leaves and small cost in the current graph, while in case (1) we would 
like to find a cheap edge between two terminals. Note that such a single edge 
also is a star with terminal leaves. The reduction rule will therefore find the 
star with minimum weight per contained terminal, which can be done in 
polynomial time. This rule is applied until the number of terminals, which 
decreases after each use, falls below a threshold depending on the input 
parameter $k$ and the desired approximation ratio $1+\eps$. Once the number of 
terminals is bounded by a function of $k$ and $\eps$, the 
\citet{DBLP:journals/networks/DreyfusW71} algorithm can be applied on the 
remaining instance, or a kernel can be computed using the PSAKS of 
Theorem~\ref{thm:SteinerTree}. It~can be shown that the reduction rule does not distort 
the optimum solution by much as long as the threshold is large enough, which 
implies the following theorem.

\begin{Theorem}[\cite{st-pas}]\label{thm:Steiner-tree}
For the \pname{Steiner Tree} problem a $(1+\eps)$-approximation can be computed 
in $2^{O(k^2/\eps^4)}\polyn$ time for any $\eps>0$, where $k$ is the number of 
non-terminals in the optimum solution. Moreover, a $(1+\eps)$-approximate kernel 
of size $(k/\eps)^{2^{O(1/\eps)}}$ can be computed in polynomial time.
\end{Theorem}

This theorem is also generalizable to the \pname{Steiner Forest} problem, where 
a list of terminal pairs is given and the task is to find a minimum weight 
forest in the input graph connecting each pair. In~this case though, the 
parameter has to be combined with the number of connected components of the 
optimum forest~\cite{st-pas}. 

A variation of the \pname{Steiner Forest} problem is the \pname{Shallow-Light 
Steiner Network (SLSN)} problem. Here a graph with both edge costs and edge 
lengths is given, together with a set of terminal pairs and a length threshold 
$L$. The task is to compute a minimum cost subgraph, which connects each 
terminal pair with a path of length at most $L$. For this problem a dichotomy 
result was shown~\cite{babay2018characterizing} in terms of the pattern given by 
the terminal pairs. More precisely, the terminal pairs are interpreted as edges 
in a graph for which the vertices are the terminals: if $\mc{C}$ is some class 
of graphs, then \pname{SLSN$_\mc{C}$} is the \pname{Shallow-Light Steiner 
Network} problem restricted to sets of terminal pairs that span some graph in 
$\mc{C}$. Let $\mc{C}^\star$ denote the class of all stars, and $\mc{C}_\lambda$ 
the class of graphs with at most $\lambda$ edges. The~\pname{SLSN$_{\mc{C}^\star}$} problem is APX-hard~\cite{chlebik2008steiner}, as 
it is a generalization of \pname{Steiner Tree} (where $L=\infty$). At~the same 
time, both the \pname{SLSN$_{\mc{C}^\star}$} and \pname{SLSN$_{\mc{C}_\lambda}$} 
problems parameterized by the number of terminals are 
paraNP\hy{}hard~\cite{hassin1992approximation}, 
since they are generalizations 
of the \pname{Restricted Shortest Path} problem (where there is exactly one 
terminal pair). A PAS can however be obtained for both of these problems 
(whenever $\lambda$ is a constant), but for no other class $\mc{C}$ of demand 
patterns~\cite{babay2018characterizing}.

\begin{Theorem}[\cite{babay2018characterizing}]\label{thm:SLSN}
For any constant $\lambda>0$, there is an FPTAS for the 
\pname{SLSN$_{\mc{C}_\lambda}$} problem. For the \pname{SLSN$_{\mc{C}^\star}$} 
problem a $(1+\eps)$-approximation can be computed  in $4^k(n/\eps)^{O(1)}$ time 
for any~$\eps>0$, where $k$ is the number of terminal pairs. Moreover, under 
\gapeth no $(5/3-\eps)$-approximation for \pname{SLSN$_{\mc{C}}$} can be 
computed in $f(k)\polyn$ time for any $\eps>0$ and computable function $f$, 
whenever $\mc{C}$ is a recursively enumerable class for which 
$\mc{C}\not\subseteq\mc{C}^\star\cup\mc{C}_\lambda$ for every constant 
$\lambda$.
\end{Theorem}

A notable special case is when all edge lengths are $1$ but edge costs are 
arbitrary. Then \pname{SLSN$_{\mc{C}_\lambda}$} is polynomial time solvable for 
any constant $\lambda$, while \pname{SLSN$_{\mc{C}^\star}$} is FPT parameterized 
by the number of terminals~\cite{babay2018characterizing}. At the same time the 
parameterized approximation lower-bound of Theorem~\ref{thm:SLSN} is still valid for 
this case. It is not known however, whether constant approximation factors can 
be obtained for \pname{SLSN$_{\mc{C}}$} when $\mc{C}$ is a class different from 
$\mc{C}_\lambda$ and $\mc{C}^\star$. More generally we may ask the following 
question.

\begin{Open Question}
Given some class of graphs $\mc{C}\not\subseteq\mc{C}^\star\cup\mc{C}_\lambda$,
which approximation factor $\alpha_\mc{C}$ can be obtained in FPT time for 
\pname{SLSN$_{\mc{C}}$} parameterized by the number of terminals?
\end{Open Question}

Turning to the \pname{TSP} problem, a generalization of \pname{TSP} introduces 
deadlines until which vertices need to be visited by the computed tour. A 
natural parameterization in this setting is the number of vertices that have 
deadlines. It can be shown~\cite{bockenhauer2007parameterized} that no 
approximation better than $2$ can be computed when using this parameter. 
Nevertheless, a $2.5$-approximation can be computed in FPT 
time~\cite{bockenhauer2007parameterized}. The algorithm will guess the order in 
which the vertices with deadlines are visited by the optimum solution. It then 
computes a $3/2$-approximation for the remaining vertices using Christofides 
algorithm~\cite{Vazirani01book}. The approximation ratio follows, since the 
optimum tour can be thought of as two tours, of which one visits only the 
deadline vertices, while the other contains all remaining vertices. The~approximation algorithm incurs a cost of $\Opt$ for the former, and a cost of 
$\frac{3}{2}\cdot\Opt$ for the latter part of the optimum tour.

\begin{Theorem}[\cite{bockenhauer2007parameterized}]
For the \pname{DlTSP} problem a $2.5$-approximation can be computed  in 
$O(k!\cdot k)+\polyn$ time, if the number of vertices with deadlines is $k$. 
Moreover, no $(2-\eps)$-approximation can be computed in $f(k)\polyn$ time for 
any $\eps>0$ and computable function~$f$, unless P=NP.
\end{Theorem}

\mypar{Low dimensional metrics}
Just as for clustering problems, another well-studied parameter in network 
design is the dimension of the underlying geometric space. A typical setting is 
when the input is assumed to be a set of points in some $k$-dimensional 
$\ell_p$-metric, where distances between points $x$ and $y$ are given by a 
function $\dist(x,y)=(\sum_{i=1}^k |x_i-y_i|^p)^{1/p}$. Two prominent examples 
are Euclidean metrics (where $p=2$) and Manhattan metrics (where $p=1$). The 
dimension $k$ of the metric space has been studied as a parameter from the 
parameterized approximation point-of-view {avant la lettre} for quite a 
while. It was shown~\cite{papadimitriou1977euclidean,garey1977complexity} that 
both \pname{Steiner Tree} and \pname{TSP} are paraNP-hard for this parameter 
(since they are NP-hard even if $k=2$), and that they are APX-hard in general 
metrics~\cite{chlebik2008steiner,karpinski2015}. However, a PAS for Euclidean 
metrics both for the \pname{Steiner Tree} and the \pname{TSP} problems were 
shown to exist in the seminal work 
of~\citet{arora1998polynomial}.\footnote{In~\cite{arora1998polynomial} the 
runtime of these algorithms is stated as $O(n(\log 
n)^{O(\sqrt{k}/\eps)^{k-1}})$, which can be shown to be upper bounded by 
$k^{O(\sqrt{k}/\eps)^{k-1}}n^2$ (see 
e.g.,~\cite[Lemma~1]{katsikarelis2019structural}.} The techniques are similar to 
those used for clustering, and we refer to Section~\ref{sec:cluster-Euclidean} 
for an overview.

\begin{Theorem}[\cite{arora1998polynomial}]
For the \pname{Steiner Tree} and \pname{TSP} problems a $(1+\eps)$-approximation 
can be computed in $k^{O(\sqrt{k}/\eps)^{k-1}}n^2$ time for any $\eps>0$, if the 
input consists of $n$ points in $k$-dimensional Euclidean space.
\end{Theorem}

This result also holds for the \pname{$t$-MST} and \pname{$t$-TSP} 
problems~\cite{arora1998polynomial}, where the cheapest tree or tour, 
respectively, on at least $t$ nodes needs to be found. In this case the runtime 
has to be multiplied by $t$ however.

A related setting is the parameterization by the {doubling dimension} of 
the underlying metric. That is, when the parameter $k$ is the smallest integer 
such that any ball in the metric can be covered by $2^k$ balls of half the 
radius. Any point set in a $k$ dimensional $\ell_p$-metric has doubling 
dimension~$O(k)$, and thus the latter parameter generalizes the former. For the 
\pname{TSP} problem the above theorem can be generalized~\cite{Gottlieb15} to a 
PAS parameterized by the doubling dimension.

\begin{Theorem}[\cite{Gottlieb15}]
For the \pname{TSP} problem a $(1+\eps)$-approximation can be computed in 
$2^{(k/\eps)^{O(k^2)}}n\log^2 n$ time for any $\eps>0$, if the input consists 
of $n$ points with doubling dimension $k$.
\end{Theorem}

Given that a PAS exists for \pname{Steiner Tree} in the Euclidean case, it is 
only natural to ask whether this is also possible for low doubling metrics. 
Only a QPTAS is known so far~\cite{talwar2004bypassing}. Moreover, a related 
parameter is the {highway dimension}, which is used to model transportation 
networks. As shown by \citet{FeldmannFKP-highway-2015} the techniques 
of \citet{talwar2004bypassing} for low doubling metrics can be generalized to 
the highway dimension to obtain a QPTAS as well. Again, it is quite 
plausible to assume that a PAS exists.

\begin{Open Question}
Is there a PAS for \pname{Steiner Tree} parameterized by the doubling 
dimension? Is there a PAS for either \pname{Steiner Tree} or \pname{TSP} 
parameterized by the highway dimension?
\label{q:steiner}
\end{Open Question}

\mypar{Directed Graphs}
When considering directed input graphs (asymmetric metrics), the \pname{Directed 
Steiner Tree} problem takes as input a terminal set and a special terminal 
called the {root}. The task is to compute a directed tree of minimum weight 
that contains a path from each terminal to the root. In~general no 
$f(k)$-approximation can be computed in FPT time for any computable 
function~$f$, when~the parameter $k$ is the number of Steiner vertices in the 
optimum solution~\cite{st-pas}. A notable special case is the {unweighted} 
\pname{Directed Steiner Tree} problem, which for this parameter admits a PAS. 
The~techniques here are the same as those used to obtain Theorem~\ref{thm:Steiner-tree} 
for the undirected case. However, in contrast to the undirected case which 
admits a PSAKS, no polynomial-sized $(2-\eps)$\hy{}approximate kernelization 
exists for \pname{Directed Steiner Tree}~\cite{st-pas}, unless \NPcoNP. It is 
an intriguing question whether a $2$-approximate kernel exists.

\begin{Open Question}
Is there a polynomial-sized $2$-approximate kernel for the unweighted 
\pname{Directed Steiner Tree} problem parameterized by the number of Steiner 
vertices in the optimum solution?
\end{Open Question}

If the parameter is the number of terminals, the (weighted) \pname{Directed 
Steiner Tree} problem is FPT, using the same algorithm as for the undirected 
version~\cite{DBLP:journals/networks/DreyfusW71,fuchs2007dynamic}. A different 
variant of \pname{Steiner Tree} in directed graphs is the \pname{Strongly 
Connected Steiner Subgraph} problem, where a terminal set needs to be strongly 
connected in the cheapest possible way. This problem is W[1]-hard parameterized 
by the number of terminals~\cite{DBLP:journals/siamdm/GuoNS11}, and no 
$O(\log^{2-\eps} n)$-approximation can be computed in polynomial 
time~\cite{approx-hardness}, unless 
NP~$\subseteq$~ZTIME$(n^{\text{polylog}(n)})$. However, a $2$\hy{}approximation 
can be computed in FPT time~\cite{DBLP:conf/iwpec/ChitnisHK13}. 

The crucial observation for this algorithm is that in any strongly connected 
solution, fixing some terminal as the root, every terminal can be reached from 
the root, while at the same time the root can be reached from each terminal. 
Thus the optimum solution is the union of two directed trees, of which one is 
directed towards the root and the other is directed away from the root, and the 
leaves of both trees are terminals. Hence it suffices to compute two solutions 
to the \pname{Directed Steiner Tree} problem, which can be done in FPT time, to 
obtain a $2$-approximation for \pname{Strongly Connected Steiner Subgraph}. 
Interestingly, no better approximation is possible with this 
runtime~\cite{chitnis2017parameterized}.

\begin{Theorem}[\cite{chitnis2017parameterized,DBLP:conf/iwpec/ChitnisHK13}]
For the \pname{Strongly Connected Steiner Subgraph} problem a $2$-approximation 
can be computed in $(2+\delta)^k\polyn$ time for any constant $\delta>0$, where 
$k$ is the number of terminals. Moreover, under \gapeth no 
$(2-\eps)$-approximation can be computed in $f(k)\polyn$ time for any $\eps>0$ 
and computable function~$f$.
\end{Theorem}

A generalization of both \pname{Directed Steiner Tree} and \pname{Strongly 
Connected Steiner Subgraph} is the \pname{Directed Steiner Network} 
problem,\footnote{sometimes also called \pname{Directed Steiner Forest}; 
note however that the optimum is not necessarily a forest.} for which an 
edge-weighted directed graph is given together with a list of ordered terminal 
pairs. The aim is to compute the cheapest subgraph that contains a path from $s$ 
to~$t$ for every terminal pair $(s,t)$. If $k$ is the number of terminals, then 
for this problem no $k^{1/4-o(1)}$-approximation can be computed in $f(k)\polyn$ 
time~\cite{DM18} for any computable function $f$, under \gapeth. Both a PAS 
and a PSAKS exist~\cite{chitnis2017parameterized} for the special case when the 
input graph is planar and {bidirected}, i.e., for every directed edge $uv$ 
the reverse edge $vu$ exists and has the same cost.

Similar to the PSAKS for the \pname{Steiner Tree} problem, these two algorithms 
are based on a generalization of \citet{borchers-du}. That is, 
\citet{chitnis2017parameterized} show that a planar solution in a bidirected 
graph can be covered by planar graphs with at most $2^{O(1/\eps)}$ terminals 
each, such that the sum of their costs is at most $1+\eps$ times the cost of the 
solution. These covering graphs may need to contain edges that are reverse to 
those in the solution, but are themselves not part of the solution. For~this the 
underlying graph needs to be bidirected. Analogous to \pname{Steiner Tree}, to 
obtain a kernel it then suffices to compute solutions for every possible list of 
ordered pairs of at most $2^{O(1/\eps)}$ terminals. In contrast to 
\pname{Steiner Tree} however, there is no FPT algorithm for this. Instead, an XP 
algorithm with runtime $2^{O(k^{3/2}\log k)}n^{O(\sqrt{k})}$ needs to be used, 
which runs in polynomial time for $k\leq 2^{O(1/\eps)}$ terminals with $\eps$ 
being a constant. After taking the union of all computed solutions, the number 
of Steiner vertices and the encoding length of the edge weights can be reduced 
in a similar way as for the \pname{Steiner Tree} problem. To obtain a PAS, the 
algorithm will guess how the planar optimum can be covered by solutions 
involving only small numbers of terminals. It will then compute solutions on 
these subsets of at most $2^{O(1/\eps)}$ terminals using the same XP algorithm.

\begin{Theorem}[\cite{chitnis2017parameterized}]
For the \pname{Directed Steiner Network} problem on planar bidirected graphs a 
\mbox{$(1+\eps)$}-approximation can be computed in 
$\max\{2^{k^{2^{O(1/\varepsilon)}}}, 
n^{2^{O(1/\varepsilon)}}\}$ time for any $\eps>0$, where $k$ is the number of 
terminals. Moreover, a $(1+\eps)$-approximate kernel of size 
$(k/\eps)^{2^{O(1/\eps)}}$ can be computed in polynomial time.
\end{Theorem}

\subsection{Cut Problems}
\label{sec:cuts}
Starting from Menger's theorem and the corresponding algorithm for \stcut, graph cut problems have always been at the heart of combinatorial optimization.
While many natural generalizations of \stcut are NP-hard, further study of 
these cut problems yielded beautiful techniques such as flow-cut gaps and
metric embeddings in approximation algorithms~\cite{leighton1999multicommodity, 
arora2009expander}, and also important separators and randomized contractions 
in parameterized algorithms~\cite{marx2006parameterized, marx2014fixed, 
chitnis2016designing, cygan2019minimum}.

\subsubsection{Multicut}
\label{sec:multicut}
An instance of \umc (resp. \dmc) is an undirected (resp. directed) graph $G = (V, E)$ with $k$ pairs of vertices $(s_1, t_1), \dots, (s_k, t_k)$. 
The goal is to remove the minimum number of edges such that there is no path from $s_i$ to $t_i$ for every $i \in [k]$. 
\umwc (resp. \dmwc) is a special case of \umc (resp. \dmwc) where $k$ vertices are given as terminals and the goal is to make sure there is no path between any pair of terminals. 
They have been actively studied from both approximation and parameterized 
algorithms perspectives.
We survey parameterized approximation algorithms for these problems with 
parameters $k$ and the solution size $\OPT$. 

\mypar{Undirected Multicut} 
\umc admits an $O(\log k)$-approximation algorithm~\cite{garg1996approximate} in polynomial time, and is NP-hard to approximate within any constant factor assuming the Unique Games Conjecture~\cite{chawla2006hardness}. 
\umwc admits an $1.2965$-approximation algorithm~\cite{sharma2013multiway} in polynomial time, and is NP-hard to approximate within a factor $1.20016$~\cite{berczi2019improving}. \umc (and thus \umwc) admits an exact algorithm parametrized by $\OPT$~\cite{marx2006parameterized, marx2014fixed}.

With $k$ as a parameter, we cannot hope for an exact algorithm or an 
approximation scheme, since even \umwc with $3$ terminals is NP-hard to 
approximate within a factor $12/11 - \eps$ for any $\eps > 0$ under the Unique 
Games Conjecture. However, for \umc with $k$ pairs $(s_1, t_1), \dots, (s_k, 
t_k)$, one can reduce it to $k^{O(k)}$ instances of \umwc with at most $2k$ 
terminals, by guessing a partition of these $s_1, t_1, \dots, s_k, t_k$ 
according to the connected components containing them in the optimal solution 
(e.g., $s_i$ and $t_i$ should be always in different groups), merging the 
vertices in the same group into one vertex, and solving \umwc with the merged 
vertices as terminals. This shows an $1.2965$-approximation algorithm for \umc 
that runs in time $k^{O(k)} n^{O(1)}$. 

Some recent results improve or generalize this observation. 
For graphs with bounded genus $g$, Cohen-Addad et al~\cite{cohen2018near-cut} gave an \epas
running in time $f(g, k, \eps)\cdot n \log n$. 
Chekuri and Madan~\cite{chekuri2017approximating} considered the { demand graph} $H$, which is the graph formed by $k$ edges $(s_1, t_1), \dots, (s_k, t_k)$. When~$t$ is the smallest integer such that $H$ does not contain $t$ disjoint edges as an induced subgraph, they presented a $2$-approximation algorithm that runs in time $k^{O(t)} n^{O(1)}$. 

\mypar{Directed Multicut}
Generally, \dmc is a much harder computational task than \umc in terms of both approximation and parameterized algorithms. 
\dmc admits a $\min(k, \tilde{O}(n^{11/23}))$-approximation 
algorithm~\cite{agarwal2007improved}. It is NP-hard to approximate within a 
factor $k - \eps$ for any $\eps > 0$ for fixed $k$~\cite{lee2017improved} under 
the Unique Games Conjecture, or $2^{\Omega(\log^{1-\eps} n)}$ for any $\eps > 
0$~\cite{chuzhoy2009polynomial} for general $k$. \dmwc admits an 
$2$-approximation algorithm~\cite{naor19972}, which is tight even when $k = 
2$~\cite{lee2017improved}. 
Parameterizing by \Opt, \dmc is FPT for $k = 2$, but \dmc is W[1]-hard even 
when $k = 4$~\cite{PilipczukW18a}. \dmwc on the other hand is in 
FPT~\cite{chitnis2016designing}.

Since it is hard to improve the trivial $k$-approximation algorithm even for fixed $k$~\cite{lee2017improved}, parameterizing by $k$ does not yield a better approximation algorithm. 
Chitnis and Feldmann~\cite{chitnis2019inapproximability} gave a $k/2$-approximation algorithm that runs in time $2^{O(\Opt^2)} n^{O(1)}$, and also proved that the problem under the same parameterization is still hard to approximate within a factor $59/58$ with $k = 4$.
\begin{Open Question}
What is the best approximation ratio (as a function of $k$) achieved by a parameterized algorithm (with parameter $\Opt$)? Will it be close to $O(1)$ or $\Omega(k)$?
\label{q:dmc}
\end{Open Question}

\subsubsection{Minimum Bisection and Balanced Separator.}
Given a graph $G = (V, E)$, \meb (resp. \mvb) asks to remove the fewest number of edges such that the graph is partitioned into two parts $A$ and $B$ with $||A| - |B|| \leq 1$.
\bes (resp. \bvs) is a more relaxed version of the problem where the goal is to bound the size of the largest component by $\alpha n$ for some $1/2 < \alpha < 1$. 
It has been actively studied from approximation algorithms, culminating in $O(\sqrt{\log n})$-approximation algorithms for both \bes and \bvs~\cite{arora2009expander, feige2008improved}, and an $O(\log n)$-approximation algorithm for \meb~\cite{racke2008optimal}. 

If we parameterize by the size of optimal separator $k$, 
\meb admits an exact parameterized algorithm~\cite{cygan2019minimum}.
While \mvb is W[1]-hard~\cite{marx2006parameterized}, \mbox{Feige and Mahdian~\cite{feige2006finding}} gave an algorithm that given $2/3 \leq \alpha < 1$ and $\eps > 0$, in time $2^{O(k)} n^{O(1)}$ returns an $(\alpha + \eps)$ separator of size at most $k$.

\subsubsection{$k$-Cut}
Given an undirected graph $G = (V, E)$ and an integer $k \in \bbn$, the \kcut 
problem asks to remove the smallest number of edges such that $G$ is 
partitioned into at least $k$ non-empty connected components. 
The edge contraction algorithm by Karger and Stein~\cite{karger1996new} yields 
a randomized exact XP algorithm running in time $O(n^{2k})$, which was made 
deterministic by Thorup~\cite{thorup2008minimum}. There were recent improvements 
to the running time~\cite{gupta2018faster, gupta2019number}. There is an 
exact parameterized algorithm with parameter $\OPT$~\cite{kawarabayashi2011minimum, 
chitnis2016designing}. For general $k$, it admits a $(2 - 2/k)$-approximation 
algorithm~\cite{saran1995finding}, and is NP-hard to approximate within a 
factor $(2 -\eps)$ for any $\eps > 0$ under the Small Set Expansion Hypothesis~\cite{manurangsi2018inapproximability}.

A simple reduction shows that \kcut captures \textsc{$(k - 1)$-Clique}, so an 
exact FPT algorithm with parameter $k$ is unlikely to exist. Gupta et 
al.~\cite{gupta2018fpt} gave an $(2 - \delta)$-approximation algorithm for a 
small universal constant $\delta > 0$ that runs in time $f(k)\cdot n^{O(1)}$. 
The approximation ratio was improved to $1.81$ 
in~\cite{gupta2018faster}, and further to $1.66$~\cite{kawarabayashi2020nearly}.
Very recently, Lokshtanov et al.~\cite{lokshtanov2020parameterized} gave a PAS that runs in time $(k/\eps)^{O(k)} n^{O(1)}$, thereby (essentially) resolving the parameterized approximability of \kcut.

\subsection{\fdeletion Problems}
\label{sec:width-reduction}

Let $\calf$ be a vertex-hereditary family of undirected graphs, 
which means that if $G \in \calf$ and $H$ is a vertex-induced subgraph of $G$, then $H \in \calf$ as well. 
\fdeletion is the problem where given a graph $G = (V, E)$, we are supposed to find $S \subseteq V$ such that the subgraph induced by $V \setminus S$ (denoted by $G \setminus S$) belongs to $\calf$. 
The goal is to minimize $|S|$. 
The natural { weighted version}, where there is a non-negative weight $w(v)$ for each vertex $v$ and the goal is to minimize the sum of the weights of the vertices in $S$, is called \wfdeletion.

\fdeletion captures numerous combinatorial optimization problems, including
\vc (when $\calf$ includes all graphs with no edges), 
\fvs (when $\calf$ is the set of all forests), and 
\oct (when $\calf$ is the set of all bipartite graphs).
There are a lot more interesting graph classes $\calf$ studied in structural and algorithmic graph theory.
Some famous examples include planar graphs, perfect graphs, chordal graphs, and graphs with bounded treewidth. 

In addition to beautiful structural results that give multiple equivalent characterizations, 
these~graph classes often admit very efficient algorithms for some tasks that are believed to be hard in general graphs.
Therefore, a systematic study of \fdeletion for more graph classes is not only 
an interesting algorithmic task by itself, but also a way to obtain better 
algorithms for other optimization problems when the given graph $G$ is { 
close} to a nice class $\calf$ (i.e., deleting few vertices from $G$ makes it 
belong to $\calf$.)
Indeed, some algorithms for \is for noisy planar/minor-free graphs discussed in Section~\ref{sec:packing} 
use an algorithm for \fdeletion as a subroutine~\cite{bansal2017lp}.

\iffalse %
A well-known example is \is, which is hard to approximate within a factor $n^{1 - \eps}$ for any $\eps > 0$ in general graphs,
but is known to admit an exact polynomial time algorithm in perfect graphs and a \ptas in planar graphs. 
\afnote{citations missing}
\enote{Cite Baker, BRU, etc?}
\fi %

For the maximization version where the goal is to {maximize} $|V \setminus 
S|$, a powerful but pessimistic characterization is known. 
Lund and Yannakakis~\cite{lund1993approximation} showed that whenever $\calf$ 
is vertex-hereditary and {nontrivial} (i.e., there are infinitely many 
graphs in $\calf$ and out of $\calf$), the maximization version is hard to 
approximate within a factor $2^{\log^{1/2 - \eps} n}$ for any $\eps > 0$. 
So no nontrivial $\calf$ is likely to admit even a polylogarithmic approximation 
algorithm. However, the situation is different for the minimization problem, since 
\vc admits a $2$-approximation algorithm, while \oct~\cite{khot2002power} and 
\perfectdeletion~\cite{heggernes2011parameterized} are NP-hard to  
approximate within any constant factor approximation algorithm. 
(The first result assumes the Unique Games Conjecture.) 
It indicates that a characterization 
of approximabilities for the minimization versions will be more complex and 
challenging.

There are two (closely related) frameworks to capture large graph classes. 

\begin{itemize}
\item Choose a {graph width parameter} (e.g., treewidth, pathwidth, cliquewidth, rankwidth, etc.) and $k \in \bbn$. Let $\calf$ be the set of graphs $G$ with the chosen width parameter at most $k$. The parameter of \fdeletion is $k$. 
\item Choose a notion of {subgraph} (e.g., subgraph, induced subgraph, minor, etc.) and a finite family of forbidden graphs $\calh$. Let $\calf$ be the set of graphs $G$ that do not have any graph in $\calh$ as the chosen notion of subgraph. The parameter of \fdeletion is $|\calh| := \sum_{H \in \calh} |V(H)|$. 
\end{itemize}

Many interesting classes are capture by the above frameworks. For example, to 
express \fvs, we can take $\calf$ to be the set of graphs with treewidth at most 
$1$, or equivalently, the~set of graphs that does not have the triangle graph 
$K_3$ as a minor. In the rest of the subsection, we~introduce known results of 
\fdeletion under the above two parameterization. Note that under these two 
parameterizations, the need for approximation is inherent since the simplest 
problem in both frameworks, \vc, already does not admit a polynomial-time
$(2-\eps)$-approximation algorithm under the Unique Games Conjecture.

Finally, we mention that the parameterization by the size of the optimal 
solution has been studied more actively from the parameterized complexity 
community, where many important problems are shown to be in FPT~\cite{fomin2012planar, marx2010chordal, cao2014interval}.

\subsubsection{Treewidth and Planar Minor Deletion}\label{sec:treewidth}
The treewidth of a graph (see Definition~\ref{def:treewidth}) is arguably the most well-studied 
graph width parameter with numerous structural and algorithmic applications. 
It is one of the most important concepts in the graph minor project of Robertson and Seymour.
Algorithmically, Courcelle's theorem~\cite{courcelle1990monadic} states that every problem expressible in the monadic second-order logic of graphs can be solved in FPT time parameterized by treewidth. 
We refer the reader to the survey of Bodlaender~\cite{bodlaender2007treewidth}. 
Computing treewidth is NP-hard in general~\cite{arnborg1987complexity}, but if we parameterize by treewidth, it can be done in FPT time~\cite{bodlaender1996linear}, and there is a faster constant-factor parameterized approximation algorithm~\cite{bodlaender2016c}.

Let $k \in \bbn$ be the parameter. \tkd (also known as \textsc{Treewidth $k$-Modulator} in the literature) is a special case of \fdeletion where $\calf$ is the set of all graphs with treewidth at most $k$. Note the case $k = 0$ yields \vc and $k = 1$ yields \fvs. 

Fomin et al.~\cite{fomin2012planar} gave a randomized $f(k)$-approximation algorithm that runs in $g(k) \cdot nm$ for some computable functions $f$ and $g$. 
The approximation ratio was improved by Gupta et al.~\cite{gupta2019losing} that gave a deterministic $O(\log k)$-approximation algorithm that runs in $f(k) \cdot n^{O(1)}$ some $f$. 

This result has immediate applications to minor deletion problems. Let $\calh$ 
be a finite set of graphs, and consider \hmd, which is a special case of 
\fdeletion when $\calf$ is the set of all graphs that do not have any graph in 
$\calh$ as a minor. 
Its parameterized and kernelization complexity (with parameter $\Opt$) for family $\calh$ has been actively studied~\cite{fomin2012planar, jansen2018polynomial, donkers2019turing}.

When $\calh$ contains a planar graph $H$ (also known 
as \phd in the literature), by~the polynomial grid-minor 
theorem~\cite{chekuri2016polynomial}, any graph $G \in \calf$ has treewidth at 
most $k := \poly(|V(H)|)$. Therefore, in order to solve \hmd, one can first 
solve \tkd to reduce the treewidth to $k$ and then solve \hmd optimally 
using Courcelle's theorem~\cite{courcelle1990monadic}. 
Combined with the above algorithm for 
\tkd~\cite{gupta2019losing}, this strategy yields an $O(\log k)$-approximation 
algorithm that runs in $f(|\calh|) \cdot n^{O(1)}$ time. 

Beyond \phd, there are not many results known for \hmd. The case $\calh = \{ 
K_5, K_{3, 3} \}$ is called \planarization and was recently shown to admit an 
$O(\log^{O(1)} n)$-approximation algorithm in $n^{O(\log n/\log \log n)}$ 
time~\cite{kawarabayashi2017polylogarithmic}. 

While the unweighted versions of \tkd and \phd admit an approximation algorithm 
whose approximation ratio only depends on $k$ not $n$, such an algorithm is not 
known for \wtkd or \wphd.
Agrawal et al.~\cite{agrawal2018polylogarithmic} gave a randomized 
$O(\log^{1.5} n)$-approximation algorithm and a deterministic $O(\log^2 
n)$-approximation algorithm that run in polynomial time for fixed $k$, i.e., 
the degree of the polynomial depends on $k$. Bansal et al.~\cite{bansal2017lp} 
gave an $O(\log n \log \log n)$-approximation algorithm for the edge deletion 
version. 
The only graphs $H$ whose weighted minor deletion problem is known to admit a constant factor approximation algorithm are single edge (\wvc), triangle (\wfvs), and diamond~\cite{fiorini2010hitting}.
For the weighted versions, no hardness beyond \vc is known. 

\begin{Open Question}
Does \wtkd admit an $f(k)$-approximation algorithm with parameter $k$ for some function $f$?
Does \tkd admit a $c$-approximation algorithm with parameter $k$ for some universal constant $c$?
\label{q:tkd}
\end{Open Question}

\mypar{Algorithms for \tkd} 
Here we present high-level ideals of~\cite{gupta2019losing,agrawal2018polylogarithmic} for \tkd and \wtkd respectively. These two algorithms share the following two important ingredients: 

\begin{enumerate}
\item Graphs with bounded treewidth admit good { separators.} 
\item There are good approximation algorithms to find such separators. 
\end{enumerate}

Given an undirected and vertex-weighted graph $G = (V, E)$ and an integer $k \in 
\bbn$, let~\wkvs be the problem whose goal is to remove the vertices of minimum 
total weight so that each connected component has at most $k$ vertices. An 
algorithm is called an $\alpha$-bicriteria approximation algorithm if it returns 
a solution whose total weight is at most $\alpha \cdot \OPT$ and each connected 
component has at most $1.1k$ vertices.\footnote{here $1.1$ can be 
replaced by $1 + \eps$ for any constant $\eps > 0$.} The case $k = 2n/3$ is 
called \balsep and has been actively studied in the approximation algorithms 
community, and the best approximation algorithm achieves $O(\sqrt{\log 
n})$-bicriteria approximation~\cite{feige2008improved}. When $k$ is small, 
$O(\log k)$-bicriteria approximation is also 
possible~\cite{lee2017partitioning}. 

\mypar{\wtkd}
\citet{agrawal2018polylogarithmic} achieves an $O(\log^{1.5} n)$-approximation 
for \wtkd in time $n^{O(k)}$. It would be interesting to see whether the running time can be made FPT with parameter $k$. 

The main structure of their algorithm is {top-down recursive}.
Deleting the optimal solution $S^*$ from $G$ reduces the treewidth of $G \setminus S^*$ to $k$, 
so from the forest decomposition of $G \setminus S^*$, there exists a set $M^* \subseteq G \setminus S^*$ with at most $k + 1$ vertices such that each connected component of $G \setminus (M^* \cup S^*)$ has at most $2n/3$ vertices. While we do not know $S^*$, we can exhaustively try every possible $M \subseteq V$ with $|M| \leq k + 1$ and use the bicriteria approximation algorithm for \balsep to find $M$ and $S$ such that (1) $|M| \leq k + 1$, (2) $w(S) \leq O(\sqrt{\log n}) \OPT$, and (3) $G \setminus (M \setminus S)$ has at most $1.1 \cdot (2n/3) \leq 3n/4$ vertices. 

Let $G_1, \dots, G_t$ be the resulting connected components of $G \setminus (S \cap M)$. We solve each $G_i$ recursively to compute $S_i$ such that each $G_i \setminus S_i$ has treewidth at most $k$. The weight of $S$ was already bounded in terms of $\OPT$, but the weight of $M$ was not, so we finally need to consider the graph induced by $M \cup V(G_1) \cup \dots \cup V(G_t)$ and delete vertices of small weight to ensure small treewidth. However, this~task is easy since since the treewidth of each $G_i$ is bounded by $k$ and $|M| \leq k + 1$, which bounds the treewidth of the considered graph by $2k + 1$. So we can fetch the algorithm for small treewidth graphs to solve the problem optimally. 
Note that the total weight of removed vetices in this recursive call is at most $(O(\sqrt{\log n}) + 1)\OPT$. Since $\sum_i \OPT(G_i) \leq \OPT(G)$ and the recursion depth is at most $O(\log n)$, the total approximation ratio is $O(\log^{1.5} n)$.

\mypar{\tkd}
\citet{gupta2019losing} give an $O(\log k)$-approximation algorithm that runs 
in time $f(k) \cdot n^{O(1)}$ for the unweighted version of \tkd. 
The main structure of this algorithm is { bottom-up iterative refinement}. 
The algorithm maintains a feasible solution $S \subseteq V$ (we can start with 
$S = V$), and iteratively uses $S$ to obtain another feasible solution $S'$. If 
the new solution is not smaller (i.e., $|S'| \geq |S|$), then $|S| \leq O(\log 
k) \cdot \OPT$. 

Let us focus on one refinement step with the current feasible solution $S$. Let $S^*$ be the optimal solution, so that $G \setminus S^*$ has treewidth at most $k$. 
We use the following simple lemma showing the existence of a good separator of $G$ in a finer scale than before. 
\begin{Lemma}[\cite{fomin2011bidimensionality, gupta2019losing}]
Let $H$ be a graph with treewidth at most $k$, $T \subseteq V(H)$ be any subset of vertices, and~$\eps > 0$. There exists $R \subseteq V(H)$ such that 
(1) $|R| \leq \eps |T|$ and (2) every connected component of $H \setminus R$ has at most $O(k/\eps)$ vertices from $T$. 
\end{Lemma}
Plugging $H \leftarrow G \setminus S^*$, $T \leftarrow S$ in the above lemma and letting $S' = R \cup S^*$, we can conclude that there exists $S' \subseteq V$ such that $|S'| \leq |R| + |S^*| \leq \eps |S| + \OPT$ and each connected component of $G \setminus S'$ has at most $O(k/\eps)$ vertices from $S$. 

How can we find such a set $S'$ efficiently? Note that if $S = V$, then $S'$ is an $O(k/\eps)$-vertex separator of $G$. Lee~\cite{lee2017partitioning} defined a generalization of \kvs called \ksvs, where the input consists of $G = (V, E)$, $S \subseteq V$, $k \in \bbn$, and the goal is to remove the smallest number of vertices so that each connected component has at most $k$ vertices from $S$, and gave an $O(\log k)$-bicriteria approximation algorithm.%

Since the above lemma guarantees that $\OPT$ for \textsc{$O(k/\eps)$-Subset 
Vertex Separator} is at most $\OPT$ for \tkd plus $\eps |S|$, applying this 
bicriteria approximation algorithm yields $S'$ such that $|S'| \leq O(\log k) ( 
\OPT + \eps |S|)$ and each connected component of $G \setminus S'$ has at most 
$O(k/\eps)$ vertices from $S$. Since $S$ is a feasible solution, it implies 
that the treewidth of each connected component is bounded by $O(k/\eps)$, so we 
can solve each component optimally in time $f(k/\eps) \cdot n^{O(1)}$. By 
setting $\eps = 0.5$, we can see the size of new solution is strictly decreased 
unless $|S| = O(\log k) \cdot \OPT$, finishing the proof. 

\subsubsection{Subgraph Deletion}
Let $H$ be a fixed {pattern graph} with $k$ vertices. Given a { host 
graph} $G$, deciding whether $H$ is a subgraph of $G$ (in the usual sense) is 
known as \si, whose parameterized complexity with various parameters (e.g., $k$, $\tw(H)$, 
$\genus(G)$, etc.) was studied by \citet{marx2014everything}. 
\iffalse
One result 
relevant to our parameterization by $H$ is the algorithm that runs in time 
$2^{O(k)} \cdot n^{O(\tw(H))}$ by the color coding 
technique~\cite{alon1995color}. 
\enote{Define tw? genus?} \afnote{tw and genus were both defined in some other 
section (covering or packing, I forget which one.)}
\fi

\citet{guruswami2015inapproximability} studied the corresponding vertex 
deletion problem \hsd (called \textsc{$H$-Transversal} in the paper),
which is a special case of \fdeletion where $\calf$ is the set of graphs that 
do not have $H$ as a subgraph. Note that the problem admits a simple 
$k$-approximation algorithm that runs in time $O(n \cdot f(n, H))$, where 
$f(n, H)$
denotes time to solve \si with the 
pattern graph $H$ and a host graph with $n$ vertices. Their main hardness result 
states that assuming the Unique Games Conjecture, whenever $H$ is $2$-vertex 
connected, for any $\eps > 0$, no polynomial time algorithm (including 
algorithms running in time $n^{f(k)}$ for any $f$) can 
achieve a $(k - \eps)$-approximation. (Without the UGC, they still ruled out 
a $(k - 1 - \eps)$-approximation.)

Among $H$ that are not $2$-vertex-connected, 
there is an
$O(\log k)$-approximation algorithm when $H$ is a star (in time $n^{O(1)}$) 
or a path (in time $f(k) n^{O(1)}$)~\cite{ebenlendrapproximation, guruswami2015inapproximability, lee2017partitioning}. 
\iffalse %
This is 
optimal up to a constant because when $G$ is $(k - 1)$-regular, the problem 
becomes \textsc{Dominating Set}, which is $\Omega(\log k)$-hard to approximate. 
\afnote{again: poly time approx}
When $H$ is a simple path, note that detecting a $k$-path is one of the famous examples in FPT. 
For the corresponding deletion problem, Lee~\cite{lee2017partitioning} gave an 
$O(\log k)$-approximation algorithm that runs in time $f(k) \cdot n^{O(1)}$. 
\fi %
The algorithm for $k$-path  follows from the result for \tkd, because any graph without a 
$k$-path has treewidth at most~$k$. 
Whenever $H$ is a tree with $k$ vertices, detecting a copy of $H$ in $G$ with $n$ vertices
can be done in $2^{O(k)} n^{O(1)}$ time~\cite{alon1995color}, and it is open whether there is an 
$O(\log k)$-approximation algorithm for \hsd in time $f(k) \cdot n^{O(1)}$.
\iffalse %
While the simple $k$-approximation algorithm for any $H$ and the $O(\log 
k)$-approximation algorithm for $k$-star also works on the weighted version, the 
$O(\log k)$-approximation algorithm for $k$-path does not extend to the weighted 
version. While there are $(k - 1)$-approximation algorithms known when $k = 3, 
4$~\cite{tu2011primal, camby2014primal}, it is an interesting open problem to 
study whether there exists an $O(\log k$)-approximation algorithm for the 
weighted version. 
\fi %

\subsubsection{Other Deletion Problems}
\iffalse
\mypar{Perfect graphs} 
Many other interesting graph classes arise by excluding (some subset) of cycles of arbitrarily length as either a subgraph or an induced subgraph. 
Recall that a hole is an induced cycle of length at least $4$. 
Perfect graphs form one of the famous examples; by the strong perfect graph theorem, a graph is perfect if and only if it does not have an odd hole or an odd anti-hole as an induced subgraph. While there are no known nontrivial approximation algorithm for \perfectdeletion, it is proved that it is at least as hard to approximate as \setcover, which is NP-hard to approximate with in a factor $(1 - \eps)\ln n$ for any $\eps > 0$~\cite{heggernes2013parameterized}. (The paper only proves W$[2]$-hardness parameterized by $\OPT$, but the reduction also preserves the approximation ratio.) 
\afnote{I don't understand what this has to do with parameterized approximation}
\fi

\mypar{Chordal graphs} 
A graph is {chordal} if it does not have an induced cycle of length $\geq 4$. 
Chordal graphs form a subclass of perfect graphs that have been actively 
studied. Initially motivated by efficient kernels, approximation algorithms for 
\chordaldeletion have been developed recently.
The current best results are a
$\poly(\OPT)$-approximation~\cite{jansen2018approximation, kim2018erdHos} and a
$O(\log^2 n)$-approximation~\cite{agrawal2018polylogarithmic}. 
%
%

\iffalse
\mypar{Bipartite graphs} 
Bipartite graphs can be characterized by set of all graphs that do not have an odd cycle as a subgraph. The corresponding deletion problem \oct admits an $O(\sqrt{\log n})$-approximation algorithm~\cite{agarwal2005log} and does not have a constant factor approximation algorithm under the Unique Games Conjecture~\cite{khot2002power}. 
\afnote{again: poly approx}
\fi

\mypar{Edge versions}
While this subsection focused on the vertex deletion problem, there are some results on the {edge deletion}, { edge addition}, and { edge modification} versions. (Edge modification allows both addition and deletion.) Cao and Sandeep~\cite{cao2017minimum} studied \textsc{Minimum Fill-In}, whose goal is to add the minimum number of edges to make a graph chordal. They gave new inapproximability results implying improved time lower bounds for parameterized algorithms. 
Giannopoulou et al.~\cite{giannopoulou2017linear} gave $O(1)$-approximation algorithms for \textsc{Planar $\calh$-Immersion Deletion} parameterized by $\calh$. 
\mbox{Bliznets et al.~\cite{bliznets2018hardness}} considered $H$-free edge modification for a forbidden induced subgraph $H$ and give an almost complete characterization on its approximability depending on $H$. %

\mypar{Directed graphs}
There is also a large body of work on parameterized algorithms for vertex 
deletion problems in { directed graphs}.
While many of the known problems (including \dfvs~\cite{chen2007directed}) 
admit an exact FPT algorithm, 
\citet{LRSZ20} studied \doct, and proved that it is 
W[1]-hard and is unlikely to admit an \pas under the Parameterized 
Inapproximability Hypothesis (or Gap-ETH). They complemented the result 
by showing a $2$-approximation algorithm running in time $f(\OPT) n^{O(1)}$.

\subsection{Faster Algorithms and Smaller Kernels via Approximation}
\label{sec:faster-smaller}

The focus of this section so far has been on problems for which its exact 
version is intractable (i.e., W[1]/W[2]-hard) and the goal is to obtain good 
approximations in FPT time. In this subsection, we~shift our focus slightly by 
asking: {does approximation allow us to find faster algorithms for 
problems already known to be in FPT?}

To illustrate this, let us consider \pname{Vertex Cover}. It is of course 
well-known that the {exact} version of the problem can be solved in FPT 
time, with the current best running time being 
$O^*(1.2738^k)$~\cite{chen2006improved}. The question here would be: if we are 
allowed to output an $(1 - \varepsilon)$-approximate solution, instead of just 
an exact one, can we speed up the algorithm?

To the best of our knowledge, such a question was tackled for the first time by 
\mbox{\citet{BourgeoisEP09-moderatelyexp} and} revisited quite a few times in the
literature~\cite{brankovic2010combining,BrankovicF13,fellows2018fidelity,
BansalCLNN19,MT18}. As one might have suspected, the answer to this question is 
a YES, as stated below.

\begin{Theorem}[\cite{fellows2018fidelity}] \label{thm:fidelity}
Let $\delta > 0$ be such that there exists an $O^*(\delta^k)$-time algorithm for 
\pname{Vertex Cover} (e.g.,~$\delta = 1.2738$). Then, for any $\varepsilon > 0$, 
there is an $(1 + \varepsilon)$-approximation for \pname{Vertex Cover} that runs 
in $O^*(\delta^{(1 - \varepsilon)k})$ time. 
\end{Theorem}

The main idea of the algorithm is inspired by the ``local ratio'' method in the 
approximation algorithms literature (see e.g.~\cite{Bar-YehudaBFR04}) and we 
sketch it here. The algorithm works in two stages. In the first stage, we run 
the greedy algorithm: as long as we have picked less than $2\varepsilon k$ 
vertices so far and not all edges are covered, pick an uncovered edge and add 
both endpoints to our solution. In the second stage, we run the exact algorithm 
on the remaining part of the graph to find a \pname{Vertex Cover} of size $(1 - 
\varepsilon)k$. Since the first stage runs in polynomial time, the running time 
of the entire algorithm is dominated by the second stage, whose running time is 
$O^*(\delta^{(1 - \varepsilon)k})$ as desired. The correctness of the algorithm 
follows from the fact that, for each selected edge in the first step, the 
optimal solution still needs to pick at least one endpoint. As a result, the 
optimal solution must pick at least $\varepsilon k$ vertices with respect to the 
first stage (compared to $2\varepsilon k$ picked by the algorithm). Thus, when 
the optimal solution is of size at most $k$, there must be a solution in the 
second stage of size at most $(1 - \varepsilon)k$, meaning that the algorithm 
finds such a solution and outputs a vertex cover of size $(1 + \varepsilon)k$ as 
claimed.

The above ``approximate a small fraction and brute force the rest'' approach of 
\mbox{\citet{fellows2018fidelity}} generalizes naturally to problems 
beyond \pname{Vertex Cover}. \citet{fellows2018fidelity} formalized the method 
in terms of {$\alpha$-fidelity kernelization} and apply it to several 
problems, including \pname{Connected Vertex Cover}, \pname{$d$-Hitting Set} and 
\pname{Steiner Tree}. For these problems, the method gives an $(1 + 
\varepsilon)$-approximation algorithm that runs in time $O^*(\delta^{(1 - 
\Omega(\varepsilon))k})$, where $\delta > 0$ denotes a constant for which a
$O^*(\delta^k)$-time algorithm is known for the exact version of the 
corresponding problem. The approach, in some form or another, is also applicable 
both to other parameterized 
problems~\cite{escoffier2015new,bonnet2016parameterized} and to 
non-parameterized problems (e.g.~\cite{BourgeoisEP09-moderatelyexp}); since the 
latter is out-of-scope for the survey, we will not discuss the specifics here. 

An intriguing question related to this line of work is whether it must be the 
case that the running time of $(1 + \varepsilon)$-approximation algorithms is of 
the form $O^*(\delta^{(1 - \Omega(\varepsilon))k})$. That is, can we get a 
\mbox{$(1 + o(1))$}-approximation for these problems in time $O^*(\lambda^k)$ 
where $\lambda$ is a constant strictly smaller than $\delta$? More specifically, 
we may ask the following:

\begin{Open Question}
Let $\delta > 0$ be the smallest (known) constant such that an 
$O^*(\delta^k)$-time exact algorithm exists for \pname{Vertex Cover}. Is there 
an algorithm that, for any $\varepsilon > 0$, runs in time $f(1/\varepsilon) 
\cdot O^*(\lambda^k)$ for some constant $\lambda < \delta$?
\end{Open Question}

Of course, the question applies not only for \pname{Vertex Cover} but other 
problems in the list as well. The informal crux of this question is whether, in 
the regime of very good approximation factors (i.e.~$1 + o(1)$), approximation 
can still be exploited in such a way that the algorithm works significantly 
better than the approach ``approximate a $o(1)$ fraction and then brute 
force''. 

Turning back once again to our running example of \pname{Vertex Cover}, it turns 
out that algorithms faster than ``approximate a small fraction and then brute 
force'' are known~\cite{brankovic2010combining,BansalCLNN19,MT18} but only for 
the regime of large approximation ratios. In particular, 
\citet{brankovic2010combining} give faster algorithms than in 
Theorem~\ref{thm:fidelity} already for approximation ratios as small as 3/2. The 
algorithms in~\cite{BansalCLNN19,MT18} focus on the case of ``barely 
non-trivial'' $(2 - \rho)$-approximation factors. (Recall the greedy algorithm 
yields a 2-approximation and, under the Unique Games Conjecture, the problem is 
NP-hard to approximate to within any constant factor less than 2.) The algorithm 
in~\cite{BansalCLNN19} has a running time of $O^*(2^{k/2^{\Omega(1/\rho)}})$, 
which was later improved in~\cite{MT18} to $O^*(2^{k/2^{\Omega(1/\rho^2)}})$. 
These running times should be contrasted with that of ``approximate a small 
fraction and then brute force'' (i.e., applying Theorem~\ref{thm:fidelity} 
directly with $\varepsilon = 1 - \rho$) which gives an algorithm with running 
time $O^*(2^{k \rho})$. In other words, Refs.~\cite{BansalCLNN19,MT18}
improve the ``saving factor'' from $1/\rho$ to $2^{\Omega(1/\rho)}$ and 
$2^{\Omega(1/\rho^2)}$ respectively. It should be noted however that, since the 
known $(2 - o(1))$-factor hardness of approximation is shown via the Unique 
Games Conjecture and unique games admit subexponential time 
algorithms~\cite{AroraBS15,BarakRS11}, it is still entirely possible that this 
regime of approximating \pname{Vertex Cover} admits subexponential time 
algorithms as well. This is perhaps the biggest open question in the ``barely 
non-trivial'' approximation range:
\begin{Open Question}
Is there an algorithm that runs in $2^{o(k)} \polyn$ time and achieves an 
approximation ratio of $(2 - \rho)$ for some absolute constant $\rho > 0$?
\end{Open Question}

Let us now briefly discuss the techiques used in some of the aforementioned 
works. The~algorithms in~\cite{brankovic2010combining,BansalCLNN19} are based on 
branching in conjunction with certain approximation techniques. (See 
also~\cite{Fernau12} where a similar technique is used for a related problem 
\pname{Total Vertex Cover}.) A key idea 
in~\cite{brankovic2010combining,BansalCLNN19} is that (i) if the (average or 
maximum) degree of the graph is small, then good polynomial-time approximation 
algorithms are known~\cite{Halperin02} and (ii) if the degree is large, then 
branching algorithms are naturally already fast. The second part 
of~\cite{brankovic2010combining} involves a delicate branching rule. However, 
for~\cite{BansalCLNN19}, it is quite simple: for some threshold $d$ (to be 
specified), as long as there exists a vertex with degree at least $d$, then (1) 
with some probability, simply add the vertex to the vertex cover, or (2) branch 
on both possibilities of it being inside the cover and outside. After~this 
branching finishes and we are left with low-degree graphs, just run the known 
polynomial-time approximation algorithms~\cite{Halperin02} on these graphs. The 
point here is that the ``error'' incurred if option (1) is chosen will be 
absorbed by the approximation. By carefully selecting $d$ and the probability, 
one~can arrive at the desired running time and approximation guarantee. This 
algorithm is randomized, but~can be derandomized using the sparsification 
lemma~\cite{IPZ01}. 

To the best of our knowledge, this ``barely non-trivial approximation'' regime has not been studied beyond \pname{Vertex Cover}. In particular, while Bansal et al.~\cite{BansalCLNN19} apply their techniques on several problems, these are not parameterized problems and we are not aware of any other parameterized study related to the regime discussed here. %

Parallel to the running time questions we have discussed so far, one may ask an 
analogous question in the kernelization regime: {does approximation allow 
us to find smaller kernels for problems that already admit polynomial-size 
kernels?} As is the case with exact algorithms, parameterized approximation 
algorithms go hand in hand with approximate kernels. Indeed, many algorithmic 
improvements mentioned can also be viewed as improvements in terms of the size 
of the kernels. In~particular, recall the proof sketch of 
Theorem~\ref{thm:fidelity} for \pname{Vertex Cover}. If we stop and do not 
proceed with brute force in the second step, then we are left with an $(1 + 
\varepsilon)$-approximate kernel. It is also not hard to argue that, by for 
instance applying the standard $2k$-size kernelization at the end, we are left 
with at most $2(1 - \varepsilon)k$ vertices. This improves upon the best known 
$2k - \Theta(\log k)$ bound for the exact 
kernel~\cite{DBLP:journals/ipl/Lampis11}. A similar improvement is known also 
for \pname{$d$-Hitting Set}~\cite{fellows2018fidelity}.

\section{Future Directions}\label{sec:open}

Although we have provided open questions along the way, we end this survey by 
zooming out and discussing some general future directions or meta-questions, 
which we find to be interesting and could be the basis for future work.

\subsection{Approximation Factors}

The quality of a polynomial-time approximation algorithm is mainly measured by 
the obtainable approximation factor $\alpha$: the smaller it is the more 
feasibly solvable the problem is. Therefore, a lot of work has been invested 
into determining the smallest obtainable approximation factor $\alpha$ for all 
kinds of computationally hard problems. %
In the non-parameterized (i.e., NP-hardness) world, a whole spectrum of 
approximability has been discovered 
(cf.~\cite{Vazirani01book,williamson2011design}): the most feasibly solvable 
NP-hard problems (e.g., the \pname{Knapsack} problem) admit a so-called 
{polynomial-time approximation scheme (PTAS)}, which is an algorithm 
computing a $(1+\eps)$-approximation for any given constant $\eps>0$. Some 
problems can be shown not to admit a PTAS (under reasonable complexity 
assumptions), but still allow constant approximation factors (e.g., the 
\pname{Steiner Tree} problem). Yet others can only be approximated within a 
polylogarithmic factor (e.g., the \pname{Set Cover} problem), while some are 
even harder than this, as the best approximation factor obtainable is polynomial 
in the input size (e.g., the \pname{Clique} problem).

In contrast to polynomial-time approximation algorithms, a full spectrum of 
obtainable approximation ratios is still missing when allowing parameterized 
runtimes. Instead, only some scattered basic results are known. In particular, 
most of parameterized approximation problems belongs to one of the following categories:

\begin{itemize}
 \item A {parameterized approximation scheme (PAS)} exists, i.e., for any 
constant $\eps>0$ a \mbox{$(1+\eps)$}-approximation can be computed in 
$f(k)\polyn$ time for some parameter $k$. These are currently the most prevalent 
types of results in the literature. To just mention one example, the 
\pname{Steiner Tree} problem is APX-hard, but admits a PAS~\cite{st-pas} when 
parameterized by the number of non-terminals (so-called {Steiner vertices}) in 
the optimum solution (cf.~Section~\ref{sec:network-design}).

 \item A lower bound excluding any non-trivial approximation factor exists. For 
example, under \eth the \pname{Dominating Set} problem has no 
$g(k)$-approximation in $f(k)n^{o(k)}$ time~\cite{dom-set} for any functions $g$ 
and~$f$, where $k$ is the size of the largest dominating set. 

 \item A polynomial-time approximation algorithm can achieve a similar 
approximation ratio, i.e., the~parameterization is not very helpful. For 
instance, for the \pname{$k$-Center} problem\footnote{Here we consider 
the version where the set of candidate centers is not separately given.} a 
$2$-approximation can be computed in polynomial 
time~\cite{hochbaum1986bottleneck}, but even when parameterizing by~$k$ no 
$(2-\eps)$-approximation is possible~\cite{DBLP:conf/icalp/Feldmann15} for any 
$\eps>0$, under standard complexity assumptions. A similar situation holds for 
\pname{Max $k$-Coverage}, which we discussed in Section~\ref{sec:covering-max}.

\item Constant or logarithmic approximation ratios can be shown, and which beat 
any approximation ratio obtainable in polynomial time.
For instance, 
\pname{Strongly Connected Steiner Subgraph} problem : under standard 
complexity assumptions, for this problem no polynomial-time $O(\log^{2-\eps} 
n)$-approximation algorithm exists~\cite{approx-hardness}, and there is no FPT 
algorithm parameterized by the number $k$ of 
terminals~\cite{DBLP:journals/siamdm/GuoNS11}. However it is not hard to compute 
a $2$-approximation in $2^{O(k)}\polyn$ time~\cite{DBLP:conf/iwpec/ChitnisHK13}, 
and no $(2-\eps)$-approximation algorithm with runtime $f(k)\polyn$ 
exists~\cite{chitnis2017parameterized} under \gapeth, for any function~$f$ and 
any~$\eps>0$ (cf.~Section~\ref{sec:network-design}).
\end{itemize}

For many problems discussed in this survey, 
including 
\pname{Densest $k$-Subgraph}, %
\pname{Steiner Tree} with bounded doubling/highway dimension,
it has not been determined which category they belong. 
There are also a lot of problems in the final category for which asymptotically tight approximation ratios have not been found,
including \pname{Directed Multicut}, \tkd (both weighted and unweighted).
The parameterized approximability of \hmd for non-planar $\calh$ is also widen open except 
\planarization ($\calh = \{ K_5, K_{3, 3} \}$)~\cite{kawarabayashi2017polylogarithmic}.
It is an immediate but still interesting direction to prove tight parameterized approximation ratios for these (and more) problems. 

Digressing, we remark that this survey does not include FPT-approximation of 
counting problems, such as approximately counting number of $k$-paths in a 
graph. The best $(1+\varepsilon)$-multiplicative factor algorithm 
known~\cite{BDH18,BLSZ19} for counting number of  $k$-paths runs in time $4^k 
f(\varepsilon) poly(n)$ for some subexponential function $f$ 
(cf.~\cite{pratt2019waring}). So a natural question is: can we count $k$-paths 
approximately in time $c^k$, where $c$ is as close to the base of running time 
of the algorithm of deciding existence of $k$-Path in a graph (the best 
currently known $c$ is roughly 1.657~\cite{Bj14,BHKK17})?

\subsection{Parameterized Running Times}
The quality of FPT algorithms is mainly measured in the obtainable runtime. 
Given a parameter~$k$, for some problems the optimum solution can be computed in 
$f(k)n^{g(k)}$ time, for some functions $f$ and~$g$ independent of the input 
size $n$ (i.e., the degree of the polynomial also depends on the parameter). If 
such an algorithm exists the problem is {slice-wise polynomial~(XP)}, and the 
algorithm is called an {XP~algorithm}. A typical example is if a solution of 
size $k$ is to be found within a data set of size~$n$, in~which case 
often an $n^{O(k)}$ time exhaustive search algorithm exists. However, an FPT 
algorithm with runtime, say, $O(2^k n)$ is a lot more efficient than an XP 
algorithm with runtime~$n^{O(k)}$, and therefore the aim is usually to find FPT 
algorithms, while XP algorithms are counted as prohibitively slow. The~discovery 
of the W-hierarchy in complexity theory has paved the way to providing evidence 
when an FPT algorithm is unlikely to exist. Assuming \eth, it is even possible 
to provide lower bounds on the runtimes obtainable by any FPT or XP algorithm. 
Similar to approximation algorithms, this~has lead to the discovery of a 
spectrum of tractability (cf.~\cite{pc-book}): starting from slightly 
sub-exponential $2^{O(\sqrt{k})}\polyn$ time, through single exponential 
$2^{O(k)}\polyn$ time, to double exponential $2^{2^{O(k)}}\polyn$ time for FPT 
algorithms with matching asymptotic lower bounds under \eth (e.g., for the 
\pname{Planar Vertex Cover}, \pname{Vertex Cover}, and \pname{Edge Clique Cover} 
problems, respectively, each parameterized by the solution size). For XP 
algorithms, asymptotically tight runtime bounds of the form $n^{O(\sqrt{k})}$ 
and $n^{O(k)}$ can be obtained under \eth (e.g., for the \pname{Clique} problem 
parameterized by the solution size, and the \pname{Planar Bidirected Steiner 
Network} problem parameterized by the number of 
terminals~\cite{chitnis2017parameterized}, respectively). Finally, problems that 
are NP-hard when the given parameter is constant do not even allow XP algorithms 
unless P=NP (e.g., the \pname{Graph Colouring} problem where the parameter is 
the number of colours).

In terms of tight runtime bounds, existing results on parameterized 
approximation algorithms are few and far between. In particular, most of them 
show that for a given parameter $k$ one of the following cases applies.
\begin{itemize}
 \item An approximation is possible in $f(k)\polyn$ time for {some} 
function $f$. Most current results are only concerned with the existence of an 
algorithm with this type of runtime, i.e., they do not provide any evidence 
that the obtained runtime is best possible, or try to optimize it. The only 
lower bounds known exclude certain types of approximation schemes when a 
hardness result for the parameterization by the solution size exists. For 
instance, it is known that if some problem does not admit a $2^{o(k)}\polyn$ 
time algorithm for this parameter $k$ then it also does not admit an EPTAS with 
runtime $2^{o(1/\eps)}\polyn$ (cf.~\cite{marx2008fpa,downey2013fpt}).

 \item A certain approximation ratio cannot be obtained in $f(k)\polyn$ time 
for {any} function~$f$. For~example, it is known that while a 
$2$-approximation for the \pname{Strongly Connected Steiner Subgraph} problem 
can be computed in $2^{O(k)}\polyn$ time~\cite{DBLP:conf/iwpec/ChitnisHK13}, 
where $k$ is the number of terminals, no $(2-\eps)$-approximation can be 
computed in $f(k)\polyn$ time~\cite{chitnis2017parameterized} for any 
function~$f$, under \gapeth (cf.~Section~\ref{sec:network-design}).
\end{itemize}

Hence, matching lower bounds on the time needed to compute an approximation are 
missing. For~example, is the runtime of $2^{O(k)}\polyn$ best possible to 
compute a $2$-approximation for the \pname{Strongly Connected Steiner Subgraph} 
problem? Could there be a $2^{O(\sqrt{k})}\polyn$ time algorithm to compute a 
$2$-approximation as well? For PASs the exact obtainable runtime is 
often elusive, even if certain types of approximation schemes can be excluded. 
For instance, for the \pname{Steiner Tree} problem parameterized by the number 
of Steiner vertices in the optimum solution a $(1+\eps)$-approximation can be 
computed in $2^{O(k^2/\eps^4)}\polyn$ time~\cite{st-pas}. Is the dependence 
on $k$ and~$\eps$ best possible? Could there be a $2^{O(k/\eps^4)}\polyn$ or 
$2^{O(k^2/\eps)}\polyn$ time algorithm as well?

We remark that, for problems for which straightforward algorithms are known to 
be (essentially) the best possible in FPT time, or for which an improvement 
over polynomial time approximation is not possible, sometimes tight running 
time lower bounds are known in conjunction with tight inapproximability ratios. 
This includes \pname{$k$-Dominating Set} (Section~\ref{sec:domset}), 
\pname{$k$-Clique} (Section~\ref{sec:clique}) and \pname{Max $k$-Coverage} 
(Section~\ref{sec:covering-max}).

\subsection{Kernel Sizes}
The development of {compositionality} has lead to a theory from which lower 
bounds on the size of the smallest possible kernel of a problem can be derived 
(under reasonable complexity assumptions). The spectrum (cf.~\cite{pc-book}) 
here reaches from polynomial-sized kernels (e.g., for any $q\geq 3$ and $\eps>0$ 
the \pname{$q$-SAT} problem parameterized by the number of variables $n$ has no 
$O(n^{q-\eps})$-sized kernel) to exponential-sized kernels (e.g., the 
\pname{Steiner Tree} problem parameterized by the number of terminals does not 
admit any polynomial-sized kernel despite being FPT).

For approximate kernels, only a small number of publications exist, and the few 
known results fall into two categories:

\begin{itemize}
 \item A {polynomial-sized approximate kernelization scheme (PSAKS)} 
exists, i.e., for any $\eps>0$ there is a $(1+\eps)$-approximate kernelization 
algorithm that computes a $(1+\eps)$-approximate kernel of size polynomial in 
the parameter $k$. For example, the \pname{Steiner Tree} problem admits a PSAKS 
for both the parameterization in the number of 
terminals~\cite{lokshtanov2017lossy} and in the number of Steiner vertices in 
the optimum~\cite{st-pas}, even though neither of these two parameters admits a 
polynomial-sized (exact)~kernel.

 \item A lower bound excluding any approximation factor for polynomial-sized 
kernels exists. For~example, the \pname{Longest Path} problem parameterized by 
the maximum path length has no $\alpha$-approximate polynomial-sized kernel for 
any $\alpha$~\cite{lokshtanov2017lossy}, despite being FPT for this 
parameter~\cite{pc-book}.
\end{itemize}

Hence again the intermediate cases, for which tight constant or logarithmic 
approximation factors can be proved for polynomial-sized kernels, are missing. 
Studying approximate kernelization algorithms however is of undeniable 
importance to the field of parameterized approximation algorithms, as witnessed 
by the importance of exact kernelization to fixed-parameter tractability.

\subsection{Completeness in Hardness Of Approximation}
A final direction we would like to highlight is to obtain more 
{completeness} in inapproximability results. Most of the results so far for 
FPT hardness of approximation either (i) rely on gap hypothesis or (ii) yield a 
hardness in terms of the W-hierarchy but the exact version of the problem is 
known to be complete on an even higher level (e.g., \pname{Dominating Set} is 
known to be W[1]-hard to approximate but its exact version is W[2]-complete). 
We have discussed (i) extensively in Section~\ref{sec:Gap} and some examples of 
(ii) in Section~\ref{sec:noGap}. There are also some examples of (ii) that are not 
covered here; for instance, \citet{Marx13} showed W[t]-hardness for certain 
monotone/anti-monotone circuit satisfiability problems and the exact versions of 
these problems are known to be complete for higher levels of the W-hierarchy. 
The situation here is unlike that in the theory of NP-hardness of approximation; 
there the PCP Theorem~\cite{ALMSS98,AS92} implies NP-completeness of 
optimization problems.\footnote{To be more precise, these problems need to be 
phrased as promise problems and NP-hardness is with respect to these. We will 
not go into details here.}

Thus, in the parameterized inapproximability arena, the main question here is 
whether we can prove completeness results for hardness of approximation for the
aforementioned problems. The~two important examples here are: is 
\pname{$k$-Clique} W[1]-hard to approximate, and is \pname{$k$-Dominating Set} 
W[2]-hard to approximate? As discussed in Section~\ref{sec:Gap}, the former is also 
closely related to resolving~PIH.

Finally, we note that, while completeness results are somewhat rare in FPT 
hardness of approximation, some are known. We give two such examples here. First 
is the \pname{$k$-Steiner Orientation} problem, discussed in 
Section~\ref{sec:steiner-orientation}; it is W[1]-complete to 
approximate~\cite{Wlo19}. Second is the \pname{Monotone Circuit Satisfiability} 
problem (without depth bound), which was proved to be W[P]-complete by 
\citet{Marx13}. However, it does not seem clear to us whether these 
techniques can be applied elsewhere, e.g., for \pname{$k$-Clique}.

\vspace{6pt} 

\authorcontributions{Writing---original draft and writing---review and 
editing, A.E.F., C.S., E.L. and P.M.}
\funding{Andreas Emil Feldmann is supported by the Czech Science Foundation GACR 
(grant \#19-27871X), and~by the Center for Foundations of Modern Computer 
Science (Charles Univ. project UNCE/SCI/004). \mbox{Karthik C. S.} is supported by 
ERC-CoG grant 772839, the Israel Science Foundation (grant number 552/16), and~from the Len Blavatnik and the Blavatnik Family foundation. Euiwoong Lee is 
supported by the Simons Collaboration on Algorithms and Geometry.}

\conflictsofinterest{The authors declare no conflict of interest.} %

\reftitle{References}


\begin{thebibliography}{-------}
\providecommand{\natexlab}[1]{#1}

\end{thebibliography}


\begin{thebibliography}{999}
\providecommand{\natexlab}[1]{#1}
\bibitem[Cobham(1964)]{cobham1964intrinsic}
Cobham, A.
\newblock The intrinsic computational difficulty of functions.
\newblock  In Proceedings of the 1964 Congress for Logic, Methodology, and the
  Philosophy of Science, Paris, France, 23--25 July 1964; pp. 24--30.


\bibitem[Edmonds(1965)]{edmonds1965paths}
Edmonds, J.
\newblock Paths, Trees, and Flowers.
\newblock {\em Can. J. Math.} {\bf 1965}, {\em 17},~449--467.

\bibitem[Vazirani(2001)]{Vazirani01book}
Vazirani, V.V.
\newblock {\em Approximation Algorithms}; Springer: Berlin, Germany,  2001.

\bibitem[Williamson and Shmoys(2011)]{williamson2011design}
Williamson, D.P.; Shmoys, D.B.
\newblock {\em The Design of Approximation Algorithms}; Cambridge University
  Press: Cambridge, UK, 2011.

\bibitem[Downey and Fellows(2013)]{downey2013fpt}
Downey, R.G.; Fellows, M.R.
\newblock {\em Fundamentals of Parameterized Complexity}; Springer: Berlin, 
Germany, 2013; Volume 4. 

\bibitem[Cygan {et~al.}(2015)Cygan, Fomin, Kowalik, Lokshtanov, Marx,
  Pilipczuk, Pilipczuk, and Saurabh]{pc-book}
Cygan, M.; Fomin, F.V.; Kowalik, L.; Lokshtanov, D.; Marx, D.; Pilipczuk, M.;
  Pilipczuk, M.; Saurabh, S.
\newblock {\em {Parameterized Algorithms}}; Springer: Berlin, Germany,  
2015.

\bibitem[Cai and Chen(1997)]{cai1997fixed}
Cai, L.; Chen, J.
\newblock On fixed-parameter tractability and approximability of NP
  optimization problems.
\newblock {\em J.~Comput. Syst. Sci.} {\bf 1997}, {\em
  54},~465--474.

\bibitem[Marx(2008)]{marx2008fpa}
Marx, D.
\newblock Parameterized complexity and approximation algorithms.
\newblock {\em  Comput. J.} {\bf 2008}, {\em 51},~60--78.

\bibitem[Flum and Grohe(2006)]{flum2006parameterized}
Flum, J.; Grohe, M.
\newblock {\em Parameterized Complexity Theory}; Springer: Berlin, Germany, 
 2006.

\bibitem[Rubinstein and Williams(2019)]{RW19}
Rubinstein, A.; Williams, V.V.
\newblock {SETH} vs. Approximation.
\newblock {\em {SIGACT} News} {\bf 2019}, {\em 50},~57--76, doi:10.1145/3374857.3374870.

\bibitem[Cesati and Trevisan(1997)]{cesati1997efficiency}
Cesati, M.; Trevisan, L.
\newblock On the efficiency of polynomial time approximation schemes.
\newblock {\em Inf. Process. Lett.} {\bf 1997}, {\em 64},~165--171.

\bibitem[Cygan {et~al.}(2016)Cygan, Lokshtanov, Pilipczuk, Pilipczuk, and
  Saurabh]{cygan2016lower}
Cygan, M.; Lokshtanov, D.; Pilipczuk, M.; Pilipczuk, M.; Saurabh, S.
\newblock Lower Bounds for Approximation Schemes for Closest String.
\newblock   In Proceedings of the 15th Scandinavian Symposium and Workshops on 
Algorithm Theory, Reykjavik, Iceland, 22--24 June 2016.

\bibitem[Cai and Chen(1993)]{cai1993fixed}
Cai, L.; Chen, J.
\newblock On fixed-parameter tractability and approximability of NP-hard
  optimization problems.
\newblock  In~Proceedings of the IEEE 2nd Israel Symposium on Theory and 
Computing Systems, Natanya, Israel, 7--9 June 1993; pp. 118--126.

\bibitem[Chen {et~al.}(2007)Chen, Huang, Kanj, and Xia]{chen2007polynomial}
Chen, J.; Huang, X.; Kanj, I.A.; Xia, G.
\newblock Polynomial time approximation schemes and parameterized complexity.
\newblock {\em Discret. Appl. Math.} {\bf 2007}, {\em 155},~180--193.

\bibitem[Kratsch(2012)]{kratsch2012polynomial}
Kratsch, S.
\newblock Polynomial kernelizations for MIN $F^+\Pi_1$ and MAX NP.
\newblock {\em Algorithmica} {\bf 2012}, {\em 63},~532--550.

\bibitem[Guo {et~al.}(2011)Guo, Kanj, and Kratsch]{guo2011safe}
Guo, J.; Kanj, I.; Kratsch, S.
\newblock Safe approximation and its relation to kernelization.
\newblock  In \emph{International Symposium on Parameterized and Exact Computation};
  Springer: Berlin, Germany,  2011; pp. 169--180.

\bibitem[Cai {et~al.}(1997)Cai, Chen, Downey, and Fellows]{CaiCDF97}
Cai, L.; Chen, J.; Downey, R.G.; Fellows, M.R.
\newblock Advice Classes of Parameterized Tractability.
\newblock {\em Ann. Pure Appl. Log.} {\bf 1997}, {\em 84},~119--138, doi:10.1016/S0168-0072(95)00020-8.

\bibitem[Lokshtanov {et~al.}(2017)Lokshtanov, Panolan, Ramanujan, and
  Saurabh]{lokshtanov2017lossy}
Lokshtanov, D.; Panolan, F.; Ramanujan, M.; Saurabh, S.
\newblock {Lossy Kernelization}.
\newblock  In Proceedings of the 49th Annual ACM SIGACT Symposium on Theory of 
Computing, Montreal, PQ, Canada, 19--23 June 2017; pp.~224--237.

\bibitem[Hermelin {et~al.}(2015)Hermelin, Kratsch, Soltys, Wahlstr{\"{o}}m,
  and Wu]{HKSWW15}
Hermelin, D.; Kratsch, S.; Soltys, K.; Wahlstr{\"{o}}m, M.; Wu, X.
\newblock A Completeness Theory for Polynomial (Turing) Kernelization.
\newblock {\em Algorithmica} {\bf 2015}, {\em 71},~702--730, doi:10.1007/s00453-014-9910-8.

\bibitem[Fellows {et~al.}(2012)Fellows, Kulik, Rosamond, and
  Shachnai]{fellows2018fidelity}
Fellows, M.R.; Kulik, A.; Rosamond, F.; Shachnai, H.
\newblock Parameterized approximation via fidelity preserving transformations.
\newblock  In \emph{International Colloquium on Automata, Languages, and 
Programming}; Springer: Berlin/Heidelberg, Germany, 2012; pp. 351--362.

\bibitem[Arora {et~al.}(1998)Arora, Lund, Motwani, Sudan, and
  Szegedy]{ALMSS98}
Arora, S.; Lund, C.; Motwani, R.; Sudan, M.; Szegedy, M.
\newblock Proof Verification and the Hardness of Approximation Problems.
\newblock {\em J. {ACM}} {\bf 1998}, {\em 45},~501--555, doi:10.1145/278298.278306.

\bibitem[Arora and Safra(1992)]{AS92}
Arora, S.; Safra, S.
\newblock Probabilistic Checking of Proofs; {A} New Characterization of {NP}.
\newblock  In Proceedings of the 33rd Annual Symposium on Foundations of Computer Science,
  Pittsburgh, PA, USA, 24--27 October 1992; pp. 2--13, doi:10.1109/SFCS.1992.267824.

\bibitem[Lin(2018)]{Lin18}
Lin, B.
\newblock The Parameterized Complexity of the \emph{K}-Biclique Probl. {\em J. {ACM}} {\bf 2018}, {\em 65},~34:1--34:23, doi:10.1145/3212622.

\bibitem[{Karthik {C. S.}} and Manurangsi(2019)]{KM19}
{Karthik {C. S.}}.; Manurangsi, P.
\newblock On Closest Pair in Euclidean Metric: Monochromatic is as Hard as
  Bichromatic.
\newblock  In Proceedings of the 10th Innovations in Theoretical Computer Science Conference {ITCS}, San Diego, CA, {USA}, 10--12 January 2019; pp.
  17:1--17:16, doi:10.4230/LIPIcs.ITCS.2019.17.

\bibitem[Chen and Lin(2019)]{chen2016constant}
Chen, Y.; Lin, B.
\newblock The Constant Inapproximability of the Parameterized Dominating Set
  Problem.
\newblock {\em {SIAM} J. Comput.} {\bf 2019}, {\em 48},~513--533.
\newblock Preliminary version in {FOCS} 2016,
  doi:{\changeurlcolor{black}\href{https://doi.org/10.1137/17M1127211}{\detokenize{10.1137/17M1127211}}}.

\bibitem[Lin(2019)]{Lin19}
Lin, B.
\newblock A Simple Gap-Producing Reduction for the Parameterized Set Cover
  Problem.
\newblock  In Proceedings of the 46th International Colloquium on Automata, Languages, and
  Programming {ICALP}, Patras, Greece, 9--12 July 2019; pp.
  81:1--81:15, doi:10.4230/LIPIcs.ICALP.2019.81.

\bibitem[Kann(1992)]{K92}
Kann, V.
\newblock On the Approximability of NP-complete Optimization Problems.
\newblock Ph.D. Thesis, Royal Institute of Technology, Stockholm, Sweden, 
1992.

\bibitem[Feige(1998)]{feige1998threshold}
Feige, U.
\newblock A threshold of $\ln n$ for approximating set cover.
\newblock {\em J. ACM (JACM)} {\bf 1998}, {\em 45},~634--652.

\bibitem[Lai(2019)]{Lai19}
Lai, W.
\newblock The Inapproximability of k-DominatingSet for Parameterized {AC} 0
  Circuits {\textdagger}.
\newblock {\em Algorithms} {\bf 2019}, {\em 12},~230, doi:10.3390/a12110230.

\bibitem[Bhattacharyya {et~al.}(2019)Bhattacharyya, Bonnet, Egri, Ghoshal,
  {Karthik {C. S.}}, Lin, Manurangsi, and Marx]{even-set}
Bhattacharyya, A.; Bonnet, {\'{E}}.; Egri, L.; Ghoshal, S.; {Karthik {C. S.}}.;
  Lin, B.; Manurangsi, P.; Marx, D.
\newblock Parameterized Intractability of Even Set and Shortest Vector Problem.
\newblock {\em Electron. Colloq. Comput. Complex. {(ECCC)}} {\bf
  2019}, {\em 26},~115.

\bibitem[Downey {et~al.}(1999)Downey, Fellows, Vardy, and Whittle]{DFVW99}
Downey, R.G.; Fellows, M.R.; Vardy, A.; Whittle, G.
\newblock The Parametrized Complexity of Some Fundamental Problems in Coding
  Theory.
\newblock {\em {SIAM} J. Comput.} {\bf 1999}, {\em 29},~545--570, doi:10.1137/S0097539797323571.

\bibitem[van Emde-Boas(1981)]{VEB}
van Emde-Boas, P.
\newblock {\em Another {NP}-Complete Partition Problem and the Complexity of
  Computing Short Vectors in a Lattice}; Report Department of Mathematics;
  University of Amsterdam: Amsterdam, The Netherlands, 1981.

\bibitem[Ajtai(1998)]{Ajt98}
Ajtai, M.
\newblock The Shortest Vector Problem in $\ell_2$ is {NP}-hard for Randomized
  Reductions (Extended Abstract).
\newblock  In Proceedings of the Thirtieth Annual ACM Symposium on Theory of 
Computing, Dallas, TX, USA, 23--26 May 1998; pp. 10--19, 
doi:10.1145/276698.276705.

\bibitem[{Karthik {C. S.}} {et~al.}(2019){Karthik {C. S.}}, Laekhanukit, and
  Manurangsi]{dom-set}
{Karthik {C. S.}}.; Laekhanukit, B.; Manurangsi, P.
\newblock On the parameterized complexity of approximating dominating set.
\newblock {\em J. {ACM}} {\bf 2019}, {\em 66},~33:1--33:38.
\newblock Preliminary version in {STOC} 2018.

\bibitem[Goldreich(2008)]{G09}
Goldreich, O.
\newblock {\em Computational Complexity: A Conceptual Perspective}, 1st ed.;
  Cambridge University Press: New York, NY, USA,  2008.

\bibitem[Chalermsook {et~al.}(2017)Chalermsook, Cygan, Kortsarz,
  Laekhanukit, Manurangsi, Nanongkai, and Trevisan]{param-inapprox}
Chalermsook, P.; Cygan, M.; Kortsarz, G.; Laekhanukit, B.; Manurangsi, P.;
  Nanongkai, D.; Trevisan, L.
\newblock From~Gap-{ETH} to {FPT}-Inapproximability: Clique, Dominating Set,
  and More.
\newblock In Proceedings of the 58th {IEEE} Annual Symposium on Foundations 
of Computer Science (FOCS), Berkeley, CA, USA, 15--17 October 2017; pp. 
743--754.

\bibitem[Abboud {et~al.}(2017)Abboud, Rubinstein, and Williams]{ARW17}
Abboud, A.; Rubinstein, A.; Williams, R.R.
\newblock Distributed {PCP} Theorems for Hardness of Approximation in {P}.
\newblock   In Proceedings of the 58th {IEEE} Annual Symposium on Foundations of Computer Science,
  {FOCS}, Berkeley, CA, USA, 15--17 October 2017; pp. 25--36.

\bibitem[Berman and Schnitger(1992)]{BermanS92}
Berman, P.; Schnitger, G.
\newblock On the Complexity of Approximating the Independent Set Problem.
\newblock {\em Inf. Comput.} {\bf 1992}, {\em 96},~77--94, doi:10.1016/0890-5401(92)90056-L.

\bibitem[Raz(1998)]{Raz98}
Raz, R.
\newblock A Parallel Repetition Theorem.
\newblock {\em {SIAM} J. Comput.} {\bf 1998}, {\em 27},~763--803, doi:10.1137/S0097539795280895.

\bibitem[Dinur(2007)]{Dinur07}
Dinur, I.
\newblock The {PCP} theorem by gap amplification.
\newblock {\em J. {ACM}} {\bf 2007}, {\em 54},~12, doi:10.1145/1236457.1236459.

\bibitem[Wlodarczyk(2019)]{Wlo19}
Wlodarczyk, M.
\newblock Inapproximability within {W[1]:} the case of Steiner Orientation.
\newblock {\em arXiv} {\bf 2019}, arXiv:1907.06529.

\bibitem[Cygan {et~al.}(2013)Cygan, Kortsarz, and Nutov]{CyganKN13}
Cygan, M.; Kortsarz, G.; Nutov, Z.
\newblock Steiner Forest Orientation Problems.
\newblock {\em {SIAM} J. Discret. Math.} {\bf 2013}, {\em 27},~1503--1513, doi:10.1137/120883931.

\bibitem[Pilipczuk and Wahlstr{\"{o}}m(2018)]{PilipczukW18a}
Pilipczuk, M.; Wahlstr{\"{o}}m, M.
\newblock Directed Multicut is W[1]-hard, Even for Four Terminal Pairs.
\newblock {\em {TOCT}} {\bf 2018}, {\em 10},~13:1--13:18, doi:10.1145/3201775.

\bibitem[Lokshtanov {et~al.}(2020)Lokshtanov, Ramanujan, Saurabh, and
  Zehavi]{LRSZ20}
Lokshtanov, D.; Ramanujan, M.S.; Saurabh, S.; Zehavi, M.
\newblock Parameterized Complexity and Approximability of Directed Odd Cycle
  Transversal.
\newblock  In Proceedings of the 2020 {ACM-SIAM} Symposium on Discrete Algorithms,
  {SODA} 2020, Salt Lake City, UT, USA, 5--8 January 2020; pp.
  2181--2200, doi:10.1137/1.9781611975994.134.

\bibitem[Feige {et~al.}(1996)Feige, Goldwasser, Lov{\'{a}}sz, Safra, and
  Szegedy]{FeigeGLSS96}
Feige, U.; Goldwasser, S.; Lov{\'{a}}sz, L.; Safra, S.; Szegedy, M.
\newblock Interactive Proofs and the Hardness of Approximating Cliques.
\newblock {\em J. {ACM}} {\bf 1996}, {\em 43},~268--292, doi:10.1145/226643.226652.

\bibitem[Chitnis {et~al.}(2018)Chitnis, Feldmann, and
  Manurangsi]{chitnis2017parameterized}
Chitnis, R.; Feldmann, A.E.; Manurangsi, P.
\newblock Parameterized Approximation Algorithms for Bidirected Steiner Network
  Problems.
\newblock In Proceedings of the 26th Annual European Symposium on Algorithms
  (ESA), Helsinki, Finland, 20--22 August 2018; pp. 20:1--20:16, 
doi:10.4230/LIPIcs.ESA.2018.20.

\bibitem[Papadimitriou and Yannakakis(1991)]{PapadimitriouY91}
Papadimitriou, C.H.; Yannakakis, M.
\newblock Optimization, Approximation, and Complexity Classes.
\newblock {\em J. Comput. Syst. Sci.} {\bf 1991}, {\em 43},~425--440, doi:10.1016/0022-0000(91)90023-X.

\bibitem[Dinur(2016)]{Dinur16}
Dinur, I.
\newblock Mildly exponential reduction from gap 3SAT to polynomial-gap
  label-cover.
\newblock {\em Electron. Colloq. Comput. Complex. {(ECCC)}} {\bf
  2016}, {\em 23},~128.

\bibitem[Manurangsi and Raghavendra(2017)]{ManurangsiR17}
Manurangsi, P.; Raghavendra, P.
\newblock A Birthday Repetition Theorem and Complexity of Approximating Dense
  CSPs.
\newblock  In Proceedings of the 44th International Colloquium on Automata, Languages, and
  Programming {ICALP}, Warsaw, Poland, 10--14 July 2017; pp.
  78:1--78:15, doi:10.4230/LIPIcs.ICALP.2017.78.

\bibitem[Chen {et~al.}(2004)Chen, Huang, Kanj, and Xia]{ChenHKX04}
Chen, J.; Huang, X.; Kanj, I.A.; Xia, G.
\newblock Linear {FPT} reductions and computational lower bounds.
\newblock In Proceedings of the 36th Annual {ACM} Symposium on Theory of 
Computing (STOC), Chicago, IL, USA, 13--16 June 2004; pp. 212--221, 
doi:10.1145/1007352.1007391.

\bibitem[Chen {et~al.}(2006)Chen, Huang, Kanj, and Xia]{ChenHKX06}
Chen, J.; Huang, X.; Kanj, I.A.; Xia, G.
\newblock Strong computational lower bounds via parameterized complexity.
\newblock {\em J.~Comput. Syst. Sci.} {\bf 2006}, {\em 72},~1346--1367, doi:10.1016/j.jcss.2006.04.007.

\bibitem[Bellare {et~al.}(1998)Bellare, Goldreich, and Sudan]{BellareGS98}
Bellare, M.; Goldreich, O.; Sudan, M.
\newblock Free Bits, PCPs, and Nonapproximability-Towards Tight Results.
\newblock {\em {SIAM} J. Comput.} {\bf 1998}, {\em 27},~804--915, doi:10.1137/S0097539796302531.

\bibitem[Zuckerman(1996)]{Zuckerman96}
Zuckerman, D.
\newblock Simulating {BPP} Using a General Weak Random Source.
\newblock {\em Algorithmica} {\bf 1996}, {\em 16},~367--391, doi:10.1007/BF01940870.

\bibitem[Dinur and Manurangsi(2018)]{DM18}
Dinur, I.; Manurangsi, P.
\newblock {ETH-Hardness of Approximating 2-CSPs and Directed Steiner Network}.
\newblock In Proceedings of the 9th Innovations in Theoretical Computer Science 
Conference (ITCS), Cambridge, MA, {USA}, 11--14 January 2018; pp. 36:1--36:20.

\bibitem[Bellare {et~al.}(1993)Bellare, Goldwasser, Lund, and
  Russeli]{BGLR93}
Bellare, M.; Goldwasser, S.; Lund, C.; Russeli, A.
\newblock Efficient probabilistically checkable proofs and applications to
  approximations.
\newblock  In Proceedings of the Twenty-Fifth Annual {ACM} Symposium on Theory of
  Computing, San Diego, CA, {USA}, 16--18 May 1993; pp. 294--304, doi:10.1145/167088.167174.

\bibitem[Moshkovitz(2015)]{Moshkovitz15}
Moshkovitz, D.
\newblock The Projection Games Conjecture and the NP-Hardness of ln
  n-Approximating Set-Cover.
\newblock {\em Theory Comput.} {\bf 2015}, {\em 11},~221--235, doi:10.4086/toc.2015.v011a007.

\bibitem[Raz and Safra(1997)]{RazS97}
Raz, R.; Safra, S.
\newblock A Sub-Constant Error-Probability Low-Degree Test, and a Sub-Constant
  Error-Probability {PCP} Characterization of {NP}.
\newblock  In Proceedings of the Twenty-Ninth Annual {ACM} Symposium on the Theory
  of Computing, El Paso, TX, USA, 4--6 May 1997; pp. 475--484, doi:10.1145/258533.258641.

\bibitem[Impagliazzo {et~al.}(2012)Impagliazzo, Kabanets, and
  Wigderson]{ImpagliazzoKW12}
Impagliazzo, R.; Kabanets, V.; Wigderson, A.
\newblock New Direct-Product Testers and 2-Query PCPs.
\newblock {\em {SIAM} J. Comput.} {\bf 2012}, {\em 41},~1722--1768, doi:10.1137/09077299X.

\bibitem[Dinur and Navon(2017)]{DinurN17}
Dinur, I.; Navon, I.L.
\newblock Exponentially Small Soundness for the Direct Product Z-Test.
\newblock  In Proceedings of the 32nd Computational Complexity Conference, {CCC}, Riga, Latvia,  6--9 July 2017; pp. 29:1--29:50, doi:10.4230/LIPIcs.CCC.2017.29.

\bibitem[Arora {et~al.}(1993)Arora, Babai, Stern, and Sweedyk]{AroraBSS93}
Arora, S.; Babai, L.; Stern, J.; Sweedyk, Z.
\newblock The Hardness of Approximate Optimia in Lattices, Codes, and Systems
  of Linear Equations.
\newblock In Proceedings of the  34th Annual Symposium on Foundations of Computer Science, Palo Alto,
  CA, USA, 3--5 November 1993; pp. 724--733, doi:10.1109/SFCS.1993.366815.

\bibitem[H{\aa}stad(2001)]{Hastad01}
H{\aa}stad, J.
\newblock Some optimal inapproximability results.
\newblock {\em J. {ACM}} {\bf 2001}, {\em 48},~798--859, doi:10.1145/502090.502098.

\bibitem[Chan(2016)]{Chan16}
Chan, S.O.
\newblock Approximation Resistance from Pairwise-Independent Subgroups.
\newblock {\em J. {ACM}} {\bf 2016}, {\em 63},~27:1--27:32, doi:10.1145/2873054.

\bibitem[Manurangsi()]{M20}
Manurangsi, P.
\newblock Tight Running Time Lower Bounds for Strong Inapproximability of
  Maximum $k$-Coverage, Unique Set Cover and Related Problems (via $t$-Wise
  Agreement Testing Theorem).
\newblock Proceedings of the 2020 {ACM-SIAM} Symposium on Discrete 
Algorithms (SODA), Salt Lake City, UT, USA, 5--8 January 2020; pp. 62--81, 
doi:10.1137/1.9781611975994.5.

\bibitem[H{\aa}stad(1996)]{Hastad96}
H{\aa}stad, J.
\newblock Clique is Hard to Approximate Within $n^{1-\eps}$.
\newblock  In Proceedings of the 37th Annual Symposium on Foundations of Computer Science, {FOCS}
  '96, Burlington, VT, USA, 14--16 October 1996; pp. 627--636, doi:10.1109/SFCS.1996.548522.

\bibitem[Khot(2006)]{Khot06}
Khot, S.
\newblock Ruling Out {PTAS} for Graph Min-Bisection, Dense k-Subgraph, and
  Bipartite Clique.
\newblock {\em {SIAM} J. Comput.} {\bf 2006}, {\em 36},~1025--1071, doi:10.1137/S0097539705447037.

\bibitem[Bhangale {et~al.}(2017)Bhangale, Gandhi, Hajiaghayi, Khandekar, and
  Kortsarz]{BhangaleGHKK17}
Bhangale, A.; Gandhi, R.; Hajiaghayi, M.T.; Khandekar, R.; Kortsarz, G.
\newblock Bi-Covering: Covering Edges with Two Small Subsets of Vertices.
\newblock {\em {SIAM} J. Discret. Math.} {\bf 2017}, {\em 31},~2626--2646, doi:10.1137/16M1082421.

\bibitem[Manurangsi(2018)]{manurangsi2018inapproximability}
Manurangsi, P.
\newblock Inapproximability of maximum biclique problems, minimum k-cut and
  densest at-least-k-subgraph from the small set expansion hypothesis.
\newblock {\em Algorithms} {\bf 2018}, {\em 11},~10.

\bibitem[Manurangsi(2017)]{Manurangsi17}
Manurangsi, P.
\newblock Almost-polynomial ratio ETH-hardness of approximating densest
  k-subgraph.
\newblock  In Proceedings of the 49th Annual {ACM} {SIGACT} Symposium on Theory of
  Computing {STOC}, Montreal, QC, Canada, 19--23 June 2017; pp.
  954--961, doi:10.1145/3055399.3055412.

\bibitem[Raghavendra and Steurer(2010)]{raghavendra2010graph}
Raghavendra, P.; Steurer, D.
\newblock Graph expansion and the unique games conjecture.
\newblock  In Proceedings of the ACM Forty-Second ACM Symposium on Theory of
  Computing, Cambridge, MA, USA, 6--8 June 2010; pp.~755--764.

\bibitem[Alon {et~al.}(2011)Alon, Arora, Manokaran, Moshkovitz, and
  Weinstein]{AAMMW11}
Alon, N.; Arora, S.; Manokaran, R.; Moshkovitz, D.; Weinstein, O.
\newblock Inapproximabilty of Densest $k$-Subgraph from Average Case Hardness.
\newblock Unpublished Manuscript.

\bibitem[K\H{o}v\'{a}ri {et~al.}(1954)K\H{o}v\'{a}ri, S\'{o}s, and
  Tur\'{a}n]{KST54}
K\H{o}v\'{a}ri, T.; S\'{o}s, V.T.; Tur\'{a}n, P.
\newblock {On a problem of K. Zarankiewicz}.
\newblock {\em Colloq. Math.} {\bf 1954}, {\em 3},~50--57.

\bibitem[Zuckerman(2006)]{zuckerman2006linear}
Zuckerman, D.
\newblock Linear degree extractors and the inapproximability of max clique and
  chromatic number.
\newblock  In Proceedings of the ACM Thirty-Eighth Annual ACM Symposium on Theory of
  Computing, Seattle, WA, USA, 21--23 May 2006; pp. 681--690.

\bibitem[Baker(1994)]{baker1994approximation}
Baker, B.S.
\newblock Approximation algorithms for NP-complete problems on planar graphs.
\newblock {\em J. ACM (JACM)} {\bf 1994}, {\em 41},~153--180.

\bibitem[Johnson and Garey(1979)]{johnson1979computers}
Johnson, D.S.; Garey, M.R.
\newblock {\em Computers and Intractability: A Guide to the Theory of
  NP-Completeness}; WH Freeman San Francisco: New York, US, 1979; Volume 1.

%
\bibitem[Demaine and Hajiaghayi(2004)]{demaine2004equivalence}
Demaine, E.D.; Hajiaghayi, M.
\newblock Equivalence of local treewidth and linear local treewidth and its
  algorithmic applications.
\newblock In Proceedings of the Fifteenth Annual ACM-SIAM Symposium on Discrete
  Algorithms, New~Orleans, LA, USA, 11--13 January 2004; pp. 840--849.

\bibitem[Grohe {et~al.}(2013)Grohe, Kawarabayashi, and
  Reed]{grohe2013simple}
Grohe, M.; Kawarabayashi, K.I.; Reed, B.
\newblock A simple algorithm for the graph minor decomposition: Logic meets
  structural graph theory.
\newblock  In Proceedings of the Twenty-Fourth Annual ACM-SIAM Symposium on
  Discrete Algorithms, New~Orleans, LA, USA, 6--8 January 2013; pp. 414--431.

\bibitem[Demaine and Hajiaghayi(2008)]{demaine2008bidimensionality}
Demaine, E.D.; Hajiaghayi, M.
\newblock The bidimensionality theory and its algorithmic applications.
\newblock {\em  Comput. J.} {\bf 2008}, {\em 51},~292--302.

\bibitem[Fomin {et~al.}(2011)Fomin, Lokshtanov, Raman, and
  Saurabh]{fomin2011bidimensionality}
Fomin, F.V.; Lokshtanov, D.; Raman, V.; Saurabh, S.
\newblock Bidimensionality and EPTAS.
\newblock  In Proceedings of the Twenty-Second Annual ACM-SIAM Symposium on
  Discrete Algorithms, San Francisco, CA, USA, 23--25 January 2011;
  pp. 748--759.

\bibitem[Demaine {et~al.}(2011)Demaine, Hajiaghayi, and
  Kawarabayashi]{demaine2011contraction}
Demaine, E.D.; Hajiaghayi, M.; Kawarabayashi, K.i.
\newblock Contraction decomposition in H-minor-free graphs and algorithmic
  applications.
\newblock  In Proceedings of the Forty-Third Annual ACM Symposium on Theory of
  Computing, San Jose, CA, USA, 6--8 June 2011; pp. 441--450.

\bibitem[Bansal {et~al.}(2017)Bansal, Reichman, and Umboh]{bansal2017lp}
Bansal, N.; Reichman, D.; Umboh, S.W.
\newblock LP-based robust algorithms for noisy minor-free and bounded treewidth
  graphs.
\newblock  In Proceedings of the Twenty-Eighth Annual ACM-SIAM Symposium on
  Discrete Algorithms, Barcelona, Spain, 16--19 January 2017; pp. 
1964--1979.

\bibitem[Magen and Moharrami(2009)]{magen2009robust}
Magen, A.; Moharrami, M.
\newblock Robust algorithms for on minor-free graphs based on the Sherali-Adams
  hierarchy. In {\em Approximation, Randomization, and Combinatorial
  Optimization. Algorithms and Techniques}; Springer: Berlin, Germany,  
2009; pp. 258--271.

%
\bibitem[Demaine {et~al.}(2019)Demaine, Goodrich, Kloster, Lavallee, Liu,
  Sullivan, Vakilian, and van~der Poel]{demaine2019structural}
Demaine, E.D.; Goodrich, T.D.; Kloster, K.; Lavallee, B.; Liu, Q.C.; Sullivan,
  B.D.; Vakilian, A.; van~der Poel, A.
\newblock Structural Rounding: Approximation Algorithms for Graphs Near an
  Algorithmically Tractable Class.
\newblock  In Proceedings of the 27th Annual European Symposium on 
Algorithms (ESA), Dagstuhl, Germany, 2019.

\bibitem[Katsikarelis {et~al.}(2018)Katsikarelis, Lampis, and
  Paschos]{KLP18}
Katsikarelis, I.; Lampis, M.; Paschos, V.T.
\newblock Structurally Parameterized d-Scattered Set.
\newblock   In Proceedings of the Graph-Theoretic Concepts in Computer Science---44th International
  Workshop {WG}, Cottbus, Germany, 27--29 June 2018; pp. 292--305, doi:10.1007/978-3-030-00256-5\_24.

\bibitem[Katsikarelis {et~al.}(2019)Katsikarelis, Lampis, and
  Paschos]{KLP19}
Katsikarelis, I.; Lampis, M.; Paschos, V.T.
\newblock Improved (In-)Approximability Bounds for d-Scattered Set.
\newblock  In Proceedings of the Approximation and Online Algorithms---17th International Workshop,
  {WAOA}, Munich, Germany, 12--13 September 2019; Revised Selected Papers; pp. 202--216, doi:10.1007/978-3-030-39479-0\_14.

\bibitem[Marx(2005)]{marx2005efficient}
Marx, D.
\newblock Efficient approximation schemes for geometric problems?
\newblock  In \emph{European Symposium on Algorithms}; Springer: Berlin, 
Germany,  2005; pp. 448--459.

\bibitem[Adamaszek and Wiese(2013)]{adamaszek2013approximation}
Adamaszek, A.; Wiese, A.
\newblock Approximation schemes for maximum weight independent set of
  rectangles.
\newblock  In Proceedings of the 2013 IEEE 54th Annual Symposium on Foundations 
of Computer Science, Berkeley, CA, USA, 27--29 October 2013; pp. 400--409.

\bibitem[Grandoni {et~al.}(2019)Grandoni, Kratsch, and
  Wiese]{grandoni2019parameterized}
Grandoni, F.; Kratsch, S.; Wiese, A.
\newblock Parameterized Approximation Schemes for Independent Set of Rectangles
  and Geometric Knapsack.
\newblock  In Proceedings of the 27th Annual European Symposium on Algorithms 
{(ESA)}, Munich/Garching, Germany, September 9-11, 2019; pp. 53:1--53:16.

\bibitem[Pilipczuk {et~al.}(2017)Pilipczuk, van Leeuwen, and
  Wiese]{pilipczuk2017approximation}
Pilipczuk, M.; van Leeuwen, E.J.; Wiese, A.
\newblock Approximation and Parameterized Algorithms for Geometric Independent
  Set with Shrinking.
\newblock  In Proceedings of the 42nd International Symposium on Mathematical 
Foundations of Computer Science {(MFCS)}, Aalborg, Denmark, 21--25 August 2017; 
42:1--42:13.

\bibitem[Clark {et~al.}(1990)Clark, Colbourn, and Johnson]{ClarkCJ90}
Clark, B.N.; Colbourn, C.J.; Johnson, D.S.
\newblock Unit disk graphs.
\newblock {\em Discret. Math.} {\bf 1990}, {\em 86},~165--177, doi:10.1016/0012-365X(90)90358-O.

\bibitem[III {et~al.}(1998)III, Marathe, Radhakrishnan, Ravi, Rosenkrantz,
  and Stearns]{HuntMRRRS98}
III, H.B.H.; Marathe, M.V.; Radhakrishnan, V.; Ravi, S.S.; Rosenkrantz, D.J.;
  Stearns, R.E.
\newblock NC-Approximation Schemes for {NP-} and PSPACE-Hard Problems for
  Geometric Graphs.
\newblock {\em J. Algorithms} {\bf 1998}, {\em 26},~238--274, doi:10.1006/jagm.1997.0903.

\bibitem[Alber and Fiala(2004)]{AF04}
Alber, J.; Fiala, J.
\newblock Geometric separation and exact solutions for the parameterized
  independent set problem on disk graphs.
\newblock {\em J. Algorithms} {\bf 2004}, {\em 52},~134--151, doi:10.1016/j.jalgor.2003.10.001.

\bibitem[Stockmeyer(1973)]{stockmeyer1973planar}
Stockmeyer, L.
\newblock Planar 3-colorability is NP-complete.
\newblock {\em ACM Sigact News} {\bf 1973}, {\em 5},~19--25.

\bibitem[Demaine {et~al.}(2005)Demaine, Hajiaghayi, and
  Kawarabayashi]{demaine2005algorithmic}
Demaine, E.D.; Hajiaghayi, M.T.; Kawarabayashi, K.i.
\newblock Algorithmic graph minor theory: Decomposition, approximation, and
  coloring.
\newblock   In Proceedings of the IEEE 46th Annual IEEE Symposium on Foundations of Computer Science
  (FOCS'05), Pittsburgh, PA, USA, 23--25 October 2005; pp. 637--646.

%
\bibitem[Belmonte {et~al.}(2018)Belmonte, Lampis, and
  Mitsou]{belmonte2018parameterized}
Belmonte, R.; Lampis, M.; Mitsou, V.
\newblock Parameterized (Approximate) Defective Coloring.
\newblock  In Proceedings of the 35th Symposium on Theoretical Aspects of 
Computer Science {(STACS)}, Caen, France, 28 February--3 March 2018; pp. 
10:1--10:15.

\bibitem[Lampis(2014)]{lampis2014fpas}
Lampis, M.
\newblock Parameterized Approximation Schemes Using Graph Widths.
\newblock  In Proceedings of Automata, Languages, and Programming---41st 
International Colloquium {(ICALP)} Copenhagen, Denmark, 8--11 July 2014;
pp. 775--786.



\bibitem[Fellows {et~al.}(2011)Fellows, Fomin, Lokshtanov, Rosamond,
  Saurabh, Szeider, and Thomassen]{fellows2011complexity}
Fellows, M.R.; Fomin, F.V.; Lokshtanov, D.; Rosamond, F.; Saurabh, S.; Szeider,
  S.; Thomassen, C.
\newblock On the complexity of some colorful problems parameterized by
  treewidth.
\newblock {\em Inf. Comput.} {\bf 2011}, {\em 209},~143--153.

\bibitem[Corneil and Rotics(2005)]{CR05}
Corneil, D.G.; Rotics, U.
\newblock On the Relationship Between Clique-Width and Treewidth.
\newblock {\em {SIAM} J. Comput.} {\bf 2005}, {\em 34},~825--847, doi:10.1137/S0097539701385351.

\bibitem[Katsikarelis {et~al.}(2019)Katsikarelis, Lampis, and
  Paschos]{katsikarelis2019structural}
Katsikarelis, I.; Lampis, M.; Paschos, V.T.
\newblock Structural parameters, tight bounds, and approximation for (k,
  r)-center.
\newblock {\em Discret. Appl. Math.} {\bf 2019}, {\em 264},~90--117.

\bibitem[Salavatipour(2003)]{salavatipour2003sum}
Salavatipour, M.R.
\newblock On sum coloring of graphs.
\newblock {\em Discret. Appl. Math.} {\bf 2003}, {\em 127},~477--488.

\bibitem[Marx(2009)]{marx2009complexity}
Marx, D.
\newblock Complexity results for minimum sum edge coloring.
\newblock {\em Discret. Appl. Math.} {\bf 2009}, {\em 157},~1034--1045.

\bibitem[Giaro and Kubale(2000)]{giaro2000edge}
Giaro, K.; Kubale, M.
\newblock Edge-chromatic sum of trees and bounded cyclicity graphs.
\newblock {\em Inf. Process. Lett.} {\bf 2000}, {\em 75},~65--69.

\bibitem[Marx(2004)]{marx2004minimum}
Marx, D.
\newblock Minimum sum multicoloring on the edges of planar graphs and partial
  k-trees.
\newblock In \emph{International Workshop on Approximation and Online Algorithms};
  Springer: Berlin, Germany,  2004, pp. 9--22.

\bibitem[Cygan(2013)]{cygan2013improved}
Cygan, M.
\newblock Improved approximation for 3-dimensional matching via bounded
  pathwidth local search.
\newblock  In~Proceedings of the 2013 IEEE 54th Annual Symposium on Foundations 
of Computer Science, Berkeley, CA, USA, 27--29 October 2013; pp. 509--518.

\bibitem[Guruswami and Lee(2015)]{guruswami2015inapproximability}
Guruswami, V.; Lee, E.
\newblock Inapproximability of H-Transversal/Packing.
\newblock  In Proceedings of Approximation, Randomization, and 
Combinatorial Optimization;
  Algorithms and Techniques (APPROX/RANDOM), Princeton, NJ, {USA}, August 24-26, 
2015; pp. 284--304.

\bibitem[Lee(2017)]{lee2017partitioning}
Lee, E.
\newblock Partitioning a graph into small pieces with applications to path
  transversal.
\newblock In Proceedings of the Twenty-Eighth Annual ACM-SIAM Symposium on
  Discrete Algorithms, Barcelona, Spain, 16--19 January 2017;
  pp. 1546--1558.

\bibitem[Fomin {et~al.}(2019)Fomin, Le, Lokshtanov, Saurabh, Thomass{\'e},
  and Zehavi]{fomin2019subquadratic}
Fomin, F.V.; Le, T.N.; Lokshtanov, D.; Saurabh, S.; Thomass{\'e}, S.; Zehavi,
  M.
\newblock Subquadratic kernels for implicit 3-hitting set and 3-set packing
  problems.
\newblock {\em ACM Trans. Algorithms (TALG)} {\bf 2019}, {\em
  15},~1--44.

\bibitem[Friggstad and Salavatipour(2007)]{friggstad2007approximability}
Friggstad, Z.; Salavatipour, M.R.
\newblock Approximability of packing disjoint cycles.
\newblock  In \emph{International Symposium on Algorithms and Computation}; 
Springer: Berlin, Germany, 2007; pp. 304--315.

\bibitem[Lokshtanov {et~al.}(2019)Lokshtanov, Mouawad, Saurabh, and
  Zehavi]{lokshtanov2019packing}
Lokshtanov, D.; Mouawad, A.E.; Saurabh, S.; Zehavi, M.
\newblock Packing Cycles Faster Than Erdos--Posa.
\newblock {\em SIAM J. Discret. Math.} {\bf 2019}, {\em
  33},~1194--1215.

\bibitem[Bodlaender {et~al.}(2011)Bodlaender, Thomass{\'e}, and
  Yeo]{bodlaender2011kernel}
Bodlaender, H.L.; Thomass{\'e}, S.; Yeo, A.
\newblock Kernel bounds for disjoint cycles and disjoint paths.
\newblock {\em Theor. Comput. Sci.} {\bf 2011}, {\em 412},~4570--4578.

%
\bibitem[Batra {et~al.}(2015)Batra, Garg, Kumar, M{\"o}mke, and
  Wiese]{batra2015new}
Batra, J.; Garg, N.; Kumar, A.; M{\"o}mke, T.; Wiese, A.
\newblock New approximation schemes for unsplittable flow on a path.
\newblock  In Proceedings of the Twenty-Sixth Annual ACM-SIAM Symposium on
  Discrete Algorithms, San Diego, CA, USA, 4--6 January 2015;
  pp. 47--58.

\bibitem[Wiese(2017)]{wiese2017unsplit}
Wiese, A.
\newblock A $(1 + \epsilon)$-approximation for Unsplittable Flow on a Path in
  fixed-parameter running time.
\newblock In Proceedings of the 44th International Colloquium on Automata, 
Languages, and Programming {(ICALP)}, Warsaw, Poland, 10--14 July 2017; pp. 
67:1--67:13.

\bibitem[Garg {et~al.}(2008)Garg, Kumar, and
  Muralidhara]{garg2008minimizing}
Garg, N.; Kumar, A.; Muralidhara, V.
\newblock Minimizing Total Flow-Time: The Unrelated Case.
\newblock In {\em International Symposium on Algorithms and Computation}; 
Springer: Berlin/Heidelberg, Germany, {2008}; pp. 424--435.

\bibitem[Kellerer {et~al.}(1999)Kellerer, Tautenhahn, and
  Woeginger]{kellerer1999approximability}
Kellerer, H.; Tautenhahn, T.; Woeginger, G.
\newblock Approximability and nonapproximability results for minimizing total
  flow time on a single machine.
\newblock {\em SIAM J. Comput.} {\bf 1999}, {\em 28},~1155--1166.

\bibitem[Wiese(2018)]{wiese2018fixed}
Wiese, A.
\newblock Fixed-Parameter approximation schemes for weighted flowtime.
\newblock  In Proceedings of Approximation, Randomization, and 
Combinatorial Optimization. Algorithms and Techniques (APPROX/RANDOM 2018), 
Princeton, NJ, {USA}, August 20-22, 2018;  pp. 28:1--28:19.

\bibitem[Buss and Goldsmith(1993)]{BussG93}
Buss, J.F.; Goldsmith, J.
\newblock Nondeterminism Within {P}.
\newblock {\em {SIAM} J. Comput.} {\bf 1993}, {\em 22},~560--572, doi:10.1137/0222038.

\bibitem[Nemhauser and Jr.(1975)]{NemhauserT75}
Nemhauser, G.L.; Jr., L.E.T.
\newblock Vertex packings: Structural properties and algorithms.
\newblock {\em Math. Program.} {\bf 1975}, {\em 8},~232--248, doi:10.1007/BF01580444.

\bibitem[Abu{-}Khzam(2010)]{Abu-Khzam10}
Abu{-}Khzam, F.N.
\newblock A kernelization algorithm for d-Hitting Set.
\newblock {\em J. Comput. Syst. Sci.} {\bf 2010}, {\em 76},~524--531, doi:10.1016/j.jcss.2009.09.002.

\bibitem[Cygan(2012)]{cygan2012deterministic}
Cygan, M.
\newblock Deterministic parameterized connected vertex cover.
\newblock  In \emph{Scandinavian Workshop on Algorithm Theory}; Springer: 
Berlin, Germany,  2012; pp.  95--106.

\bibitem[Dom {et~al.}(2014)Dom, Lokshtanov, and Saurabh]{dom}
Dom, M.; Lokshtanov, D.; Saurabh, S.
\newblock Kernelization Lower Bounds Through Colors and {IDs}.
\newblock {\em {ACM} Trans. Algorithms} {\bf 2014}, {\em 11},~1--20, doi:10.1145/2650261.

\bibitem[Krithika {et~al.}(2018)Krithika, Majumdar, and
  Raman]{krithika2018revisiting}
Krithika, R.; Majumdar, D.; Raman, V.
\newblock Revisiting connected vertex cover: FPT algorithms and lossy kernels.
\newblock {\em Theory Comput. Syst.} {\bf 2018}, {\em 62},~1690--1714.

\bibitem[Diptapriyo~Majumdar(2020)]{majumdar2020compress}
Diptapriyo~Majumdar, M. S.~Ramanujan, S.S.
\newblock On the Approximate Compressibility of Connected Vertex Cover.
\newblock \emph{arXiv} \textbf{2019}, arXiv:1905.03379.

\bibitem[Eiben {et~al.}(2018)Eiben, Kumar, Mouawad, Panolan, and
  Siebertz]{eiben2017lossy}
Eiben, E.; Kumar, M.; Mouawad, A.E.; Panolan, F.; Siebertz, S.
\newblock Lossy Kernels for Connected Dominating Set on Sparse Graphs;
\newblock In Proceedings of STACS: Birmingham, England, 2018; Volume 96, pp. 
29:1--29:15.

\bibitem[Angel {et~al.}(2016)Angel, Bampis, Escoffier, and
  Lampis]{angel2016parameterized}
Angel, E.; Bampis, E.; Escoffier, B.; Lampis, M.
\newblock Parameterized power vertex cover.
\newblock  In \emph{International Workshop on Graph-Theoretic Concepts in Computer
  Science}; Springer: Berlin, Germany,  2016; pp. 97--108.

\bibitem[Dom {et~al.}(2008)Dom, Lokshtanov, Saurabh, and
  Villanger]{dom2008capacitated}
Dom, M.; Lokshtanov, D.; Saurabh, S.; Villanger, Y.
\newblock Capacitated domination and covering: A parameterized perspective.
\newblock  In \emph{International Workshop on Parameterized and Exact Computation};
  Springer: Berlin, Germany,  2008; pp. 78--90.

\bibitem[Erd{\H{o}}s and P{\'o}sa(1965)]{erdos1965}
Erd{\H{o}}s, P.; P{\'o}sa, L.
\newblock On Independent Circuits Contained in a Graph.
\newblock {\em Can. J. Math.} {\bf 1965}, {\em
  17},~347–352, doi:10.4153/CJM-1965-035-8.

\bibitem[Raymond and Thilikos(2017)]{raymond2017recent}
Raymond, J.F.; Thilikos, D.M.
\newblock Recent techniques and results on the Erd{\H{o}}s--P{\'o}sa property.
\newblock {\em Discret. Appl. Math.} {\bf 2017}, {\em 231},~25--43.

\bibitem[Kim and Kwon(2018)]{kim2018erdHos}
Kim, E.J.; Kwon, O.j.
\newblock Erd{\H{o}}s-P{\'o}sa property of chordless cycles and its
  applications.
\newblock  In Proceedings of the Twenty-Ninth Annual ACM-SIAM Symposium on
  Discrete Algorithms, New Orleans, LA, USA, 7--10 January 2018; pp. 
1665--1684.

\bibitem[Van~Batenburg {et~al.}(2019)Van~Batenburg, Huynh, Joret, and
  Raymond]{van2019tight}
Van~Batenburg, W.C.; Huynh, T.; Joret, G.; Raymond, J.F.
\newblock A tight Erd{\H{o}}s-P{\'o}sa function for planar minors.
\newblock  In~Proceedings of the Thirtieth Annual ACM-SIAM Symposium on Discrete
  Algorithms, San Diego, CA, USA, 6--9 January 2019; pp. 1485--1500.

\bibitem[Cornuejols {et~al.}(1980)Cornuejols, Nemhauser, and
  Wolsey]{cornuejols1980worst}
Cornuejols, G.; Nemhauser, G.L.; Wolsey, L.A.
\newblock Worst-case and probabilistic analysis of algorithms for a location
  problem.
\newblock {\em Oper. Res.} {\bf 1980}, {\em 28},~847--858.

\bibitem[{Cohen-Addad} {et~al.}(2019){Cohen-Addad}, Gupta, Kumar, Lee, and
  Li]{cohenaddad2019tight}
{Cohen-Addad}, V.; Gupta, A.; Kumar, A.; Lee, E.; Li, J.
\newblock {Tight FPT Approximations for k-Median and k-Means}.
\newblock  In \emph{46th International Colloquium on Automata, Languages, and
  Programming (ICALP 2019)}; Baier, C., Chatzigiannakis, I., Flocchini, P.,
  Leonardi, S., Eds.; Schloss Dagstuhl--Leibniz-Zentrum fuer Informatik:
  Dagstuhl, Germany,  2019; Volume 132, {Leibniz International Proceedings in
  Informatics (LIPIcs)}; pp.~42:1--42:14, doi:10.4230/LIPIcs.ICALP.2019.42.

\bibitem[Badanidiyuru {et~al.}(2012)Badanidiyuru, Kleinberg, and
  Lee]{badanidiyuru2012approximating}
Badanidiyuru, A.; Kleinberg, R.; Lee, H.
\newblock Approximating low-dimensional coverage problems.
\newblock  In Proceedings of the Twenty-Eighth Annual Symposium on Computational
  Geometry, Chapel Hill, NC, USA, 17--20 June 2012; pp. 161--170.

\bibitem[Guo {et~al.}(2007)Guo, Niedermeier, and
  Wernicke]{guo2007parameterized}
Guo, J.; Niedermeier, R.; Wernicke, S.
\newblock Parameterized complexity of vertex cover variants.
\newblock {\em Theory Comput. Syst.} {\bf 2007}, {\em 41},~501--520.

\bibitem[Skowron and Faliszewski(2017)]{skowron2017chamberlin}
Skowron, P.; Faliszewski, P.
\newblock Chamberlin--Courant Rule with Approval Ballots: Approximating the
  MaxCover Problem with Bounded Frequencies in FPT Time.
\newblock {\em J. Artif. Intell. Res.} {\bf 2017}, {\em
  60},~687--716.

\bibitem[Manurangsi(2018)]{manurangsi2018note}
Manurangsi, P.
\newblock A Note on Max k-Vertex Cover: Faster FPT-AS, Smaller Approximate
  Kernel and Improved Approximation.
\newblock In \emph{2nd Symposium on Simplicity in Algorithms (SOSA 2019)}; Schloss
  Dagstuhl-Leibniz-Zentrum fuer Informatik:  2018.

\bibitem[Petrank(1994)]{petrank1994hardness}
Petrank, E.
\newblock The hardness of approximation: Gap location.
\newblock {\em Comput. Complex.} {\bf 1994}, {\em 4},~133--157.

\bibitem[Chlamt{\'{a}}c {et~al.}(2018)Chlamt{\'{a}}c, Dinitz, Konrad,
  Kortsarz, and Rabanca]{ChlamtacDKKR18}
Chlamt{\'{a}}c, E.; Dinitz, M.; Konrad, C.; Kortsarz, G.; Rabanca, G.
\newblock The Densest k-Subhypergraph Problem.
\newblock {\em {SIAM} J. Discret. Math.} {\bf 2018}, {\em 32},~1458--1477, doi:10.1137/16M1096402.

\bibitem[Chlamt{\'{a}}c {et~al.}(2017)Chlamt{\'{a}}c, Dinitz, and
  Makarychev]{ChlamtacDM17}
Chlamt{\'{a}}c, E.; Dinitz, M.; Makarychev, Y.
\newblock Minimizing the Union: Tight Approximations for Small Set Bipartite
  Vertex Expansion.
\newblock  In Proceedings of the Twenty-Eighth Annual {ACM-SIAM} Symposium on
  Discrete Algorithms SODA, Barcelona, Spain, 16--19 January 2017; pp. 881--899, doi:10.1137/1.9781611974782.56.

\bibitem[Gupta {et~al.}(2018{\natexlab{a}})Gupta, Lee, and Li]{gupta2018fpt}
Gupta, A.; Lee, E.; Li, J.
\newblock An FPT algorithm beating 2-approximation for k-cut.
\newblock  In \emph{Twenty-Ninth Annual ACM-SIAM Symposium on
  Discrete Algorithms}, New Orleans, LA, USA, 7--10 January 2018;
  pp.~2821--2837.

\bibitem[Gupta {et~al.}(2018{\natexlab{b}})Gupta, Lee, and
  Li]{gupta2018faster}
Gupta, A.; Lee, E.; Li, J.
\newblock Faster exact and approximate algorithms for k-cut.
\newblock  In Proceedings of the 2018 IEEE 59th Annual Symposium on Foundations of Computer Science
  (FOCS), Philadelphia, PA, USA, 18--21 October 2018; pp. 113--123.

\bibitem[Byrka {et~al.}(2014)Byrka, Pensyl, Rybicki, Srinivasan, and
  Trinh]{byrka2014improved}
Byrka, J.; Pensyl, T.; Rybicki, B.; Srinivasan, A.; Trinh, K.
\newblock An improved approximation for k-median, and positive correlation in
  budgeted optimization.
\newblock  In Proceedings of the Twenty-Sixth Annual ACM-SIAM Symposium on
  Discrete Algorithms, Budapest, Hungary, 26--29 August 2014; pp. 737--756.

\bibitem[Kanungo {et~al.}(2004)Kanungo, Mount, Netanyahu, Piatko, Silverman,
  and Wu]{kanungo2004local}
Kanungo, T.; Mount, D.M.; Netanyahu, N.S.; Piatko, C.D.; Silverman, R.; Wu,
  A.Y.
\newblock A local search approximation algorithm for k-means clustering.
\newblock {\em Comput. Geom.} {\bf 2004}, {\em 28},~89--112.

\bibitem[Gonzalez(1985)]{gonzalez1985clustering}
Gonzalez, T.F.
\newblock Clustering to minimize the maximum intercluster distance.
\newblock {\em Theor. Comput. Sci.} {\bf 1985}, {\em 38},~293--306.

\bibitem[Guha and Khuller(1999)]{guha1999greedy}
Guha, S.; Khuller, S.
\newblock Greedy strikes back: Improved facility location algorithms.
\newblock {\em J. Algorithms} {\bf 1999}, {\em 31},~228--248.

\bibitem[Chen(2006)]{chen2006k}
Chen, K.
\newblock On k-median clustering in high dimensions.
\newblock  In \emph{Seventeenth Annual ACM-SIAM Symposium on Discrete
  Algorithm}, Miami, Florida, USA, 22--26 January 2006; pp.
  1177--1185.

\bibitem[Feldman and Langberg(2011)]{feldman2011unified}
Feldman, D.; Langberg, M.
\newblock A unified framework for approximating and clustering data.
\newblock  In Proceedings of the Forty-Third Annual ACM Symposium on Theory of
  Computing, San Jose, CA, USA, 6--8 June 2011; pp.~569--578.

\bibitem[Calinescu {et~al.}(2011)Calinescu, Chekuri, P{\'a}l, and
  Vondr{\'a}k]{calinescu2011maximizing}
Calinescu, G.; Chekuri, C.; P{\'a}l, M.; Vondr{\'a}k, J.
\newblock Maximizing a monotone submodular function subject to a matroid
  constraint.
\newblock {\em SIAM J. Comput.} {\bf 2011}, {\em 40},~1740--1766.

\bibitem[Haussler(1992)]{haussler1992decision}
Haussler, D.
\newblock Decision theoretic generalizations of the PAC model for neural net
  and other learning applications.
\newblock {\em Inf. Comput.} {\bf 1992}, {\em 100},~78--150.

\bibitem[Lee {et~al.}(2017)Lee, Schmidt, and
  Wright]{lee2017improved-k-means}
Lee, E.; Schmidt, M.; Wright, J.
\newblock Improved and simplified inapproximability for k-means.
\newblock {\em Inf. Process. Lett.} {\bf 2017}, {\em 120},~40--43.

\bibitem[{Cohen-Addad} and {Karthik {C. S.}}(2019)]{cohen2019inapproximability}
{Cohen-Addad}, V.; {Karthik {C. S.}}.
\newblock Inapproximability of Clustering in $L_p$-metrics.
\newblock   In Proceedings of the 2019 IEEE 60th Annual Symposium on Foundations 
of Computer Science, Baltimore, Maryland, 9--12 November 2019.

\bibitem[Arora {et~al.}(1998)Arora, Raghavan, and
  Rao]{arora1998approximation}
Arora, S.; Raghavan, P.; Rao, S.
\newblock Approximation Schemes for Euclidean k-Medians and Related Problems.
\newblock In Proceedings of Proceedings of the Thirtieth Annual {ACM} Symposium 
on the Theory of Computing, Dallas, Texas, USA, 23--26 May 1998; Volume 98, pp. 
106--113.

\bibitem[Arora(1998)]{arora1998polynomial}
Arora, S.
\newblock Polynomial time approximation schemes for Euclidean traveling
  salesman and other geometric problems.
\newblock {\em J. ACM (JACM)} {\bf 1998}, {\em 45},~753--782.

\bibitem[Kolliopoulos and Rao(1999)]{kolliopoulos1999nearly}
Kolliopoulos, S.G.; Rao, S.
\newblock A nearly linear-time approximation scheme for the Euclidean k-median
  problem.
\newblock  European Symposium on Algorithms. Springer: Berlin, Germany,  
1999, pp. 378--389.

\bibitem[Matou{\v{s}}ek(2000)]{matouvsek2000approximate}
Matou{\v{s}}ek, J.
\newblock On approximate geometric k-clustering.
\newblock {\em Discret. Comput. Geom.} {\bf 2000}, {\em
  24},~61--84.

\bibitem[B{\=a}doiu {et~al.}(2002)B{\=a}doiu, Har-Peled, and
  Indyk]{badoiu2002approximate}
B{\=a}doiu, M.; Har-Peled, S.; Indyk, P.
\newblock Approximate clustering via core-sets.
\newblock  In Proceedings of the thiry-fourth annual ACM symposium on Theory of
  computing. ACM,  2002, pp. 250--257.

\bibitem[De~La~Vega {et~al.}(2003)De~La~Vega, Karpinski, Kenyon, and
  Rabani]{de2003approximation}
De~La~Vega, W.F.; Karpinski, M.; Kenyon, C.; Rabani, Y.
\newblock Approximation schemes for clustering problems.
\newblock  In Proceedings of the Thirty-Fifth Annual ACM Symposium on Theory of
  Computing, San Diego, CA, USA, 9--11 June 2003; pp. 50--58.

\bibitem[Har-Peled and Mazumdar(2004)]{har2004coresets}
Har-Peled, S.; Mazumdar, S.
\newblock On coresets for k-means and k-median clustering.
\newblock  In Proceedings of the Thirty-Sixth Annual ACM Symposium on Theory of
  Computing, Chicago, IL, USA, 13--15 June 2004; pp. 291--300.

\bibitem[Kumar {et~al.}(2004)Kumar, Sabharwal, and Sen]{kumar2004simple}
Kumar, A.; Sabharwal, Y.; Sen, S.
\newblock A simple linear time $(1+\varepsilon)$-approximation algorithm for
  $k$-means clustering in any dimensions.
\newblock   In Proceedings of the 45th Annual IEEE Symposium on Foundations of 
Computer Science, Rome, Italy, 17--19 October 2004;  pp. 454--462.

\bibitem[Kumar {et~al.}(2005)Kumar, Sabharwal, and Sen]{kumar2005linear}
Kumar, A.; Sabharwal, Y.; Sen, S.
\newblock Linear time algorithms for clustering problems in any dimensions.
\newblock  In~\emph{International Colloquium on Automata, Languages, and Programming};
  Springer: Berlin, Germany,  2005; pp.~1374--1385.

\bibitem[Feldman {et~al.}(2007)Feldman, Monemizadeh, and
  Sohler]{feldman2007ptas}
Feldman, D.; Monemizadeh, M.; Sohler, C.
\newblock A PTAS for k-means clustering based on weak coresets.
\newblock  In~Proceedings of the Twenty-Third Annual Symposium on Computational
  Geometry, Gyeongju, South Korea, 6--8 June 2007; pp. 11--18.

\bibitem[Sohler and Woodruff(2018)]{sohler2018strong}
Sohler, C.; Woodruff, D.P.
\newblock Strong coresets for k-median and subspace approximation: Goodbye
  dimension.
\newblock  In~Proceedings of the 2018 IEEE 59th Annual Symposium on Foundations of Computer Science
  (FOCS), Paris, France, 7--9 October 2018; pp. 802--813.

\bibitem[Becchetti {et~al.}(2019)Becchetti, Bury, {Cohen-Addad}, Grandoni,
  and Schwiegelshohn]{becchetti2019oblivious}
Becchetti, L.; Bury, M.; {Cohen-Addad}, V.; Grandoni, F.; Schwiegelshohn, C.
\newblock Oblivious dimension reduction for k-means: Beyond subspaces and the
  Johnson-Lindenstrauss lemma.
\newblock  In Proceedings of the 51st Annual ACM SIGACT Symposium on Theory of
  Computing, Arizona Sunday, 23--26 June  2019; pp. 1039--1050.

\bibitem[Huang and Vishnoi(2020)]{huang2020coresets}
Huang, L.; Vishnoi, N.K.
\newblock Coresets for Clustering in Euclidean Spaces: Importance Sampling is
  Nearly Optimal.
\newblock {\em arXiv} {\bf 2020}, arXiv:2004.06263.

\bibitem[Braverman {et~al.}(2020)Braverman, Jiang, Krauthgamer, and
  Wu]{braverman2020coresets}
Braverman, V.; Jiang, S.H.C.; Krauthgamer, R.; Wu, X.
\newblock Coresets for Clustering in Excluded-minor Graphs and Beyond.
\newblock {\em arXiv} {\bf 2020}, arXiv:2004.07718.

\bibitem[{Cohen-Addad} {et~al.}(2019){Cohen-Addad}, Klein, and
  Mathieu]{cohen2019local}
{Cohen-Addad}, V.; Klein, P.N.; Mathieu, C.
\newblock Local search yields approximation schemes for k-means and k-median in
  euclidean and minor-free metrics.
\newblock {\em SIAM J. Comput.} {\bf 2019}, {\em 48},~644--667.

\bibitem[Friggstad {et~al.}(2019)Friggstad, Rezapour, and
  Salavatipour]{friggstad2019local}
Friggstad, Z.; Rezapour, M.; Salavatipour, M.R.
\newblock Local search yields a PTAS for k-means in doubling metrics.
\newblock {\em SIAM J. Comput.} {\bf 2019}, {\em 48},~452--480.

\bibitem[{Cohen-Addad}(2018)]{cohen2018fast}
{Cohen-Addad}, V.
\newblock A fast approximation scheme for low-dimensional k-means.
\newblock  In \emph{Twenty-Ninth Annual ACM-SIAM Symposium on
  Discrete Algorithms}; Society for Industrial and Applied Mathematics:  2018;
  pp.~430--440.

\bibitem[{Cohen-Addad} {et~al.}(2019){Cohen-Addad}, Feldmann, and
  Saulpic]{cohen2018near}
{Cohen-Addad}, V.; Feldmann, A.E.; Saulpic, D.
\newblock Near-Linear Time Approximation Schemes for Clustering in Doubling
  Metrics. \emph{arXiv} {\bf 2019}, arXiv:1812.08664.

\bibitem[Feldmann and Marx(2018)]{DBLP:conf/swat/FeldmannM18}
Feldmann, A.E.; Marx, D.
\newblock The Parameterized Hardness of the k-Center Problem in Transportation
  Networks.
\newblock  In Proceedings of the 16th Scandinavian Symposium and Workshops on
  Algorithm Theory (SWAT), Malmö, Sweden, 18--20 June 2018; pp. 
19:1--19:13, doi:10.4230/LIPIcs.SWAT.2018.19.

\bibitem[Fox-Epstein {et~al.}(2019)Fox-Epstein, Klein, and
  Schild]{planar-EPTAS}
Fox-Epstein, E.; Klein, P.N.; Schild, A.
\newblock Embedding Planar Graphs into Low-Treewidth Graphs with Applications
  to Efficient Approximation Schemes for Metric Problems.
\newblock In Proceedings of the Thirtieth Annual {ACM-SIAM} Symposium on 
Discrete Algorithms {(SODA)}, San Diego, California, USA, 6--9 January 2019; pp. 1069--1088.

\bibitem[Becker {et~al.}(2018)Becker, Klein, and
  Saulpic]{becker2018polynomial}
Becker, A.; Klein, P.N.; Saulpic, D.
\newblock Polynomial-time approximation schemes for k-center, k-median, and
  capacitated vehicle routing in bounded highway dimension.
\newblock  In Proceedings of the 26th Annual European Symposium on 
Algorithms (ESA), Helsinki, Finland, 20--22 August 2018; pp. 8:1--8:15.

\bibitem[Feldmann(2015)]{DBLP:conf/icalp/Feldmann15}
Feldmann, A.E.
\newblock Fixed Parameter Approximations for k-Center Problems in Low Highway
  Dimension Graphs.
\newblock  In 4\emph{2nd International Colloquium on Automata,
  Languages, and Programming (ICALP)}; Springer: Berlin/Heidelberg, Germany,  2015; pp.
  588--600, doi:10.1007/978-3-662-47666-6\_47.

\bibitem[Li(2016)]{li2016approximating}
Li, S.
\newblock Approximating capacitated $k$-median with $(1+ \varepsilon) k$ open
  facilities.
\newblock  In Proceedings of the Twenty-Seventh Annual ACM-SIAM Symposium 
on Discrete Algorithms, Arlington, VA, USA, 10--12 January 2016;
  pp. 786--796.

\bibitem[Demirci and Li(2016)]{demirci2016constant}
Demirci, G.; Li, S.
\newblock Constant Approximation for Capacitated $k$-Median with
  $(1+\varepsilon)$-Capacity Violation.
\newblock {\em arXiv} {\bf 2016}, arXiv:1603.02324.

\bibitem[Adamczyk {et~al.}(2018)Adamczyk, Byrka, Marcinkowski, Meesum, and
  W{\l}odarczyk]{adamczyk2018constant}
Adamczyk, M.; Byrka, J.; Marcinkowski, J.; Meesum, S.M.; W{\l}odarczyk, M.
\newblock Constant factor FPT approximation for capacitated k-median.
\newblock {\em arXiv} {\bf 2018}, arXiv:1809.05791.

\bibitem[{Cohen-Addad} and Li(2019)]{cohen2019fixed}
{Cohen-Addad}, V.; Li, J.
\newblock On the Fixed-Parameter Tractability of Capacitated Clustering.
\newblock  In Proceedings of the 46th International Colloquium on 
Automata, Languages, and
  Programming (ICALP), Patras, Greece, 9--12 July 2019; pp. 41:1--41:14.

\bibitem[Xu {et~al.}(2019)Xu, Zhang, and Zou]{xu2019constant}
Xu, Y.; Zhang, Y.; Zou, Y.
\newblock A constant parameterized approximation for hard-capacitated k-means.
\newblock {\em arXiv} {\bf 2019}, arXiv:1901.04628.

\bibitem[Krishnaswamy {et~al.}(2018)Krishnaswamy, Li, and
  Sandeep]{krishnaswamy2018constant}
Krishnaswamy, R.; Li, S.; Sandeep, S.
\newblock Constant approximation for k-median and k-means with outliers via
  iterative rounding.
\newblock  In Proceedings of the 50th Annual ACM SIGACT Symposium on Theory of
  Computing, Los Angeles, CA, USA, 25--29 June 2018; pp. 646--659.

\bibitem[Swamy(2016)]{swamy2016improved}
Swamy, C.
\newblock Improved approximation algorithms for matroid and knapsack median
  problems and applications.
\newblock {\em ACM Trans. Algorithms (TALG)} {\bf 2016}, {\em 12},~49.

\bibitem[Dreyfus and Wagner(1971)]{DBLP:journals/networks/DreyfusW71}
Dreyfus, S.E.; Wagner, R.A.
\newblock The {S}teiner problem in graphs.
\newblock {\em Networks} {\bf 1971}, {\em 1},~195--207.

\bibitem[Fuchs {et~al.}(2007)Fuchs, Kern, Molle, Richter, Rossmanith, and
  Wang]{fuchs2007dynamic}
Fuchs, B.; Kern, W.; Molle, D.; Richter, S.; Rossmanith, P.; Wang, X.
\newblock Dynamic programming for minimum Steiner trees.
\newblock {\em Theory Comput. Syst.} {\bf 2007}, {\em 41},~493--500.

\bibitem[Nederlof(2009)]{DBLP:conf/icalp/Nederlof09}
Nederlof, J.
\newblock Fast Polynomial-Space Algorithms Using M{\"{o}}bius Inversion:
  Improving on Steiner Tree and Related Problems.
\newblock In Proceedings of the Automata, Languages and Programming, 36th International Colloquium,
  {ICALP}, Rhodes, Greece, 5--12 July 2009; pp. 713--725.

\bibitem[Borchers and Du(1997)]{borchers-du}
Borchers, A.; Du, D.Z.
\newblock {The $k$-{S}teiner Ratio in Graphs}.
\newblock {\em SIAM J. Comput.} {\bf 1997}, {\em 26},~857--869.

\bibitem[Byrka {et~al.}(2013)Byrka, Grandoni, Rothvoss, and
  Sanit{\`{a}}]{DBLP:journals/jacm/ByrkaGRS13}
Byrka, J.; Grandoni, F.; Rothvoss, T.; Sanit{\`{a}}, L.
\newblock Steiner Tree Approximation via Iterative Randomized Rounding.
\newblock {\em J. {ACM}} {\bf 2013}, {\em 60},~1--33.

\bibitem[Chlebík and Chlebíková(2008)]{chlebik2008steiner}
Chlebík, M.; Chlebíková, J.
\newblock The Steiner tree problem on graphs: Inapproximability results.
\newblock {\em Theor. Comput. Sci.} {\bf 2008}, {\em 406},~207--214.

\bibitem[Dvořák {et~al.}(2018)Dvořák, Feldmann, Knop, Masařík, Toufar,
  and Veselý]{st-pas}
Dvořák, P.; Feldmann, A.E.; Knop, D.; Masařík, T.; Toufar, T.; Veselý, P.
\newblock Parameterized Approximation Schemes for Steiner Trees with Small
  Number of Steiner Vertices.
\newblock  In Proceedings of the 35th Symposium on Theoretical Aspects of Computer
  Science (STACS), Caen, France, 28 February--3 March 2018; pp. 
26:1--26:15, doi:10.4230/LIPIcs.STACS.2018.26.

\bibitem[Babay {et~al.}(2018)Babay, Dinitz, and
  Zhang]{babay2018characterizing}
Babay, A.; Dinitz, M.; Zhang, Z.
\newblock Characterizing Demand Graphs for (Fixed-Parameter) Shallow-Light
  Steiner Network.
\newblock  In Proceedings of the 38th IARCS Annual Conference on 
Foundations of Software Technology
  and Theoretical Computer Science (FSTTCS), Ahmedabad, India, 11--13 December 2018; pp. 33:1--33:22.

\bibitem[Hassin(1992)]{hassin1992approximation}
Hassin, R.
\newblock Approximation schemes for the restricted shortest path problem.
\newblock {\em Math. Oper. Res.} {\bf 1992}, {\em
  17},~36--42.

\bibitem[Bockenhauer {et~al.}(2007)Bockenhauer, Hromkovic, Kneis, and
  Kupke]{bockenhauer2007parameterized}
Bockenhauer, H.J.; Hromkovic, J.; Kneis, J.; Kupke, J.
\newblock The parameterized approximability of TSP with deadlines.
\newblock {\em Theory Comput. Syst.} {\bf 2007}, {\em 41},~431--444.

\bibitem[Papadimitriou(1977)]{papadimitriou1977euclidean}
Papadimitriou, C.H.
\newblock The Euclidean travelling salesman problem is NP-complete.
\newblock {\em Theor. Comput. Sci.} {\bf 1977}, {\em 4},~237--244.

\bibitem[Garey {et~al.}(1977)Garey, Graham, and
  Johnson]{garey1977complexity}
Garey, M.R.; Graham, R.L.; Johnson, D.S.
\newblock The complexity of computing Steiner minimal trees.
\newblock {\em SIAM J. Appl. Math.} {\bf 1977}, {\em
  32},~835--859.

\bibitem[Karpinski {et~al.}(2015)Karpinski, Lampis, and
  Schmied]{karpinski2015}
Karpinski, M.; Lampis, M.; Schmied, R.
\newblock New inapproximability bounds for {TSP}.
\newblock {\em JCSS} {\bf 2015}, {\em 81},~1665--1677, doi:10.1016/j.jcss.2015.06.003.

\bibitem[Gottlieb(2015)]{Gottlieb15}
Gottlieb, L.
\newblock A Light Metric Spanner.
\newblock In Proceedings of the 56th Annual Symposium on Foundations of 
Computer Science, {FOCS}, Berkeley, CA, USA, 17--20 October 2015; pp. 759--772.

\bibitem[Talwar(2004)]{talwar2004bypassing}
Talwar, K.
\newblock Bypassing the embedding: Algorithms for low dimensional metrics.
\newblock In Proceedings of the 36th Annual {ACM} Symposium on Theory of 
Computing, Chicago, IL, USA, 13--16 June 2004; pp. 281--290.

\bibitem[Feldmann {et~al.}(2018)Feldmann, Fung, K{\"o}nemann, and
  Post]{FeldmannFKP-highway-2015}
Feldmann, A.E.; Fung, W.S.; K{\"o}nemann, J.; Post, I.
\newblock A (1+$\varepsilon$)-Embedding of Low Highway Dimension Graphs into
  Bounded Treewidth Graphs.
\newblock {\em SIAM J. Comput.} {\bf 2018}, {\em 47},~1667--1704, doi:10.1137/16M1067196.

\bibitem[Guo {et~al.}(2011)Guo, Niedermeier, and
  Such{\'{y}}]{DBLP:journals/siamdm/GuoNS11}
Guo, J.; Niedermeier, R.; Such{\'{y}}, O.
\newblock Parameterized Complexity of Arc-Weighted Directed {S}teiner Problems.
\newblock {\em {SIAM} J. Discret. Math.} {\bf 2011}, {\em 25},~583--599.

%
\bibitem[Halperin and Krauthgamer(2003)]{approx-hardness}
Halperin, E.; Krauthgamer, R.
\newblock Polylogarithmic inapproximability.
\newblock In Proceedings of the 35th Annual {ACM} Symposium on Theory of 
Computing, 9--11 June 2003, San Diego, CA, {USA}; pp. 585--594.

\bibitem[Chitnis {et~al.}(2013)Chitnis, Hajiaghayi, and
  Kortsarz]{DBLP:conf/iwpec/ChitnisHK13}
Chitnis, R.; Hajiaghayi, M.; Kortsarz, G.
\newblock Fixed-Parameter and Approximation Algorithms: A New Look.
\newblock In Proceedings of Parameterized and Exact Computation---8th 
International Symposium, {IPEC}, Sophia Antipolis, France, 4--6 September 2013; pp. 110--122.

\bibitem[Leighton and Rao(1999)]{leighton1999multicommodity}
Leighton, T.; Rao, S.
\newblock Multicommodity max-flow min-cut theorems and their use in designing
  approximation algorithms.
\newblock {\em J. ACM (JACM)} {\bf 1999}, {\em 46},~787--832.

\bibitem[Arora {et~al.}(2009)Arora, Rao, and Vazirani]{arora2009expander}
Arora, S.; Rao, S.; Vazirani, U.
\newblock Expander flows, geometric embeddings and graph partitioning.
\newblock {\em J. ACM (JACM)} {\bf 2009}, {\em 56},~5.

\bibitem[Marx(2006)]{marx2006parameterized}
Marx, D.
\newblock Parameterized graph separation problems.
\newblock {\em Theor. Comput. Sci.} {\bf 2006}, {\em 351},~394--406.

\bibitem[Marx and Razgon(2014)]{marx2014fixed}
Marx, D.; Razgon, I.
\newblock Fixed-parameter tractability of multicut parameterized by the size of
  the cutset.
\newblock {\em SIAM J. Comput.} {\bf 2014}, {\em 43},~355--388.

\bibitem[Chitnis {et~al.}(2016)Chitnis, Cygan, Hajiaghayi, Pilipczuk, and
  Pilipczuk]{chitnis2016designing}
Chitnis, R.; Cygan, M.; Hajiaghayi, M.; Pilipczuk, M.; Pilipczuk, M.
\newblock Designing FPT algorithms for cut problems using randomized
  contractions.
\newblock {\em SIAM J. Comput.} {\bf 2016}, {\em 45},~1171--1229.

\bibitem[Cygan {et~al.}(2019)Cygan, Lokshtanov, Pilipczuk, Pilipczuk, and
  Saurabh]{cygan2019minimum}
Cygan, M.; Lokshtanov, D.; Pilipczuk, M.; Pilipczuk, M.; Saurabh, S.
\newblock Minimum Bisection is fixed-parameter tractable.
\newblock {\em SIAM J. Comput.} {\bf 2019}, {\em 48},~417--450.

\bibitem[Garg {et~al.}(1996)Garg, Vazirani, and
  Yannakakis]{garg1996approximate}
Garg, N.; Vazirani, V.V.; Yannakakis, M.
\newblock Approximate max-flow min-(multi) cut theorems and their applications.
\newblock {\em SIAM J. Comput.} {\bf 1996}, {\em 25},~235--251.

\bibitem[Chawla {et~al.}(2006)Chawla, Krauthgamer, Kumar, Rabani, and
  Sivakumar]{chawla2006hardness}
Chawla, S.; Krauthgamer, R.; Kumar, R.; Rabani, Y.; Sivakumar, D.
\newblock On the hardness of approximating multicut and sparsest-cut.
\newblock {\em Comput. Complex.} {\bf 2006}, {\em 15},~94--114.

\bibitem[Sharma and Vondr{\'a}k(2013)]{sharma2013multiway}
Sharma, A.; Vondr{\'a}k, J.
\newblock Multiway cut, pairwise realizable distributions, and descending
  thresholds.
\newblock {\em arXiv} {\bf 2013}, arXiv:1309.2729.

\bibitem[B{\'e}rczi {et~al.}(2019)B{\'e}rczi, Chandrasekaran, Kir{\'a}ly,
  and Madan]{berczi2019improving}
B{\'e}rczi, K.; Chandrasekaran, K.; Kir{\'a}ly, T.; Madan, V.
\newblock Improving the Integrality Gap for Multiway Cut.
\newblock  In~\emph{International Conference on Integer Programming and Combinatorial
  Optimization}; Springer: Berlin, Germany,  2019; pp. 115--127.

\bibitem[{Cohen-Addad} {et~al.}(2018){Cohen-Addad}, De~Verdi{\`e}re, and
  De~Mesmay]{cohen2018near-cut}
{Cohen-Addad}, V.; De~Verdi{\`e}re, {\'E}.C.; De~Mesmay, A.
\newblock A near-linear approximation scheme for multicuts of embedded graphs
  with a fixed number of terminals.
\newblock  In Proceedings of the Twenty-Ninth Annual ACM-SIAM Symposium on
  Discrete Algorithms, New Orleans, LA, USA, 7--10 January 2018; pp. 
1439--1458.

\bibitem[Chekuri and Madan(2017)]{chekuri2017approximating}
Chekuri, C.; Madan, V.
\newblock Approximating multicut and the demand graph.
\newblock  In Proceedings of the Twenty-Eighth Annual ACM-SIAM Symposium on
  Discrete Algorithms, Barcelona, Spain, 16--19 January 2017; pp. 855--874

\bibitem[Agarwal {et~al.}(2007)Agarwal, Alon, and
  Charikar]{agarwal2007improved}
Agarwal, A.; Alon, N.; Charikar, M.S.
\newblock Improved approximation for directed cut problems.
\newblock  In Proceedings of the Thirty-Ninth Annual ACM Symposium on Theory of
  Computing, San Diego, CA, USA, 11--13 June 2007; pp. 671--680.

\bibitem[Lee(2017)]{lee2017improved}
Lee, E.
\newblock Improved Hardness for Cut, Interdiction, and Firefighter Problems.
\newblock  In \emph{44th International Colloquium on Automata, Languages, and
  Programming (ICALP 2017)}; Schloss Dagstuhl-Leibniz-Zentrum fuer Informatik:
  2017.

\bibitem[Chuzhoy and Khanna(2009)]{chuzhoy2009polynomial}
Chuzhoy, J.; Khanna, S.
\newblock Polynomial flow-cut gaps and hardness of directed cut problems.
\newblock {\em J. ACM (JACM)} {\bf 2009}, {\em 56},~6.

\bibitem[Naor and Zosin(1997)]{naor19972}
Naor, J.; Zosin, L.
\newblock A 2-approximation algorithm for the directed multiway cut problem.
\newblock  In Proceedings 38th Annual Symposium on Foundations of Computer
  Science, Miami Beach, FL, USA, 19--22 October 1997; pp.~548--553.

\bibitem[Chitnis and Feldmann(2019)]{chitnis2019inapproximability}
Chitnis, R.; Feldmann, A.E.
\newblock FPT Inapproximability of Directed Cut and Connectivity Problems.
\newblock  IPEC,  2019.

\bibitem[Feige {et~al.}(2008)Feige, Hajiaghayi, and Lee]{feige2008improved}
Feige, U.; Hajiaghayi, M.; Lee, J.R.
\newblock Improved approximation algorithms for minimum weight vertex
  separators.
\newblock {\em SIAM J. Comput.} {\bf 2008}, {\em 38},~629--657.

\bibitem[R{\"a}cke(2008)]{racke2008optimal}
R{\"a}cke, H.
\newblock Optimal hierarchical decompositions for congestion minimization in
  networks.
\newblock  In Proceedings of the Fortieth Annual ACM Symposium on Theory of
  Computing, Budapest, Hungary, 26--29 August 2008; pp. 255--264.

\bibitem[Feige and Mahdian(2006)]{feige2006finding}
Feige, U.; Mahdian, M.
\newblock Finding small balanced separators.
\newblock  In Proceedings of the Thirty-Eighth Annual ACM Symposium on Theory of
  Computing, Seattle, WA, USA, 21--23 May 2006; pp.~375--384.

\bibitem[Karger and Stein(1996)]{karger1996new}
Karger, D.R.; Stein, C.
\newblock A new approach to the minimum cut problem.
\newblock {\em J. ACM (JACM)} {\bf 1996}, {\em 43},~601--640.

\bibitem[Thorup(2008)]{thorup2008minimum}
Thorup, M.
\newblock Minimum k-way cuts via deterministic greedy tree packing.
\newblock  In Proceedings of the Fortieth Annual ACM Symposium on Theory of
  Computing, Budapest, Hungary, 26--29 August 2008, pp. 159--166.

\bibitem[Gupta {et~al.}(2019)Gupta, Lee, and Li]{gupta2019number}
Gupta, A.; Lee, E.; Li, J.
\newblock The number of minimum k-cuts: Improving the Karger-Stein bound.
\newblock  In Proceedings of the 51st Annual ACM SIGACT Symposium on Theory of
  Computing, Phoenix, AZ USA, 23--26 June 2019; pp. 229--240.

\bibitem[Kawarabayashi and Thorup(2011)]{kawarabayashi2011minimum}
Kawarabayashi, K.i.; Thorup, M.
\newblock The minimum k-way cut of bounded size is fixed-parameter tractable.
\newblock   In Proceedings of the 2011 IEEE 52nd Annual Symposium on Foundations 
of Computer Science, Palm Springs, CA, USA, 22--25 October 2011; pp. 
160--169.

\bibitem[Saran and Vazirani(1995)]{saran1995finding}
Saran, H.; Vazirani, V.V.
\newblock Finding k cuts within twice the optimal.
\newblock {\em SIAM J. Comput.} {\bf 1995}, {\em 24},~101--108.

\bibitem[Kawarabayashi and Lin(2020)]{kawarabayashi2020nearly}
Kawarabayashi, K.I.; Lin, B.
\newblock A nearly 5/3-approximation {FPT} Algorithm for Min-k-Cut.
\newblock In roceedings of the 2020 {ACM-SIAM} Symposium on Discrete Algorithms,
{SODA} 2020, Salt Lake City, UT, USA, 5--8 anuary 2020; pp. 990--999.

\bibitem[Lokshtanov {et~al.}(2020)Lokshtanov, Saurabh, and
  Surianarayanan]{lokshtanov2020parameterized}
Lokshtanov, D.; Saurabh, S.; Surianarayanan, V.
\newblock A Parameterized Approximation Scheme for Min $k$-Cut.
\newblock {\em arXiv} {\bf 2020}, arXiv:2005.00134.

\bibitem[Lund and Yannakakis(1993)]{lund1993approximation}
Lund, C.; Yannakakis, M.
\newblock The approximation of maximum subgraph problems.
\newblock  In \emph{International Colloquium on Automata, Languages, and Programming};
  Springer: Berlin, Germany, 1993; pp. 40--51.

\bibitem[Khot(2002)]{khot2002power}
Khot, S.
\newblock On the power of unique 2-prover 1-round games.
\newblock  In Proceedings of the Thiry-Fourth Annual ACM Symposium on Theory of
  Computing, Montreal, QC, Canada, day May 2002; pp. 767--775.

\bibitem[Heggernes {et~al.}(2011)Heggernes, Van’t~Hof, Jansen, Kratsch,
  and Villanger]{heggernes2011parameterized}
Heggernes, P.; Van’t~Hof, P.; Jansen, B.M.; Kratsch, S.; Villanger, Y.
\newblock Parameterized complexity of vertex deletion into perfect graph
  classes.
\newblock  In \emph{International Symposium on Fundamentals of Computation Theory};
  Springer: Berlin, Germany,  2011, pp. 240--251.

\bibitem[Fomin {et~al.}(2012)Fomin, Lokshtanov, Misra, and
  Saurabh]{fomin2012planar}
Fomin, F.V.; Lokshtanov, D.; Misra, N.; Saurabh, S.
\newblock Planar {F}-deletion: Approximation, kernelization and optimal {FPT}
  algorithms.
\newblock   In Proceedings of the 2012 IEEE 53rd Annual Symposium on Foundations 
of Computer Science, Brunswick, NJ, USA, 20--23 October 2012; pp. 470--479.

\bibitem[Marx(2010)]{marx2010chordal}
Marx, D.
\newblock Chordal deletion is fixed-parameter tractable.
\newblock {\em Algorithmica} {\bf 2010}, {\em 57},~747--768.

\bibitem[Cao and Marx(2014)]{cao2014interval}
Cao, Y.; Marx, D.
\newblock Interval deletion is fixed-parameter tractable.
\newblock  In Proceedings of the Twenty-Fifth Annual ACM-SIAM Symposium on
  Discrete Algorithms, Portland, Oregon, 5--7 January 2014; pp. 122--141.

\bibitem[Courcelle(1990)]{courcelle1990monadic}
Courcelle, B.
\newblock The monadic second-order logic of graphs. I. Recognizable sets of
  finite graphs.
\newblock {\em Inf. Comput.} {\bf 1990}, {\em 85},~12--75.

\bibitem[Bodlaender(2007)]{bodlaender2007treewidth}
Bodlaender, H.L.
\newblock Treewidth: Structure and algorithms.
\newblock  In \emph{International Colloquium on Structural Information and Communication
  Complexity}; Springer: Berlin, Germany,  2007; pp. 11--25.

\bibitem[Arnborg {et~al.}(1987)Arnborg, Corneil, and
  Proskurowski]{arnborg1987complexity}
Arnborg, S.; Corneil, D.G.; Proskurowski, A.
\newblock Complexity of finding embeddings in ak-tree.
\newblock {\em SIAM J. Algebr. Discret. Methods} {\bf 1987}, {\em
  8},~277--284.

\bibitem[Bodlaender(1996)]{bodlaender1996linear}
Bodlaender, H.L.
\newblock A linear-time algorithm for finding tree-decompositions of small
  treewidth.
\newblock {\em SIAM J. Comput.} {\bf 1996}, {\em 25},~1305--1317.

\bibitem[Bodlaender {et~al.}(2016)Bodlaender, Drange, Dregi, Fomin,
  Lokshtanov, and Pilipczuk]{bodlaender2016c}
Bodlaender, H.L.; Drange, P.G.; Dregi, M.S.; Fomin, F.V.; Lokshtanov, D.;
  Pilipczuk, M.
\newblock A $c^{k}n$ 5-Approximation Algorithm for Treewidth.
\newblock {\em SIAM J. Comput.} {\bf 2016}, {\em 45},~317--378.

\bibitem[Gupta {et~al.}(2019)Gupta, Lee, Li, Manurangsi, and
  W{\l}odarczyk]{gupta2019losing}
Gupta, A.; Lee, E.; Li, J.; Manurangsi, P.; W{\l}odarczyk, M.
\newblock Losing treewidth by separating subsets.
\newblock  In Proceedings of the Thirtieth Annual {ACM-SIAM} Symposium on 
Discrete Algorithms, {SODA}, San Diego, California, USA, 6--9 January 2019; pp. 1731--1749.

\bibitem[Jansen and Pieterse(2018)]{jansen2018polynomial}
Jansen, B.M.; Pieterse, A.
\newblock Polynomial Kernels for Hitting Forbidden Minors under Structural
  Parameterizations.
\newblock In Proceedings of 26th Annual European Symposium on Algorithms (ESA), Helsinki, Finland, 20--22 August 2018; pp. 48:1--48:15.

\bibitem[Donkers and Jansen(2019)]{donkers2019turing}
Donkers, H.; Jansen, B.M.
\newblock A Turing Kernelization Dichotomy for Structural Parameterizations of
  $F$-Minor-Free Deletion.
\newblock   In \emph{International Workshop on Graph-Theoretic Concepts in Computer
  Science}; Springer: Berlin, Germany,  2019; pp. 106--119.

\bibitem[Chekuri and Chuzhoy(2016)]{chekuri2016polynomial}
Chekuri, C.; Chuzhoy, J.
\newblock Polynomial bounds for the grid-minor theorem.
\newblock {\em J. ACM (JACM)} {\bf 2016}, {\em 63},~40.

\bibitem[Kawarabayashi and
  Sidiropoulos(2017)]{kawarabayashi2017polylogarithmic}
Kawarabayashi, K.i.; Sidiropoulos, A.
\newblock Polylogarithmic approximation for minimum planarization (almost).
\newblock   In Proceedings of the 2017 IEEE 58th Annual Symposium on Foundations of Computer Science
  (FOCS), Berkeley, CA, USA, 15--17 October 2017; pp. 779--788.

\bibitem[Agrawal {et~al.}(2018)Agrawal, Lokshtanov, Misra, Saurabh, and
  Zehavi]{agrawal2018polylogarithmic}
Agrawal, A.; Lokshtanov, D.; Misra, P.; Saurabh, S.; Zehavi, M.
\newblock Polylogarithmic Approximation Algorithms for Weighted-F-Deletion
  Problems.
\newblock  In Proceedings of Approximation, Randomization, and 
Combinatorial Optimization. Algorithms and Techniques (APPROX/RANDOM 2018), 
Princeton, NJ, {USA}, August 20-22, 2018; pp. 1:1--1:15.

\bibitem[Fiorini {et~al.}(2010)Fiorini, Joret, and
  Pietropaoli]{fiorini2010hitting}
Fiorini, S.; Joret, G.; Pietropaoli, U.
\newblock Hitting diamonds and growing cacti.
\newblock  In \emph{International Conference on Integer Programming and Combinatorial
  Optimization}; Springer: Berlin, Germany,  2010; pp. 191--204.

\bibitem[Marx and Pilipczuk(2014)]{marx2014everything}
Marx, D.; Pilipczuk, M.
\newblock Everything you always wanted to know about the parameterized
  complexity of Subgraph Isomorphism (but were afraid to ask).
\newblock  In Proceedings of 31st International Symposium on Theoretical Aspects 
of Computer Science (STACS), March 5-8, 2014, Lyon, France; pp. 542--553.

\bibitem[Ebenlendr {et~al.}()Ebenlendr, Kolman, and
  Sgall]{ebenlendrapproximation}
Ebenlendr, T.; Kolman, P.; Sgall, J.
\newblock An Approximation Algorithm for Bounded Degree Deletion.
\newblock {Preprint}.

\bibitem[Alon {et~al.}(1995)Alon, Yuster, and Zwick]{alon1995color}
Alon, N.; Yuster, R.; Zwick, U.
\newblock Color-coding.
\newblock {\em J. ACM} {\bf 1995}, {\em 42},~844--856.

\bibitem[Jansen and Pilipczuk(2018)]{jansen2018approximation}
Jansen, B.M.; Pilipczuk, M.
\newblock Approximation and kernelization for chordal vertex deletion.
\newblock {\em SIAM J. Discret. Math.} {\bf 2018}, {\em
  32},~2258--2301.

\bibitem[Cao and Sandeep(2017)]{cao2017minimum}
Cao, Y.; Sandeep, R.
\newblock Minimum fill-in: Inapproximability and almost tight lower bounds.
\newblock  In Proceedings of the Twenty-Eighth Annual ACM-SIAM Symposium on
  Discrete Algorithms, Barcelona, Spain, 16--19 January 2017; pp. 875--880.

\bibitem[Giannopoulou {et~al.}(2017)Giannopoulou, Pilipczuk, Raymond,
  Thilikos, and Wrochna]{giannopoulou2017linear}
Giannopoulou, A.C.; Pilipczuk, M.; Raymond, J.F.; Thilikos, D.M.; Wrochna, M.
\newblock Linear Kernels for Edge Deletion Problems to Immersion-Closed Graph
  Classes.
\newblock  In Proceedings of the 44th International Colloquium on Automata, 
Languages, and Programming {ICALP}, Warsaw, Poland, 10--14 July 2017; pp. 
57:1--57:15.

\bibitem[Bliznets {et~al.}(2018)Bliznets, Cygan, Komosa, and
  Pilipczuk]{bliznets2018hardness}
Bliznets, I.; Cygan, M.; Komosa, P.; Pilipczuk, M.
\newblock Hardness of approximation for $H$-free edge modification problems.
\newblock {\em ACM Trans. Comput. Theory (TOCT)} {\bf 2018}, {\em
  10},~9.

%
\bibitem[Chen {et~al.}(2007)Chen, Liu, and Lu]{chen2007directed}
Chen, J.; Liu, Y.; Lu, S.
\newblock Directed feedback vertex set problem is {FPT}.
\newblock In Proceedings of Structure Theory and {FPT} Algorithmics for Graphs, 
Digraphs and Hypergraphs, Dagstuhl, Germany, 8--13 July 2007.


\bibitem[Chen {et~al.}(2006)Chen, Kanj, and Xia]{chen2006improved}
Chen, J.; Kanj, I.A.; Xia, G.
\newblock Improved parameterized upper bounds for vertex cover.
\newblock  In International Symposium on Mathematical Foundations of Computer
  Science; Springer: Berlin, Germany,  2006; pp.~238--249.

\bibitem[Bourgeois {et~al.}(2009)Bourgeois, Escoffier, and
  Paschos]{BourgeoisEP09-moderatelyexp}
Bourgeois, N.; Escoffier, B.; Paschos, V.T.
\newblock Efficient Approximation of Combinatorial Problems by Moderately
  Exponential Algorithms.
\newblock In Proceedings of Algorithms and Data Structures, 11th International 
Symposium, {WADS}, Banff, Canada, 21--23 August 2009; pp. 
507--518, doi:10.1007/978-3-642-03367-4\_44.

\bibitem[Brankovic and Fernau(2010)]{brankovic2010combining}
Brankovic, L.; Fernau, H.
\newblock Combining Two Worlds: Parameterised Approximation for Vertex Cover.
\newblock In~{\em International Symposium on Algorithms and Computation}; 
Springer: Berlin/Heidelberg, German, {2010}; pp.~390--402.

\bibitem[Brankovic and Fernau(2013)]{BrankovicF13}
Brankovic, L.; Fernau, H.
\newblock A novel parameterised approximation algorithm for minimum vertex
  cover.
\newblock {\em Theor. Comput. Sci.} {\bf 2013}, {\em 511},~85--108, doi:10.1016/j.tcs.2012.12.003.

\bibitem[Bansal {et~al.}(2019)Bansal, Chalermsook, Laekhanukit, Nanongkai,
  and Nederlof]{BansalCLNN19}
Bansal, N.; Chalermsook, P.; Laekhanukit, B.; Nanongkai, D.; Nederlof, J.
\newblock New Tools and Connections for Exponential-Time Approximation.
\newblock {\em Algorithmica} {\bf 2019}, {\em 81},~3993--4009, doi:10.1007/s00453-018-0512-8.

\bibitem[Manurangsi and Trevisan(2018)]{MT18}
Manurangsi, P.; Trevisan, L.
\newblock Mildly Exponential Time Approximation Algorithms for Vertex Cover,
  Balanced Separator and Uniform Sparsest Cut.
\newblock   In Proceedings of the Approximation, Randomization, and Combinatorial Optimization.
  Algorithms and Techniques, {APPROX/RANDOM}, Princeton, NJ, {USA},  20--22 August 2018; pp. 20:1--20:17, doi:10.4230/LIPIcs.APPROX-RANDOM.2018.20.

\bibitem[Bar{-}Yehuda {et~al.}(2004)Bar{-}Yehuda, Bendel, Freund, and
  Rawitz]{Bar-YehudaBFR04}
Bar{-}Yehuda, R.; Bendel, K.; Freund, A.; Rawitz, D.
\newblock Local ratio: {A} unified framework for approxmation algrithms.
\newblock {\em {ACM} Comput. Surv.} {\bf 2004}, {\em 36},~422--463, doi:10.1145/1041680.1041683.

\bibitem[Escoffier {et~al.}(2015)Escoffier, Monnot, Paschos, and
  Xiao]{escoffier2015new}
Escoffier, B.; Monnot, J.; Paschos, V.T.; Xiao, M.
\newblock New results on polynomial inapproximability and fixed parameter
  approximability of edge dominating set.
\newblock {\em Theory Comput. Syst.} {\bf 2015}, {\em 56},~330--346.

\bibitem[Bonnet {et~al.}(2016)Bonnet, Paschos, and
  Sikora]{bonnet2016parameterized}
Bonnet, {\'E}.; Paschos, V.T.; Sikora, F.
\newblock Parameterized exact and approximation algorithms for maximum k-set
  cover and related satisfiability problems.
\newblock {\em RAIRO-Theor. Informatics Appl.} {\bf 2016}, {\em
  50},~227--240.

\bibitem[Arora {et~al.}(2015)Arora, Barak, and Steurer]{AroraBS15}
Arora, S.; Barak, B.; Steurer, D.
\newblock Subexponential Algorithms for Unique Games and Related Problems.
\newblock {\em J. {ACM}} {\bf 2015}, {\em 62},~42:1--42:25, doi:10.1145/2775105.

\bibitem[Barak {et~al.}(2011)Barak, Raghavendra, and Steurer]{BarakRS11}
Barak, B.; Raghavendra, P.; Steurer, D.
\newblock Rounding Semidefinite Programming Hierarchies via Global Correlation.
\newblock  In Proceedings of the {IEEE} 52nd Annual Symposium on Foundations of Computer Science,
  {FOCS},  Palm Springs, CA, USA, 22--25 October 2011; pp. 472--481, 
doi:10.1109/FOCS.2011.95.

\bibitem[Fernau(2012)]{Fernau12}
Fernau, H.
\newblock Saving on Phases: Parameterized Approximation for Total Vertex Cover.
\newblock  In Proceedings of the Combinatorial Algorithms, 23rd International Workshop, {IWOCA}, Tamil Nadu, India, 19--21 July 2012; Revised Selected Papers; pp.
  20--31, doi:10.1007/978-3-642-35926-2\_3.

\bibitem[Halperin(2002)]{Halperin02}
Halperin, E.
\newblock Improved Approximation Algorithms for the Vertex Cover Problem in
  Graphs and Hypergraphs.
\newblock {\em {SIAM} J. Comput.} {\bf 2002}, {\em 31},~1608--1623, doi:10.1137/S0097539700381097.

\bibitem[Impagliazzo {et~al.}(2001)Impagliazzo, Paturi, and Zane]{IPZ01}
Impagliazzo, R.; Paturi, R.; Zane, F.
\newblock Which Problems Have Strongly Exponential Complexity?
\newblock {\em J. Comput. Syst. Sci.} {\bf 2001}, {\em 63},~512--530, doi:10.1006/jcss.2001.1774.

\bibitem[Lampis(2011)]{DBLP:journals/ipl/Lampis11}
Lampis, M.
\newblock A kernel of order 2 k-c log k for vertex cover.
\newblock {\em Inf. Process. Lett.} {\bf 2011}, {\em 111},~1089--1091, doi:10.1016/j.ipl.2011.09.003.

\bibitem[Hochbaum and Shmoys(1986)]{hochbaum1986bottleneck}
Hochbaum, D.S.; Shmoys, D.B.
\newblock A unified approach to approximation algorithms for bottleneck
  problems.
\newblock {\em J.~ACM} {\bf 1986}, {\em 33},~533--550.

\bibitem[Brand {et~al.}(2018)Brand, Dell, and Husfeldt]{BDH18}
Brand, C.; Dell, H.; Husfeldt, T.
\newblock Extensor-coding.
\newblock  In Proceedings of the 50th Annual {ACM} {SIGACT} Symposium on Theory of
  Computing, {STOC} 2018, Los Angeles, CA, USA, 25--29 June 2018; pp.
  151--164, doi:10.1145/3188745.3188902.

\bibitem[Bj{\"{o}}rklund {et~al.}(2019)Bj{\"{o}}rklund, Lokshtanov, Saurabh,
  and Zehavi]{BLSZ19}
Bj{\"{o}}rklund, A.; Lokshtanov, D.; Saurabh, S.; Zehavi, M.
\newblock Approximate Counting of k-Paths: Deterministic and in Polynomial
  Space.
\newblock  In Proceedings of the 46th International Colloquium on Automata, Languages, and
  Programming, {ICALP}, Patras, Greece,  9--12 July 2019; pp.
  24:1--24:15, doi:10.4230/LIPIcs.ICALP.2019.24.

\bibitem[Pratt(2019)]{pratt2019waring}
Pratt, K.
\newblock Waring Rank, Parameterized and Exact Algorithms.
\newblock  In Proceedings of the 2019 IEEE 60th Annual Symposium on Foundations of Computer Science
  (FOCS), Baltimore, MI, USA, 9--12 November 2019; pp.~806--823.

\bibitem[Bj{\"{o}}rklund(2014)]{Bj14}
Bj{\"{o}}rklund, A.
\newblock Determinant Sums for Undirected Hamiltonicity.
\newblock {\em {SIAM} J. Comput.} {\bf 2014}, {\em 43},~280--299, doi:10.1137/110839229.

\bibitem[Bj{\"{o}}rklund {et~al.}(2017)Bj{\"{o}}rklund, Husfeldt, Kaski, and
  Koivisto]{BHKK17}
Bj{\"{o}}rklund, A.; Husfeldt, T.; Kaski, P.; Koivisto, M.
\newblock Narrow sieves for parameterized paths and packings.
\newblock {\em J.~Comput. Syst. Sci.} {\bf 2017}, {\em 87},~119--139, doi:10.1016/j.jcss.2017.03.003.

\bibitem[Marx(2013)]{Marx13}
Marx, D.
\newblock Completely inapproximable monotone and antimonotone parameterized
  problems.
\newblock {\em J. Comput. Syst. Sci.} {\bf 2013}, {\em 79},~144--151, doi:10.1016/j.jcss.2012.09.001.








\end{thebibliography}
\end{document}